\documentclass[11pt,a4paper]{article}
\pdfoutput=1
\usepackage{jheppub}

\usepackage{amsmath}
\usepackage{amssymb}
\usepackage{MnSymbol}
\usepackage{pifont}
\usepackage{graphicx}
\usepackage[squaren]{SIunits}

\usepackage{verbatim}   
\usepackage{multirow}   

\allowdisplaybreaks[1]

\renewcommand{\[}{\left\lbrack}
\renewcommand{\]}{\right\rbrack}

\newcommand{\cf}{cf.\,}
\newcommand{\refeq}[1]{Eq.~(\ref{eq:#1})}

\newcommand{\reffig}[1]{Fig.~\ref{fig:#1}}
\newcommand{\refsec}[1]{Section \ref{sec:#1}}
\newcommand{\reftab}[1]{Table~\ref{tab:#1}}
\newcommand{\order}[1]{\mathcal{O}\({#1}\)}

\newcommand\tabvsptop{\rule{0pt}{2.6ex}}
\newcommand\tabvspbot{\rule[-1.5ex]{0pt}{0pt}}

\newcommand{\GeV}{\,\mathrm{GeV}}
\newcommand{\MeV}{\,\mathrm{MeV}}

\newcommand{\wilson}[2][{}]{\mathcal{C}_{#2}^{\mathrm{#1}}}

\def \refeq#1{(\ref{#1})}
\def \refsec#1{Sec.~\ref{#1}}

\def \refapp#1{App.~\ref{#1}}
\def \reffig#1{Fig.~\ref{#1}}
\def \reftab#1{Tab.~\ref{#1}}

\def \aver#1{\langle #1 \rangle}

\def \order#1{ {\cal O} \left( #1 \right) }

\def \LogGamma{\textrm{LogGamma}}
\def \Amoroso{\textrm{Amoroso}}
\def \vecth{\vec{\theta}}
\def \vecnu{\vec{\nu}}

\def \cLdB1{{{\cal L}_{\Delta B = 1}^{\rm EW}}} 

\def \Op{{\cal O}}

\def \One{\leavevmode\hbox{\small1\kern-3.6pt\normalsize1}} 

\def \MeV{{\rm \; MeV}}
\def \GeV{{\rm \; GeV}}

\def \rmd{\mbox{d}}

\def \thl {{\theta_\ell}}
\def \thK {{\theta_{K}}}

\def \B0toK0ast{{ \bar{B}^0 \to K^{\ast 0}}}
\def \B0toK0mumu{{ \bar{B}^0 \to K^0 \bar{\mu} \mu}}
\def \barB0toKKpill{{ \bar{B}^0 \to \bar{K}^{\ast 0} (\to K^- \pi^+) \bar{l}l}}
\def \B0toKKpill{{ B^0 \to K^{\ast 0} (\to K^+ \pi^-) \bar{l}l}}

\def\D0{D\O}  \def\d0{D\O}

\title{Bayesian Fit of Exclusive $b \to s\,\bar\ell\ell$ Decays:
       The Standard Model Operator Basis}

\author[a]{Frederik Beaujean}
\author[b]{Christoph Bobeth}
\author[c]{Danny van Dyk}
\author[c]{Christian Wacker}

\affiliation[a]{
  Max-Planck-Institut f\"ur Physik,
  80805 M\"unchen, Germany
}
\affiliation[b]{
  Institute for Advanced Study \& Excellence Cluster Universe,
  Technische Universit\"at M\"unchen, 85748 Garching, Germany
}
\affiliation[c]{
  Institut f\"ur Physik, Technische Universit\"at Dortmund, 
  44221 Dortmund, Germany
}

\emailAdd{beaujean@mpp.mpg.de}
\emailAdd{bobeth@ph.tum.de}
\emailAdd{danny.dyk@tu-dortmund.de}
\emailAdd{christian.wacker@tu-dortmund.de}

\date{\today}

\abstract{We perform a model-independent fit of the short-distance couplings
  $\wilson[]{7,9,10}$ within the Standard Model set of $b\to s\gamma$ and $b\to
  s\bar\ell\ell$ operators. Our analysis of $B \to K^* \gamma$, $B \to K^{(*)}
  \bar\ell\ell$ and $B_s \to \bar\mu\mu$ decays is the first to harness the full
  power of the Bayesian approach: all major sources of theory uncertainty explicitly
  enter as nuisance parameters. Exploiting the latest measurements, the fit
  reveals a flipped-sign solution in addition to a Standard-Model-like
  solution for the couplings $\wilson[]{i}$. Each solution contains about half
  of the posterior probability, and both
  have nearly equal goodness of fit. The Standard Model prediction is close to
  the best-fit point. No New Physics contributions are necessary to describe the
  current data. Benefitting from the improved posterior knowledge of the nuisance
  parameters, we predict ranges for currently unmeasured, optimized observables
  in the angular distributions of $B\to K^*(\to K\pi)\,\bar\ell\ell$.
 }

\keywords{}

\preprint{DO-TH 12/14, EOS-2012-01, MPP-2012-66}

\begin{document}

%
%
\maketitle

%
%
\section{Introduction}

In the course of the last decade, rare $B$-meson decays were
discovered that are mediated at the parton level by flavor-changing
neutral-current (FCNC) transitions $b \to s\,\bar\ell\ell$ and $b\to
s\gamma$. They allow one to test Standard-Model (SM) predictions at
the loop level and to conduct searches for  indirect signals of physics
beyond the SM (BSM), providing strong constraints on the
corresponding fundamental parameters, especially in the quark flavor
sector.

The radiative FCNC decay $B\to K^*\gamma$ was first observed by the
CLEO collaboration at the Cornell Electron Storage Ring \cite{Coan:1999kh}.
The first-generation $B$ factory experiments BaBar \cite{Aubert:2004it,
  Aubert:2006vb, :2008ju, Aubert:2008gy, Aubert:2008ps, Aubert:2009ak,
  Sun:2012, Poireau:2012by} and Belle \cite{Nakao:2004th,
  Iwasaki:2005sy, Ushiroda:2006fi, :2009zv} observed rare radiative
and semileptonic FCNC decays of the $B$ meson with branching fractions
of $10^{-4}$ to $10^{-7}$.  They measured branching ratios and
spectral information for a set of inclusive $B \to X_s\,
\bar{\ell}\ell$ and exclusive $B \to K^{(*)}\,\bar{\ell}\ell$ ($\ell =
e, \mu$) decays.  Recently, additional exclusive $B \to
K^{(*)}\,\bar{\mu}\mu$ decay modes were measured by the hadron
collider experiments CDF \cite{Aaltonen:2011cn, Aaltonen:2011qs,
  Aaltonen:2011ja} at the Tevatron and LHCb \cite{Aaij:2011aa, LHCbnote08}
  at the Large Hadron Collider (LHC).  The
complete analyses of the full BaBar, Belle and CDF data sets is
expected to be published soon. LHCb is about to significantly
improve the accuracy of  measurements  of exclusive decays,
eventually dominating the other experiments in terms of collected
numbers of events by the end of 2012. The multipurpose LHC experiments
ATLAS and CMS are expected to perform similar searches.

In the last decade D{\O} \cite{Abazov:2010fs} and CDF \cite{Aaltonen:2011fi,
  CDFBsmumu:Moriond2012} significantly improved the upper bound on the branching
ratio of the very rare leptonic decay $B_s \to \bar\mu\mu$ by several orders of
magnitude. Currently, LHCb, CMS, and ATLAS continue this search
\cite{LHCb:2011ac, Aaij:2012ac, Chatrchyan:2011kr, Chatrchyan:2012rg,
  ATLAS:2012:moriond}, and a future discovery at SM rates of about $3
\times 10^{-9}$ is possible with sufficient luminosity \cite{Adeva:2009ny}.

Theory predictions of the inclusive decay $B\to X_s\,\bar\ell\ell$ have reached the
next-to-next-to-leading order (NNLO) \cite{Chetyrkin:1996vx, Misiak:2006zs,
  Misiak:2006ab, Bobeth:2003at, Huber:2005ig, Huber:2007vv,
  Greub:2008cy}. Contrary to exclusive decays, it only depends on
nonperturbative hadronic matrix elements at subleading order in the Heavy Quark
Expansion. But the current measurements of its branching fraction are still
very uncertain \cite{Aubert:2004it, Iwasaki:2005sy} and provide only very
limited spectral information in the dilepton invariant mass. This situation is
not likely to improve until the end of the run of the superflavor factories
Belle~II~\cite{Aushev:2010bq} and possibly SuperB \cite{O'Leary:2010af} around
the year 2020. Therefore, it is desirable to include exclusive decays in tests
of the SM and searches for BSM signals, especially in view of the high number of
events expected at LHCb.  Moreover, the angular distribution of the exclusive
decay $B \to {K}^* (\to K \pi) \, \bar\ell\ell$ with a 4-body final state offers
a multitude of optimized observables --- see \cite{Bobeth:2011qn} for a short
summary.

Both inclusive and exclusive decays are described by the $\Delta
B{=}1$ effective theory of electroweak interactions of the
SM and its extensions. They provide constraints on the effective
short-distance couplings which are known precisely in
the SM and are the main objects of interest due to their 
sensitivity to BSM effects at the electroweak scale.

Exclusive decays typically require the inclusion of final-state-specific
nonperturbative (hadronic) QCD effects, which complicate the extraction of the
short-distance couplings. Analyses are further complicated by the background
processes $b \to s + (\bar{q}q) \to s + \bar\ell\ell$ induced by 4-quark
operators $b\to s\,\bar{q}q$ ($q = u,d,s,c$). In particular, the narrow $J/\psi$
and $\psi'$ resonances constitute huge backgrounds to the
short-distance-dominated $b \to s \bar\ell\ell$ processes. Consequently,
theory predictions focus on the $q^2$ regions below and above both
resonances and are usually referred to as being in the low- or high-$q^2$
regions, or at large and low hadronic recoil. QCD factorization (QCDF)
\cite{Beneke:2001at, Bosch:2001gv, Beneke:2004dp, Khodjamirian:2010vf} or Soft
Collinear Effective Theory (SCET) \cite{Ali:2006ew, Lee:2006gs} is applied at
large recoil, $E \sim m_b$, of the $K^{(*)}$ system, typically in the range
$1$ GeV$^2 \lesssim q^2 \lesssim 6$ GeV$^2$.  The expansion parameter
$\lambda = \Lambda/E$ is of the order $\Lambda/m_b$, resulting in a double
expansion in $\lambda$ and the QCD coupling constant $\alpha_s$. Here $\Lambda$
denotes a scale associated with nonperturbative QCD dynamics, typically
$\lesssim 500$ MeV. The exact interpretation is process dependent and specific
to the expansion. An operator product expansion (OPE) of the 4-quark
contributions can be performed \cite{Grinstein:2004vb, Beylich:2011aq} at low
hadronic recoil for $q^2 \gtrsim (14-15)$ GeV$^2$ with the expansion parameter
$\lambda = \Lambda/\sqrt{q^2} \sim \Lambda/m_b$. Moreover, form-factor relations
\cite{Charles:1998dr, Beneke:2000wa, Isgur:1990kf, Grinstein:2002cz} from the
symmetries of QCD dynamics guide the construction of observables with
reduced hadronic uncertainties in both kinematic regions.

A large amount of phenomenological studies considering form-factor
symmetries have focused mainly on the decay $B\to K^{*} (\to K \pi)\,
\bar\ell\ell$. The angular distribution of its 4-body final state
\cite{Kruger:1999xa, Bobeth:2011qn} comprises an order of ten
observables that provide complementary information at low- and
high-$q^2$. Suitable combinations of these observables have also been
identified that have either $i)$ reduced hadronic uncertainties and
possibly higher sensitivities to BSM contributions
\cite{Kruger:2005ep, Bobeth:2008ij, Egede:2008uy,
  Altmannshofer:2008dz, Bharucha:2010bb, Egede:2010zc, Bobeth:2010wg,
  Alok:2010zd, Bobeth:2011gi, Becirevic:2011bp, Matias:2012xw}; or
$ii)$ become short-distance independent, allowing one to gain
information on form factors \cite{Bobeth:2010wg}. The decay $B\to K\,
\bar\ell\ell$ offers fewer observables, some of which are sensitive to
scalar and pseudoscalar \cite{Bobeth:2001sq, Bobeth:2007dw}
interactions; and in the high-$q^2$ region the same short-distance
dependence as in $B \to K^{*}\, \bar\ell\ell$ can be tested
\cite{Bobeth:2011nj}.

We perform a model-independent fit of the short-distance
couplings $ \wilson[]{7,9,10}$ to the experimental data
for exclusive $B \to K^* \gamma$, $B \to K^{(*)}\, \bar\ell\ell$, and
$B_s \to \bar\mu\mu$ decays considering the standard set of operators
described in more detail in \refsec{sec:effTh:expData}.  Improving on
our previous analyses \cite{Bobeth:2010wg, Bobeth:2011gi, Bobeth:2011nj},
we include the latest experimental data and add $B \to
K^* \gamma$ and $B_s\to \bar\mu\mu$, as collected in
\refsec{sec:effTh:expData}.  We go beyond \cite{DescotesGenon:2011yn}
 by including high-$q^2$ data for $B \to K^{*} \bar\ell\ell$
and the measurements of $B \to K \bar\ell\ell$; however, we do not
consider inclusive measurements. In comparison to the very recent
analysis \cite{Altmannshofer:2011gn}, we use updated data and include
$B \to K\, \bar\ell\ell$, but again do not consider the inclusive
decays. For our analysis and for all numerical evaluations we use EOS \cite{EOS:2011}.

Our analysis differs from all previous works \cite{Bobeth:2010wg,
  Bobeth:2011gi, DescotesGenon:2011yn, Altmannshofer:2011gn,
  Bobeth:2011nj} in its application of Bayesian inference with the
help of Monte-Carlo techniques to treat theory uncertainties in
the form of nuisance parameters.
The statistical treatment and the
choice of priors, as well as the determination of credibility
intervals, goodness of fit, pull values, and Bayes factors for model
comparison are described in \refsec{sec:stat:fit}. In
\refsec{sec:results}, we present the results of the fit for
short-distance couplings and discuss those nuisance parameters that
are affected by data. We present updated predictions based on the fit
results for unmeasured, optimized observables in the angular analysis
of $B\to K^{*} (\to K \pi)\, \bar\ell\ell$ in the given scenario. The
numerical input and details of the implementation of observables and
nuisance parameters are summarized in \refapp{app:num:input} and
\refapp{app:nuisance:pmrs}, respectively. We also present updated SM
predictions in \refapp{app:SM:predictions}. \refapp{app:distributions}
contains definitions of distributions that have been used to model priors.

%
%
\section{$\Delta B{=}1$ Decays: Conventions, Observables
  and Experimental Input \label{sec:effTh:expData}}

Rare $\Delta B{=}1$ decays are described by the effective theory of electroweak
interactions. In the SM, the short-distance effects of heavy degrees of freedom
of the order of the electroweak scale, due to the $W$ and $Z$ bosons and the $t$
quark, are contained in the Wilson coefficients $\wilson[]{i}$. The dynamics of
the light-quark $(q = u, d,s,c,b)$ and leptonic $(\ell = e, \mu, \tau)$ degrees
of freedom at the scale of the $b$ quark are described by operators $\Op_i$ of
dimension 5 and 6 for the parton transitions $b\to s + (\gamma,\, g,\,
\bar{q}q,\, \bar{\ell}\ell)$.  The SM Wilson coefficients $\wilson[]{i}$ ($i =
1,\ldots , 10$) are presently known up to NNLO (and partially NNNLO) in
QCD \cite{Chetyrkin:1996vx, Bobeth:1999mk, Misiak:2004ew, Gorbahn:2004my,
  Gorbahn:2005sa, Czakon:2006ss} and NLO in QED \cite{Buras:1999st,
  Gambino:2001au, Bobeth:2003at, Huber:2005ig}.  This includes the
renormalization group evolution (RGE) from the electroweak scale $\mu_W \sim
M_W$ down to $\mu_b \sim m_b$, the $b$-quark mass, which resums sizable
logarithmic corrections to all orders in the QCD coupling $\alpha_s$. Beyond the
SM, the effects due to new heavy degrees of freedom can be included
systematically as additional contributions to the short-distance couplings,
possibly giving rise to operators beyond the SM with a different chiral nature
or additional light degrees of freedom.

%
\subsection{$\Delta B{=}1$ Effective Theory}

The effective Hamiltonian of $\Delta
B{=}1$ decays reads \cite{Chetyrkin:1996vx, Bobeth:1999mk}
\begin{align}
  \label{eq:Heff}
  {\cal{H}}_{\rm eff} & = -\frac{4 G_F}{\sqrt{2}} V_{tb}^{} V_{ts}^\ast
     \left( {\cal{H}}_{\rm eff}^{(t)} + \hat{\lambda}_u {\cal{H}}_{\rm eff}^{(u)} \right), &
    \hat{\lambda}_u & = V_{ub}^{} V_{us}^\ast/V_{tb}^{} V_{ts}^\ast ,
\end{align}
\begin{align}
  \label{eq:Heff:parts}
  {\cal{H}}_{\rm eff}^{(t)} & =
    \wilson[]{1} \Op_1^c + \wilson[]{2} \Op_2^c + \sum_{3 \leq i} \wilson[]{i} \Op_i^{}, &
  {\cal{H}}_{\rm eff}^{(u)} & =
    \wilson[]{1} (\Op_1^c - \Op_1^u) + \wilson[]{2} (\Op_2^c - \Op_2^u)
\end{align}
where $V_{ij}$ denotes an element of the Cabibbo-Kobayashi-Maskawa
(CKM) quark-mixing matrix, and its unitarity relations have been used.
Above and throughout, the Wilson coefficients are understood to be
$\overline{\rm MS}$ renormalized and taken at the reference scale $\mu
= 4.2$ GeV.\footnote{Note that the actual low-energy renormalization
  scale $\mu_b$ might differ from $\mu$, and the corresponding RGE effect
  $\wilson{i}(\mu_b) = U(\mu_b, \mu)_{ij} \wilson{j}(\mu)$ should be
  taken into account in renormalization-scale variations when
  determining the related uncertainty. Throughout we use a central
  value $\mu_b = \mu$.}  In the SM, all CP-violating effects in $b\to
s$ transitions are governed by $\hat{\lambda}_u$ which is doubly
Cabibbo suppressed and leads to tiny CP violation. The operators due
to $b\to s\,\bar{q}q$ transitions are the current-current operators
$\Op_{1,2}^{u,c}$, the QCD-penguin operators for $i=3,4,5,6$, and the
chromomagnetic dipole operator $i = 8$. Effects of QED-penguin
operators are neglected since they are small for the decays under
consideration. Following the studies of QED corrections to the
inclusive decay, we choose the QED coupling $\alpha_e$ at the low
scale $\mu_b$, capturing most effects of QED corrections
\cite{Bobeth:2003at,Huber:2005ig} and removing the main uncertainty due to the choice
of the renormalization scheme at LO in QED.  The electromagnetic
dipole operator
\begin{align}
  \label{eq:SM:ops}
  \Op_7 & =
    \frac{e}{(4\pi)^2} m_b \[\bar{s} \sigma_{\mu\nu} P_R b\] F^{\mu\nu}
\end{align}
governs $b \to s \gamma$ transitions. The semileptonic operators
\begin{align}
  \Op_9 & =
    \frac{\alpha_e}{4\pi} \[\bar{s} \gamma_\mu P_L b\] \[\bar{\ell} \gamma^\mu \ell\], &
  \Op_{10} & =
    \frac{\alpha_e}{4\pi} \[\bar{s} \gamma_\mu P_L b\] \[\bar{\ell} \gamma^\mu \gamma_5 \ell\]
\end{align}
govern $b \to s\,\bar\ell\ell$ transitions, in combination with less important
contributions from ${\cal O}_7$.

In this study, we fit the Wilson coefficients $\wilson[]{7,9,10}$ at
the reference scale $\mu = 4.2$ GeV using experimental data. We assume
them to be real valued and refrain from the frequently used
decomposition into SM and BSM contributions $\wilson[]{i} =
\wilson[\rm SM]{i} + \wilson[\rm BSM]{i}$.  The Wilson coefficients $i
\leq 6$ and $i = 8$ contribute numerically only at the subleading
level in the observables of interest and are fixed to the
corresponding SM values at NNLO in QCD.

Whereas our scenario corresponds to the SM or extensions that do not
introduce new CP violation nor new operators, more general scenarios
have been investigated in the literature. The extension of this
scenario with complex Wilson coefficients --- i.e., CP violation beyond
the SM --- but no additional operators was studied in
\cite{Bobeth:2011gi, Altmannshofer:2011gn, Bobeth:2011nj}. An extended
operator basis with real Wilson coefficients, including
chirality-flipped operators $i = 7', 9', 10'$, has been analyzed in
\cite{DescotesGenon:2011yn, Altmannshofer:2011gn}. Finally, the
combination of both can be found in \cite{Altmannshofer:2011gn}.
Beyond these scenarios, it is conceivable that scalar, pseudoscalar,
and tensor $b\to s\,\bar\ell\ell$ ($\ell = e, \mu$) operators can also
contribute to the observables under consideration
\cite{Altmannshofer:2008dz, Alok:2010zd, Bobeth:2007dw}.  Beyond such
direct contributions, additional ones can arise due to operator mixing
from $b\to s\,\bar{q}q$ operators \cite{Borzumati:1999qt,
  Hiller:2003js} as well as $b\to s\,\bar\tau\tau$
\cite{Bobeth:2011st}.

%
\subsection{Observables and Experimental Input \label{sec:expData}}

Phenomenological studies have analyzed and proposed a large number of CP-symmetric
and -asymmetric observables. We summarize observables that either have
been measured and therefore impose constraints on the Wilson coefficients or
observables which are $i)$ sensitive to the operators of interest and $ii)$ exhibit
a reduced hadronic uncertainty. For the latter, we compute the ranges that are
still allowed by the data within the chosen scenario. Throughout, experimental
numbers refer to CP-averaged quantities.

\subsubsection{$B \to K^* \gamma$ \label{sec:BKstarGamma:constr}}

\begin{table}
\centering
\begin{tabular}{c|ccc}
\hline \hline
  observable & value & correlation &
\\
\hline
\multirow{3}{*}{$\mathcal{B} \times 10^{5}$}
\tabvsptop
  & $4.55^{+0.72}_{-0.68}\pm0.34$    & &  \cite{Coan:1999kh}
\\
  & $4.47\pm 0.10 \pm 0.16$          & &  \cite{Aubert:2009ak}
\\
\tabvspbot
  & $4.01\pm 0.21\pm 0.17$           & &  \cite{Nakao:2004th}
\\
\hline
\tabvsptop
  $S$
  & $-0.03\pm 0.29 \pm 0.03$         & \multirow{2}{*}{$5\%$} &  \multirow{2}{*}{\cite{Aubert:2008gy}}
\\
  $C$
  & $-0.14\pm 0.16 \pm 0.03$         & &
\tabvspbot
\\
\tabvsptop
  $S$
  & $-0.32^{+0.36}_{-0.33}\pm 0.05$  & \multirow{2}{*}{$8\%$} &  \multirow{2}{*}{\cite{Ushiroda:2006fi}}
\\
  $C$
  & $+0.20\pm 0.24\pm 0.05$          & &
\tabvspbot
\\
\hline \hline
\end{tabular}
\caption{\label{tab:BKgam:expData} Experimental results for CP-averaged
  $B^0 \to K^{*0} \gamma$ 
  observables:  branching fraction $\mathcal{B}$ (CLEO, BaBar, Belle) and
  time-dependent CP asymmetries $S$ and $C$ (BaBar, Belle), including their
  correlations. Throughout, statistical errors are given first, followed by
  the systematic errors.
}
\end{table}

For $B \to K^* \gamma$, several observables have been measured, such as the
branching ratio $\mathcal{B}$, the time-dependent CP asymmetries $S$ and $C$,
and the isospin asymmetry $A_I$. Their impact on the scenario of real
$\wilson[]{7, 7'}$ has been studied in \cite{DescotesGenon:2011yn} using the
inclusive $\mathcal{B}$ instead of the exclusive one. The measurement of $B_s
\to \phi\, \gamma$ can provide similar information and allows a third CP
asymmetry $H$ to be studied \cite{Muheim:2008vu}. The angular distribution in
the decay $B\to K_1(1270) \gamma \to (K\pi\pi) \gamma$ is sensitive to the
photon polarization and tests $\wilson[]{7, 7'}$; however, the feasibility of an
analysis remains uncertain \cite{Kou:2010kn, Tayduganov:2011ui}. In our analysis
we use  $\mathcal{B}$ and the CP asymmetries $S$ and $C$ of $B \to K^*
\gamma$ with their measurements and correlations compiled in
\reftab{tab:BKgam:expData}, and follow the calculations outlined in
\cite{Beneke:2004dp, Feldmann:2002iw}.  More details on the numerical input and
nuisance parameters can be found in \refapp{app:num:input} and
\refapp{app:nuisance:pmrs}.

\subsubsection{$B \to K\, \bar\ell\ell$ \label{sec:BKll:constr}}

In principle, the exclusive decay $B \to K\, \bar\ell\ell$ with a 3-body final
state offers three (CP-averaged) observables: the branching ratio
$\mathcal{B}(q^2)$, the lepton forward-backward asymmetry $A_{\rm FB}(q^2)$, and
the flat term $F_H(q^2)$. The latter two arise in the double-differential decay
rate when differentiating with respect to the dilepton invariant mass $q^2$ and
$\cos\thl$ \cite{Bobeth:2007dw}
\begin{align}
  \frac{1}{\rmd\Gamma/\rmd q^2} \frac{\rmd^2\Gamma}{\rmd q^2\, \rmd\!\cos\thl} &
  = \frac{3}{4} \left( 1 - F_H \right) \sin^2\!\thl + \frac{1}{2}
  F_H + A_{\rm FB} \cos\thl,
\end{align}
where $\thl$ is the angle between the 3-momenta of the negatively
charged lepton and the $\bar{B}$ meson in the dilepton center of mass
system. Two further interesting observables are the rate CP asymmetry
$A_{\rm CP}$ and the ratio of decay rates for the $\ell{=}e$ and
$\ell{=}\mu$ modes $R_K$. $A_{\rm FB}$ is nonzero only in the presence
of scalar or tensor BSM contributions, and $F_H$ is helicity suppressed
by $m_\ell/\sqrt{q^2}$ in the scenario under consideration, but is
sensitive to scalar and tensor contributions \cite{Bobeth:2001sq,
  Bobeth:2007dw}.  In view of this, available measurements of $A_{\rm
  FB}$, $F_H$, and $R_K$ are not considered, and we include only the
$\mathcal{B}$ measurements for one low-$q^2$ and two high-$q^2$ bins
as listed in \reftab{tab:BKll:expData}. Our theory evaluation at
low and high $q^2$ follows \cite{Bobeth:2007dw,
  Bobeth:2011nj}. Details concerning numerical input and nuisance
parameters are given in \refapp{app:num:input} and
\refapp{app:nuisance:pmrs}.

\begin{table}
\centering
\begin{tabular}{c|ccc c}
\hline \hline
  \tabvsptop \tabvspbot
  $q^2$-bin $[$GeV$^2]$ & $[1.00,\, 6.00]$ & $[14.18,\, 16.00]$ & $[> 16.00]$ &
\\
\hline
\multirow{3}{*}{$\aver{\mathcal{B}} \times 10^{7}$}
  & \tabvsptop \tabvspbot
    $2.05^{+0.53}_{-0.48} \pm 0.07$
  & $1.46^{+0.41}_{-0.36} \pm 0.06$
  & $1.02^{+0.47}_{-0.42} \pm 0.06$
  & \cite{Sun:2012}
\\[0.1cm]
  & \tabvsptop \tabvspbot
    $1.36^{+0.23}_{-0.21} \pm 0.08$
  & $0.38^{+0.19}_{-0.12} \pm 0.02$
  & $0.98^{+0.20}_{-0.18} \pm 0.06$
  & \cite{:2009zv}
\\[0.1cm]
  & \tabvsptop \tabvspbot
    $1.41 \pm 0.20 \pm 0.09$
  & $0.53 \pm 0.10 \pm 0.03$
  & $0.48 \pm 0.11 \pm 0.03$
  & \cite{Aaltonen:2011qs}
\\
\hline \hline
\end{tabular}
\caption{\label{tab:BKll:expData} Experimental results for the CP-averaged
  branching fraction of charged $B^\pm \to K^\pm \bar\mu\mu$ decays from
  BaBar \cite{Sun:2012}, Belle \cite{:2009zv}, and CDF \cite{Aaltonen:2011qs}, 
  integrated in bins of $q^2$. The publicly available
  results of BaBar and Belle are unknown admixtures of charged and neutral
  $B$ decays. The difference between interpreting the data as coming from
  either purely charged
  or purely neutral $B$ decays is negligible \cite{Bobeth:2011nj}.
}
\end{table}

\subsubsection{$B \to K^{*}(\to K \pi)\, \bar\ell\ell$
  \label{sec:BKstarll:constr}}

Phenomenologically, the angular analysis of the 4-body final state
$B \to K^{*}(\to K \pi)\, \bar\ell\ell$ offers a large set of ``angular''
observables
\begin{align}
  \aver{J_i}\,{[q^2_\mathrm{min},\, q^2_\mathrm{max}]} & =
    \int_{q^2_\mathrm{min}}^{q^2_\mathrm{max}} \rmd q^2 J_i(q^2)\,, &
  i & = 1,\ldots, 9\,,
\end{align}
where the boundaries of the $q^2$ bin (throughout in units of GeV$^2$)
will not be explicitly shown when they are not relevant. Throughout,
we assume that the experimental measurements are given for a certain
$q^2$ binning that requires $q^2$ integration for theory
predictions. Consequently, whenever a $q^2$-dependent
observable $X(q^2)$ is defined in a functional form $X(q^2) =
f[J_i](q^2)$ in terms of the angular observables, we define the
corresponding $q^2$-integrated quantity as follows \cite{Bobeth:2010wg}
\begin{align}
  \label{eq:aver:def}
  \aver{X} & = f\left[\aver{J_i}\right]\,.
\end{align}

\begin{table}
\centering
\begin{tabular}{c|ccc c}
\hline \hline
  \tabvsptop \tabvspbot
  $q^2$-bin $[$GeV$^2]$  & $[1.00,\, 6.00]$ & $[14.18,\, 16.00]$ & $[> 16.00]$ &
\\
\hline
\multirow{4}{*}{$\aver{\mathcal{B}} \times 10^{7}$}
  & \tabvsptop $2.05 ^{+0.53}_{-0.48} \pm 0.07$
  & $1.46 ^{+0.41}_{-0.36} \pm 0.06$
  & $1.02 ^{+0.47}_{-0.42} \pm 0.06$
  & \cite{Sun:2012}
\\[0.1cm]
  & $1.49^{+0.45}_{-0.40} \pm 0.12$
  & $1.05^{+0.29}_{-0.26} \pm 0.08$
  & $2.04^{+0.27}_{-0.24} \pm 0.16$
  & \cite{:2009zv}
\\[0.1cm]
  & $1.42 \pm 0.41 \pm 0.08$
  & $1.34 \pm 0.26 \pm 0.08$
  & $0.97 \pm 0.26 \pm 0.06$
  & \cite{Aaltonen:2011qs}
\\[0.1cm]
  & \tabvspbot $2.10 \pm 0.20 \pm 0.20 $
  & $1.0738 \pm 0.1274 \pm 0.0728$
  & $1.32 \pm 0.15 \pm 0.09$
  & \cite{LHCbnote08}
\\
\hline
\multirow{4}{*}{$\aver{A_{\rm FB}}$}
  & \tabvsptop $-0.02 ^{+0.18}_{-0.16} \pm 0.07$
  & $-0.31 ^{+0.19}_{-0.11} \pm 0.13$
  & $-0.34 ^{+0.26}_{-0.17} \pm 0.08$
  & \cite{Poireau:2012by}
\\[0.1cm]
  & $-0.26^{+0.30}_{-0.27} \pm 0.07$
  & $-0.70^{+0.22}_{-0.16} \pm 0.10$
  & $-0.66^{+0.16}_{-0.11} \pm 0.04$
  & \cite{:2009zv}
\\[0.1cm]
  & $-0.36^{+0.28}_{-0.46} \pm 0.11$
  & $-0.40^{+0.21}_{-0.18} \pm 0.07$
  & $-0.66^{+0.26}_{-0.18} \pm 0.19$
  & \cite{Aaltonen:2011ja}
\\[0.1cm]
  & \tabvspbot $0.18 \pm 0.06 ^{+0.02}_{-0.01}$
  & $-0.49^{+0.06}_{-0.04}  \, ^{+0.05}_{-0.02}$
  & $-0.30 \pm 0.07  \, ^{+0.01}_{-0.04}$
  & \cite{LHCbnote08}
\\
\hline
\multirow{4}{*}{$\aver{F_{L}}$}
  & \tabvsptop $0.47 \pm 0.13 \pm 0.04$
  & $0.42 ^{+0.12}_{-0.16} \pm 0.11$
  & $0.47 ^{+0.18}_{-0.20} \pm 0.13$
  & \cite{Poireau:2012by}
\\[0.1cm]
  & $0.67 \pm 0.23 \pm 0.05$
  & $-0.15^{+0.27}_{-0.23} \pm 0.07$
  & $0.12^{+0.15}_{-0.13} \pm 0.02$
  & \cite{:2009zv}
\\[0.1cm]
  & $0.60^{+0.21}_{-0.23} \pm 0.09$
  & $0.32 \pm 0.14 \pm 0.03$
  & $0.16^{+0.22}_{-0.18} \pm 0.06$
  & \cite{Aaltonen:2011ja}
\\[0.1cm]
  & \tabvspbot $0.66 \pm 0.06 ^{+0.04}_{-0.03} $
  & $0.35^{+0.07}_{-0.06} \, ^{+0.07}_{-0.02}$
  & $0.37^{+0.06}_{-0.07} \, ^{+0.03}_{-0.04}$
  & \cite{LHCbnote08}
\\
\hline
\tabvsptop \tabvspbot
$\aver{A_T^{(2)}}$
  & $1.6^{+1.8}_{-1.9} \pm 2.2$
  & $0.4 \pm 0.8 \pm 0.2$
  & $-0.9 \pm 0.8 \pm 0.4$
  & \cite{Aaltonen:2011ja}
\\
\hline
\tabvsptop \tabvspbot
$\aver{2S_3}$
  & $0.10^{+0.15}_{-0.16} \, ^{+0.02}_{-0.01}$
  & $0.04 ^{+0.15}_{-0.19} \, ^{+0.04}_{-0.02}$
  & $-0.47 ^{+0.21}_{-0.10} \, ^{+0.03}_{-0.05} $
  & \cite{LHCbnote08}
\\
\hline \hline
\end{tabular}
\caption{\label{tab:BKstll:expData} Experimental results used for
  $B^0 \to K^{*0} \bar\ell\ell$ for the CP-averaged branching fraction
  $\mathcal{B}$, lepton forward-backward asymmetry $A_{\rm FB}$, longitudinal
  $K^*$-polarization fraction $F_L$, the transversity observable $A_T^{(2)}$
  and $(2 S_3)$ from BaBar \cite{Sun:2012,Poireau:2012by}, Belle \cite{:2009zv},
  CDF \cite{Aaltonen:2011qs, Aaltonen:2011ja}, and LHCb \cite{LHCbnote08}.
  Note that the sign of $A_{\rm FB}$ is reversed due to a different definition
  of $\thl$ in the experimental community.
}
\end{table}

The angular observables $\aver{J_i}$ are defined in the 3-fold angular distribution
\begin{align}
  \frac{32 \pi}{9} & \frac{\rmd^3\aver{\Gamma}}{\rmd\!\cos\thl\, \rmd\!\cos\thK\, \rmd\phi}
  =
  \label{eq:3fold-ang-dist}
\\[0.2cm]
   &\big[\aver{J_{1s}} + \aver{J_{2s}} \cos 2\thl + \aver{J_{6s}} \cos\thl \big] \sin^2\!\thK +
  \big[\aver{J_{1c}} + \aver{J_{2c}} \cos 2\thl + \aver{J_{6c}} \cos\thl \big] \cos^2\!\thK
  \nonumber
\\[0.2cm]
  & + \aver{J_3} \sin^2\!\thK \sin^2\!\thl \cos 2\phi
  + \aver{J_4} \sin 2\thK \sin 2\thl \cos\phi
  + \aver{J_5} \sin 2\thK \sin\thl \cos\phi
    \nonumber
\\[0.2cm]
  & + \aver{J_7} \sin 2\thK \sin\thl \sin\phi
  + \aver{J_8} \sin 2\thK \sin 2\thl \sin\phi
  + \aver{J_9} \sin^2\!\thK \sin^2\!\thl \sin 2\phi,
    \nonumber
\end{align}
which accounts for all possible $(\bar{s} \ldots b)(\bar{\ell}\ldots
\ell)$ Lorentz structures of chirality-flipped, scalar, pseudoscalar,
and tensor operators \cite{Altmannshofer:2008dz, Alok:2010zd}. 
The angles are: $i)$ $\thl$ between the $\ell^-$ and the $K^*$ direction
of flight in the $(\ell^+ \ell^-)$ center of mass, $ii)$ $\thK$ between the $K$ and
the $K^*$ in the $(K\pi)$ center of mass and $iii)$ $\phi$ between the $(\ell^+ \ell^-)$
and $(K\pi)$ decay planes \cite{Kruger:2005ep}.
Here the normalization of the $J_i$ from \cite{Kruger:2005ep, Egede:2008uy,
  Altmannshofer:2008dz} is used and differs by a factor 4/3 from
\cite{Bobeth:2008ij, Bobeth:2010wg, Bobeth:2011gi}.  The following
simplifications arise in the limit $m_\ell \to 0$ and in the absence of
scalar and tensor operators \cite{Altmannshofer:2008dz, Alok:2010zd}:
\begin{align}
  J_{1s} & = 3\, J_{2s}, &
  J_{1c} & = - J_{2c}, &
  J_{6c} & = 0,
  \label{eq:ml-zero-limit}
\end{align}
and a fourth more complicated relation \cite{Egede:2010zc}. It is straightforward
to obtain the decay rate and the three single-differential angular distributions
from \refeq{eq:3fold-ang-dist}
\begin{align}
  \label{eq:Gint}
  \aver{\Gamma} & =
    \frac{3}{4} \big[2 \aver{J_{1s}} + \aver{J_{1c}} \big]
  - \frac{1}{4} \big[2 \aver{J_{2s}} + \aver{J_{2c}} \big], &
\\[0.2cm]
  \label{eq:dG:dphi}
  \frac{\rmd\aver{\Gamma}}{\rmd\phi} & =
    \frac{1}{2\pi} \Big[
      \aver{\Gamma} + \aver{J_3} \cos 2\phi + \aver{J_9} \sin 2\phi
   \Big], &
\\[0.2cm]
  \label{eq:dG:dcosthK}
  \frac{\rmd\aver{\Gamma}}{\rmd\!\cos\thK} & =
    \frac{3}{8} \Big[
       \big(3 \aver{J_{1s}} - \aver{J_{2s}} \big) \sin^2\!\thK
     + \big(3 \aver{J_{1c}} - \aver{J_{2c}} \big) \cos^2\!\thK
    \Big], &
\\[0.2cm]
  \label{eq:dG:dcosthL}
  \frac{\rmd\aver{\Gamma}}{\rmd\!\cos\thl} & =
    \frac{3}{8} \Big[
      2 \aver{J_{1s}} + \aver{J_{1c}}
    + \big(2 \aver{J_{6s}} + \aver{J_{6c}} \big) \cos\thl
    + \big(2 \aver{J_{2s}} + \aver{J_{2c}} \big) \cos 2\thl
    \Big].
\end{align}

The branching ratio $\aver{\mathcal{B}}$, the
lepton forward-backward asymmetry $\aver{A_{\rm FB}}$, and the longitudinal
$K^*$-polarization fraction $\aver{F_L}$
\begin{align}
  \aver{\mathcal{B}}& = \tau_{B^0} \aver{\Gamma}, &
  \aver{A_{\rm FB}} & = \frac{3}{8} \frac{2 \aver{J_{6s}} + \aver{J_{6c}}}{\aver{\Gamma}}, &
  \aver{F_L} & = \frac{3 \aver{J_{1c}} - \aver{J_{2c}}}{4 \aver{\Gamma}},
\end{align}
have been measured by BaBar \cite{Sun:2012, Poireau:2012by}, Belle \cite{:2009zv},
CDF \cite{Aaltonen:2011qs, Aaltonen:2011ja}, and LHCb \cite{LHCbnote08}. The
angular observable $\aver{A_T^{(2)}}$ \cite{Kruger:2005ep} has been
measured by CDF \cite{Aaltonen:2011ja}; and $\aver{S_3}$
\cite{Altmannshofer:2008dz} has been determined by LHCb \cite{LHCbnote08}:
\begin{align}
  \aver{A_T^{(2)}} & = \frac{\aver{J_3}}{2 \aver{J_{2s}}}, &
  \aver{S_3} & = \frac{\aver{J_3}}{\aver{\Gamma}}.
\end{align}
 All
are summarized in \reftab{tab:BKstll:expData}.  Note that $\aver{A_{\rm FB}}$
and $\aver{F_L}$ are determined from a combined fit to the single-differential
angular distributions
\begin{align}
  \frac{1}{\aver{\Gamma}} \frac{\rmd\aver{\Gamma}}{\rmd\!\cos\thK} & =
    \frac{3}{4} \big[1 - \aver{F_L}\big]\, \sin^2\!\thK
    + \frac{3}{2} \aver{F_L} \cos^2\!\thK,
\\[0.3cm]
  \label{eq:dG:dcosthL:approx}
  \frac{1}{\aver{\Gamma}} \frac{\rmd\aver{\Gamma}}{\rmd\!\cos\thl} & =
    \frac{3}{4} \aver{F_L} \sin^2\!\theta_\ell
   + \frac{3}{8} \big[1 - \aver{F_L}\big]\, (1 + \cos^2\!\thl)
   + \aver{A_{\rm FB}} \cos\thl.
\intertext{The observables $\aver{A_T^{(2)}}$ and
$\aver{A_\mathrm{im}} = \aver{J_9}/\aver{\Gamma}$ are determined from}
  \label{eq:dG:dphi:approx}
  \frac{2\pi}{\aver{\Gamma}} \frac{\rmd\aver{\Gamma}}{\rmd\phi} & =
  1 + \frac{1}{2}\big[1 - \aver{F_L} \big]\, \aver{A_T^{(2)}} \cos 2\phi
    + \aver{A_\mathrm{im}} \sin 2\phi,
\end{align}
implying $S_3 = (1 - \aver{F_L})\, \aver{A_T^{(2)}}/2$. Note that
\refeq{eq:dG:dcosthL:approx} and \refeq{eq:dG:dphi:approx} are based on
the approximation \refeq{eq:ml-zero-limit}, which is well justified
within our scenario.

The angular observables $\aver{J_i}$ and the branching ratio
$\aver{\mathcal{B}}$ are proportional to the square of hadronic form
factors, the main source of theory uncertainty. In normalized
combinations of the angular observables, for example $A_{\rm FB}$ and
$F_L$, these uncertainties partially cancel. The most prominent
example is the position $q_0^2[A_{\rm FB}]$ of the zero crossing of
$A_{\rm FB}$. A number of suitable combinations have been found for
both low- and high-$q^2$ regions. At low $q^2$ \cite{Kruger:2005ep,
  Egede:2008uy, Egede:2010zc, Becirevic:2011bp, Matias:2012xw}
\begin{align}
  \aver{A_T^{(2)}} & = \frac{\aver{J_3}}{2\, \aver{J_{2s}}}, &
  \aver{A_T^{(\rm re)}} & = \frac{\aver{J_{6s}}}{4\, \aver{J_{2s}}}, &
  \aver{A_T^{(\rm im)}} & = \frac{\aver{J_9}}{2\, \aver{J_{2s}}},
\end{align}
\begin{align}
  \label{eq:def:AT34}
  \aver{A_T^{(3)}} & = \sqrt{\frac{\aver{2\, J_4}^2 + \aver{J_7}^2}
                                  {-2\, \aver{J_{2c}}^{}\, \aver{2\, J_{2s} + J_3}}^{}}, &
  \aver{A_T^{(4)}} & = \sqrt{\frac{\aver{J_5}^2 + \aver{2\, J_8}^2}
                                  {\aver{2\, J_4}^2 + \aver{J_7}^2}},
\end{align}
\begin{align}
  \aver{A_T^{(5)}} & =
    \frac{\sqrt{\aver{4 J_{2s}}^2 - \aver{J_{6s}}^2 - 4 \big(\aver{J_{3}}^2 + \aver{J_{9}}^2\big)}}
         {8\, \aver{J_{2s}}};
\end{align}
whereas at high $q^2$ \cite{Bobeth:2010wg}
\begin{align}
  \aver{H_T^{(1)}} & = \frac{\sqrt{2}\,\aver{J_{4}}}{\sqrt{-2\, \aver{J_{2c}}^{} \aver{2 J_{2s} - J_3}}},
\end{align}
\begin{align}
  \aver{H_T^{(2)}} & = \frac{\aver{J_5}}{\sqrt{-2\, \aver{J_{2c}}^{} \aver{2 J_{2s} + J_3}}}, &
  \aver{H_T^{(3)}} & = \frac{\aver{J_{6s}}}{2\, \sqrt{\aver{2\, J_{2s}}^2 - \aver{J_3}^2}}.
\end{align}
For brevity, factors of $\beta_\ell = \sqrt{1 - 4\, m_\ell^2/q^2}$
have been set to unity, since they are negligible in our scenario for
the considered range $q^2 \gtrsim 1$ GeV$^2$. Recently, it was found
that $H_T^{(1)}$ and $H_T^{(2)}$ are also optimized observables at low $q^2$
\cite{Matias:2012xw}.

We note that at low $q^2$, $J_3$ and $J_9$ vanish at leading order in
QCDF \cite{Bobeth:2008ij}, making them ideal probes of
chirality-flipped operators $i = 7',9',10'$ because leading terms
in QCDF are $\sim\mbox{Re}[\wilson[]{i}\wilson[*]{i'}]$ and
$\sim\mbox{Im}[\wilson[]{i}\wilson[*]{i'}]$. $J_9$ (and also
$J_{7,8}$) vanishes for real Wilson coefficients, and therefore the
measurements of $\aver{A_T^{(\rm im)}}$ and $\aver{A_\mathrm{im}}$ are
not of interest for our scenario. Only partial results of the
subleading corrections exist \cite{Beneke:2004dp, Feldmann:2002iw} and
only those of kinematic origin are included in the numerical
evaluation. Nevertheless, $\aver{A_T^{(2)}}$ and $(2 S_3)$ are
included in our fit because they might allow us to obtain information
on the nuisance parameters used to model yet-unknown subleading
contributions (see \refapp{app:SL:corrections}).

At high $q^2$, $F_L$ and $A_T^{(2)}$ become short-distance independent
\cite{Bobeth:2010wg} and the experimental data allow us to constrain
the form-factor-related nuisance parameters; see
\refapp{app:form:factors}.  This has been exploited recently
\cite{Hambrock:2012dg} to extract the $q^2$ dependence of form factors
from data and, comparing with preliminary lattice results, to find
overall agreement within the still large uncertainties.

In our predictions, we therefore focus on the yet-unmeasured optimized
observables $\aver{A_T^{({\rm re}, 3, 4, 5)}}$ and
$\aver{H_T^{(1,2,3)}}$.

\subsubsection{$B_s \to \bar\mu\mu$ \label{sec:Bsmumu:constr}}

The rare decay $B_s \to \bar\mu\mu$ is helicity suppressed in the SM,
making it an ideal probe of contributions from scalar and pseudoscalar
operators. Its branching ratio depends only on $\wilson[]{10}$ in the
scenario under consideration
\begin{align}
  \label{eq:BR:Bsll}
  \mathcal{B}[B_s(t=0) \to \bar\mu\mu]& =
    \frac{G_F^2\, \alpha_e^2\, M^3_{B_s}\, f_{B_s}^2 \tau_{B_s}}{64\, \pi^3}
    \big|V_{tb}^{} V_{ts}^{\ast}\big|^2 \sqrt{1 - \frac{4\, m_\mu^2}{M_{B_s}^2}}
    \,\, \frac{4\, m_\mu^2}{M_{B_s}^2} \, \big| \wilson[]{10} \big|^2
\end{align}
and is predicted in the SM to be around $3\times 10^{-9}$. The main
uncertainties are due to the decay constant $f_{B_s}$ and the CKM
factor $|V_{tb}^{} V_{ts}^{\ast}|$. 

Above the mixing of $B_s$-meson has not been taken into account, 
i.e., the branching ratio refers to time $t = 0$. However, experimentally 
the time-integrated branching ratio is determined. Both are related in our
SM-like scenario as \cite{deBruyn:2012wk}
\begin{align}
  \mathcal{B}[B_s \to \bar\mu\mu] & =
    \frac{1}{1 - y_s} \mathcal{B}[B_s(t = 0) \to \bar\mu\mu], &
  y_s & = \frac{\Delta \Gamma_s}{2\,\Gamma_s}.
\end{align}
Lately, the most precise measurement of the life-time difference $\Delta \Gamma_s$ 
became available from LHCb \cite{Aaij:2012DeltaGammas} and moreover LHCb 
succeeded to determine the sign of $\Delta \Gamma_s$ \cite{Aaij:2012eq} 
which turned out to be SM-like. In view of this, we will use the numerical 
value from LHCb $y_s = 0.088 \pm 0.014$ \cite{Aaij:2012DeltaGammas}.

In the last decade, the Tevatron experiments D{\O} \cite{Abazov:2010fs}
and CDF \cite{Aaltonen:2011fi, CDFBsmumu:Moriond2012} lowered the
upper bound on the branching ratio by several orders of magnitude to
a value close to $1 \times 10^{-8}$; and CDF announced the first direct evidence
based on a $2\,\sigma$ fluctuation over the background-only hypothesis
\cite{Aaltonen:2011fi, CDFBsmumu:Moriond2012}.  This year the LHC
experiments LHCb, CMS, and ATLAS provided their results based on the
complete 2011 run \cite{LHCb:2011ac, Aaij:2012ac, Chatrchyan:2011kr,
  Chatrchyan:2012rg, ATLAS:2012:moriond}. In our analysis we use the most
stringent result $\mathcal{B}(B_s \to \bar\mu\mu) < 4.5 \times
10^{-9}\, (3.8 \times 10^{-9})$ at 95\% (90\%) CL, obtained by LHCb
 \cite{Aaij:2012ac}. Details of the implementation of this
bound are given in \refsec{sec:stat:exp:results}.

%
%
\section{Statistical Method \label{sec:stat:fit}}

We have decided to use the full Bayesian approach in this analysis. It allows us
to incorporate all available experimental results, to obtain probability
statements about the parameters of interest $\vecth$ --- the Wilson coefficients
--- and to compare different models using the Bayes factor.

In the Bayesian approach, we describe theory uncertainties by adding nuisance
parameters $\vecnu$. It is straightforward to incorporate existing knowledge ---
say from power counting, symmetry arguments, or even other dedicated Bayesian
analyses --- about these theory uncertainties by specifying informative priors.
As a cross validation, it is useful to employ different priors and compare the
posterior inference. Any significant discrepancy based on two different prior
choices implies that more accurate experimental or theoretical input is needed
before conclusive statements can be made.  This can be seen as a feature of the
Bayesian methodology. Throughout, we assume that parameters are independent a
priori,
\begin{equation}
P(\vecth,\,\vecnu) = \prod_i P(\theta_i) \cdot \prod_j P(\nu_j).
\end{equation}
The experimental data $D$ are used in the likelihood $P
(D\,|\,\vecth,\,\vecnu)$, and Bayes' theorem yields the posterior
knowledge about the parameters after learning from the data $D$
\begin{align}
  P \left(\vecth,\,\vecnu\,|\,D\right) & =
  \frac{P \left(D\,|\,\vecth,\,\vecnu\right)
        P\left(\vecth,\,\vecnu\right)}{Z} \, ,
\end{align}
with the normalization given by the evidence
\begin{equation}
  \label{eq:evidence}
  Z = \int \!\rmd \vecth\, \rmd\vecnu \, P \left(D\,|\,\vecth,\,\vecnu\right)
  P\left(\vecth,\,\vecnu\right).
\end{equation}
In case we want to remove the dependence on $\vecnu$ in the posterior, we simply marginalize:
\begin{align}
  P \left(\vecth\,|\,D\right) & =
  \int \rmd \vecnu\, P\left(\vecth,\,\vecnu\,|\,D\right)
  \,.
\end{align}
The integrations are performed with the Monte Carlo algorithm described next.

%
\subsection{Monte Carlo Algorithm \label{sec:MCalgorithm}}

The presence of multiple, well separated modes, the large dimensionality of the
parameter space, and the costly evaluation of the likelihood require a
sophisticated algorithm \cite{Beaujean:2012??}. We sketch the main steps of
this new algorithm:
\begin{enumerate}
\item A sufficiently large number of Markov chains are run in parallel for
  $\mathcal{O}(50000)$ iterations to explore the parameter space with an
  adaptive local random walk. The chains need a burn-in phase, thus we discard the
  first $15\%$ of the iterations.
\item Chains whose common $R$-value \cite{Gelman:1992zz} is reasonably small,
  say $R<2$, are combined into groups.
\item We create patches with a length of $\mathcal{O}(1000)$ points from the
  individual chains and define a multivariate density from the mean and covariance
  of each patch.
\item Using hierarchical clustering \cite{Roweis:2004}, we combine the patches
  into a smaller number of clusters. As the initial guess for the clustering, we
  construct a fixed number of about 30 patches of length $\mathcal{O}(5000)$
  from each group of chains.
\item We define a multivariate mixture density from the output of the clustering
  by assigning equal weights to each cluster. This mixture density serves
  as the initial proposal density for the  Population Monte Carlo (PMC)
  algorithm \cite{Wraith:2009if,Cappe:2008sd}.
\item Using a computing cluster with a few hundred cores, we draw importance
  samples and adapt the proposal density to the posterior until convergence is
  achieved; i.e., until the difference in perplexity between two consecutive
  steps is less than $2\%$.
\item Given the resulting proposal density, we collect $2\cdot 10^6$
  importance samples to compute marginal distributions and the evidence.
\end{enumerate}
The biggest advantages of this approach are the automatic adaptation to the
complicated posterior shape, and the ability to massively parallelize the costly
evaluation of the likelihood.

%
\subsection{Priors \label{sec:priors}}

We use flat priors for the Wilson coefficients. This is not done because we want
to imply complete prior ignorance, but, instead, we want a convenient,
sufficiently diffuse density, with the expectation that the posterior is
dominated by the likelihood.

For the nuisance parameters, the choice of prior depends on the parameter's
nature. There are the four quark-mixing matrix (CKM) parameters, the $b$ and $c$
quark masses, the decay constant $f_{B_s}$ entering $B_s \toÊ \bar\mu\mu$, and
most dominantly the $B\to K^{(*)}$ form factors. In addition, unknown subleading
contributions in the two distinct $\Lambda/m_b$ expansions at large and low
recoil are parametrized as nuisance parameters. The complete list of almost 30
nuisance parameters  along with the choice of the prior densities is
presented in \refapp{app:nuisance:pmrs}.  Note that most
nuisance parameters only affect a subset of the observables.

Where possible, the posterior distributions of the nuisance parameters from
previous analyses fitting different data are used as the prior distributions in
our fit. As an example, we use the output for the quark masses
\cite{Nakamura:2010zzi} in the form of LogGamma distributions; see
\refapp{sec:logg-distr}.

For the CKM parameters $\lambda,\, A,\, \bar{\rho},\, \bar{\eta}$, we choose the
results of the UTfit Collaboration \cite{Bona:2006ah}.  When allowing for BSM
contributions in our fit, we use the results of the so-called CKM
\textit{tree-level fit} for $\lambda,\, A,\, \bar{\rho},\, \bar{\eta}$.  The
tree-level fit represents only the basic constraints from SM tree-level
processes, which every extension of the SM must include and we assume that
BSM contributions are negligible. Thus no information
from rare $B$ decays and $B$--$\bar{B}$ mixing is used indirectly through the
priors.  For the SM predictions, we use the results of the SM-CKM fit
instead. The posterior distributions from either fit are assumed to be symmetric
Gaussian distributions, which was found to be a good approximation.

We model theory uncertainties with Gaussian distributions in cases
where authors only report an estimate of the magnitude. This is
justified by the principle of maximum entropy
\cite{jaynes_probability_2003}. As an example, suppose the quoted
uncertainty of a QCD form factor $f$ is 15\%.  We introduce a nuisance
parameter, $\zeta_{f}$, as a scaling factor, such that $f \to
\zeta_{f} \cdot f$, and we vary $\zeta_{f} \sim \mathcal{N} \left( \mu
  = 1,\sigma = 0.15\right)$, with an allowed range $\zeta_{f}\in
\left[ 1 - 3\cdot\sigma, 1 + 3\cdot\sigma\right]$; i.e., neglecting
the tails of the Gaussian beyond $3\,\sigma$. If necessary, we modify
the range to avoid unphysical values of $f$. Subleading phases are
incorporated with flat priors covering the full range.

%
\subsection{Experimental Results \label{sec:stat:exp:results}}

We form the total log likelihood, $\log P(D\,|\,\vecth, \vecnu)$, by summing
over the individual contributions. The complete list of experimental results
used is given in tables \ref{tab:BKgam:expData} to \ref{tab:BKstll:expData}. The
majority of results is incorporated as 1-dimensional Gaussian distributions,
whose variances are obtained by adding statistical and systematic uncertainties
in quadrature, $\sigma^{2}=\sigma_{stat}^{2}+\sigma_{syst}^{2}$.  In the case of
asymmetric uncertainties, we use a piecewise function constructed from two
Gaussian distributions around the central value with different variances. With
the exception of the upper bound on $\mathcal{B}(B_s \to \bar\mu\mu)$, we limit ourselves to
the Gaussian distribution in the likelihood, despite the discontinuities arising
from asymmetric uncertainties, to obtain results that are comparable with the
existing literature.  Known correlations between observations, e.g., between the
time-dependent CP asymmetries $S$ and $C$ in $B\to K^*\gamma$, are represented
by multivariate Gaussian distributions. Note that in $B\to K^*\bar\ell\ell$
decays the observables $A_{\rm FB}$ and $F_{L}$ are extracted from a
simultaneous fit to the double differential decay rate.  Requiring physical
values, i.e.,
\begin{align}
  \label{eq:double-diff}
  \frac{\mathrm{d}^{2}\Gamma}{\mathrm{d}q^{2}\, \mathrm{d}\!\cos\theta_{\ell, K}} > 0,
\end{align}
cuts out an allowed region in the $\left(A_{\rm FB},\, F_{L}\right)$ - plane.  
Since the fit for $\left(A_{\rm FB}, \,F_{L}\right)$ typically converges near 
the unphysical region, the resulting contribution to the likelihood would be
distinctly non-Gaussian. Unfortunately, the resulting 2D likelihood is not publicly
available, thus we assume $A_{\rm FB}$ and $F_{L}$ independent and Gaussian
distributed.

We include the results of direct searches for the decay $B_s \to \bar\mu\mu$
into the likelihood. Often, only the $90\%$ and $95\%$ limits on the branching
ratio $\mathcal{B}$, obtained with the $CL_S$ method \cite{Read:2002hq}, are
published.  However, there is no single best way to translate these limits into
a useful contribution to the likelihood, and several schemes of varying
sophistication exist in the literature \cite{de
  Austri:2006pe,Flacher:2008zq,Reece:2010}.  It is preferable to directly use
the Bayesian posterior on the branching ratio $P(\mathcal{B}\,|\,D)$, computed
by a general algorithm for multichannel search experiments
\cite{Heinrich:2005}. This posterior is often produced to compute Bayesian
limits for cross checks with $CL_S$ results. The input numbers --- expected signal
yields, background yields --- that are needed to compute $P(\mathcal{B}\,|\,D)$
are publicly available from LHCb \cite{Aaij:2012ac}; only the correlations of the
yields are not published.
By reinterpreting the
function $P(\mathcal{B}\,|\,D)$ as $P(\mathcal{B}\, |\,\vecth,\, \vecnu)$ the
desired contribution to the likelihood is found. For a convenient approximation
to $P(\mathcal{B}\,|\,D)$, we use the four-parameter $\Amoroso$-distribution
\cite{Crooks:2010} with relevant details given in
\refapp{sec:amoroso-distribution}. For the data supplied by LHCb
\cite{Aaij:2012ac}, the relative error of this interpolation is at most 2\%.

%
\subsection{Uncertainties of Theory Predictions \label{sec:theo:unc:method}}

Within the Bayesian framework, the procedure to calculate the uncertainty of an
observable's prediction within a given theory, say the SM, is essentially
uncertainty propagation. In this case, an observable $A$ depends on Wilson
coefficients and on additional nuisance parameters. We fix the values of the
Wilson coefficients, $\vecth=\vecth_{SM}$, so the value of $A$ is uniquely
determined by $\vecnu$; i.e., $A=f(\vecnu)$. We vary the nuisance parameters
according to their prior, $P(\vecnu)$. The distribution of the random variable
$A$, $P(A)$, is given by
\begin{align}
  P(A) &
  = \int\rmd\vecnu\, P\left(A,\,\vecnu\right)
  = \int\rmd\vecnu\, P\left(A\,|\,\vecnu\right)P(\vecnu)
  = \int\rmd\vecnu\, \delta\left(A-f(\vecnu)\right)P(\vecnu)\, ,
\end{align}
where we used the Dirac $\delta$-distribution.
Numerically, one only needs to draw parameter samples $\vecnu_{i} \sim
P(\vecnu)$ and calculate $A$ for each sample $\vecnu_{i}$.  We collect the
resulting samples $A_{i}$ in one dimensional histograms to extract $68\%$
intervals. As before, we assume $P(\vecnu) = \prod_{j} P\left(\nu_{j}\right)$
and use the priors listed in \refapp{app:nuisance:pmrs}. As a welcome side
effect, this form of $P(\vecnu)$ allows us to efficiently sample $\vecnu$
from the joint prior by sampling from simple, 1D priors directly without the
need to resort to MCMC or PMC.

If we take this approach one step further, we can ask: what are the likely values,
or informally speaking ``the allowed ranges,'' of $A$, given the set of measurements
$D$ listed in Section \ref{sec:effTh:expData}? Using the full posterior on both Wilson
coefficients and nuisance parameters,
$P(\vecth,\, \vecnu\,|\,D)$, we obtain
\begin{align}
  P(A\,|\,D) & = \int \rmd\vecth\, \rmd\vecnu\,
    \delta\left(A-f(\vecth,\,\vecnu)\right) P(\vecth,\,\vecnu\,|\,D)\,.
\end{align}
We simply take the posterior samples produced by the PMC algorithm, compute $A_i$
for each sample $(\vecth,\,\vecnu)_i $, and finally fill the
sample $A_i$ with its importance weight into a histogram.

%
\subsection{Goodness of Fit and Model Comparison}
\label{sec:gof}

To check that the assumed model with three real Wilson coefficients provides a
good description of the experimental observations, we determine the goodness of
fit. We follow the standard procedure: first we choose a test statistic
$T(D\,|\,\vecth,\, \vecnu)$ with the parameter values chosen at a local mode of
the posterior, then calculate its distribution, and finally determine the value
of the test statistic for the actual data set. For more details on $p$-values
and how we interpret them in this work, we refer to \cite{Beaujean:2011zz}. We
make two closely related choices for $T$, defined as follows.

For each observable $x$, we compare its theory prediction $x_{pred}(\vecth,\,
\vecnu)$ with the mode of the experimental distribution (central value) of
$x$, denoted by $x^{*}$. We compute the frequency $f$ that a value of $x$ less
extreme than $x_{pred}$ would be observed. Using the inverse of the Gaussian
cumulative distribution function, $\Phi^{-1}(\cdot)$, we define the
pull:
\begin{align}
  \label{eq:pull:def}
  \delta & = \Phi^{-1}\left[ \frac{f+1}{2} \right].
\end{align}
Note that for a 1-dimensional Gaussian, this reduces to the usual
$\delta = ( x^{*} - x_{pred})/\sigma$.
In the 1-dimensional case, the (Gaussian, Amoroso)
distributions yield a signed $\delta$ (positive if $ x^{*} > x_{pred}$, else
negative), while for the multivariate Gaussian, $\delta$ is positive semidefinite.
We define the test statistic $T_{\rm pull}$ as
\begin{equation}
  T_{\rm pull}=\sum_i \delta_i^{2} \; ,
\end{equation}
where $i$ extends over all experimental data. As a cross check, we also consider
$T_{\rm like}$, defined as the value of the log likelihood, $T_{\rm like} = \log
P(D\,|\,\vecth,\,\vecnu)$. Its frequency distribution is approximated by
generating $10^5$ pseudo experiments $D \sim P(D\,|\,\vecth^*, \vecnu^*)$, where
$\vecth^*, \vecnu^*$ are fixed values at a local maximum of the posterior.
Since we do not have the raw data --- events, detector simulations etc. ---
available, we generate pseudo experiments. Consider the case of a single
measurement with Gaussian uncertainties $\mathcal{N}(\mu=x^*, \sigma)$: we fix
the theory prediction, shifting the maximum of the Gaussian to
$x_{pred}(\vecth^*, \vecnu^*)$, but keep the uncertainties reported by the
experiment. Then we generate $x \sim \mathcal{N}(\mu=x_{pred}(\vecth^*,
\vecnu^*), \sigma)$, and proceed analogously for all observables to sample $D$.
The $p$-value is computed by counting the fraction of experiments with a
likelihood value smaller than that for the observed data set and corrected for
the number of degrees of freedom; see Section III.D.5 in \cite{Beaujean:2011zz}.
Although the generation of pseudo data is far from perfect, we emphasize that,
on the one hand, it is fast and, on the other hand, we will not consider the
actual value of $p$ too rigorously. Two models with $p$-values of $40\%$ and
$60\%$ both describe the data well, and that is all the information we need.

If we used the maximum likelihood parameters and ignored the $B_s \to
\bar\mu\mu$ contribution, both statistics would be equivalent to $\chi^2$ and
thus yield the same $p$-value.  The parameter values at the global mode of the
posterior differ only little from the maximum likelihood values, and $B_s \to
\bar\mu\mu$ presents only one of 59 inputs.  We therefore consider it reasonable
to approximate the distribution of $T_{\rm pull}$ by the $\chi^2$-distribution
in order to compute a $p$-value.

If there are several local modes with reasonably high $p$-values, it is
necessary to assess which of them is favored by the data; i.e., to perform
a model comparison.
Suppose the full parameter space is decomposed into disjoint subsets $V_i$,
$i=1\dots n$, where $V_i$ contains only a single mode of the posterior. Then we
compute the local evidence $Z_{i}$ by integrating over $V_i$ in
\eqref{eq:evidence}. In fact, the integral is available as the average weight of
all importance samples in $V_i$, with an accuracy of roughly $5\%$. 
The Bayes factor $B_{ij} = Z_i / Z_j$ is the data-dependent part in the
posterior odds of two statistical models $M_i, M_j$:
\begin{equation}
\frac{P(M_i \, | \, D)}{P(M_j \, | \, D)} = B_{ij} \cdot \frac{P(M_i)}{P(M_j)}
\end{equation}
In addition to the local evidence, we compute the evidence $Z_{SM}$
for the SM case with fixed values for the Wilson coefficients, but
with $\vecnu$ allowed to vary.  Computing the Bayes factor with
$Z_{SM}$ and the local evidence for the region with SM-like signature
allows us to assess if the data are in favor of adding three degrees of
freedom for the Wilson coefficients to achieve a better agreement with
the theory predictions, or if the SM is preferred due to its
simplicity.

%
%
\section{Results \label{sec:results}}

In the following, we discuss the results of our analysis of the experimental
data described in \refsec{sec:expData} using the statistical tools explained in
\refsec{sec:stat:fit}. First, the results of the global fit of the three Wilson
coefficients and 28 nuisance parameters to 59 experimental inputs are presented,
including marginal distributions and best fit points.  As for goodness of fit,
we list $p$-values and evidence for each of the arising solutions. Furthermore,
we show pull values for the included measurements. And we discuss the fit results of
nuisance parameters, if the posterior differs significantly from
the prior. Second, we provide predictions of yet-unmeasured, optimized
observables in $B\to K^*(\to K\pi)\,\bar\ell\ell$ at low and high $q^2$ within
our scenario, taking into account the experimental data. Finally, we give SM
predictions for measured and yet-unmeasured observables including theory
uncertainties determined using the Monte Carlo method as explained in
\refsec{sec:theo:unc:method}.

%
\subsection{Fit Results}\label{sec:fit-results}

\begin{figure}[t]
  \begin{minipage}[t]{0.49\textwidth}
    \phantom{x} \vskip -0.8cm
    \includegraphics[width=1.05\textwidth]{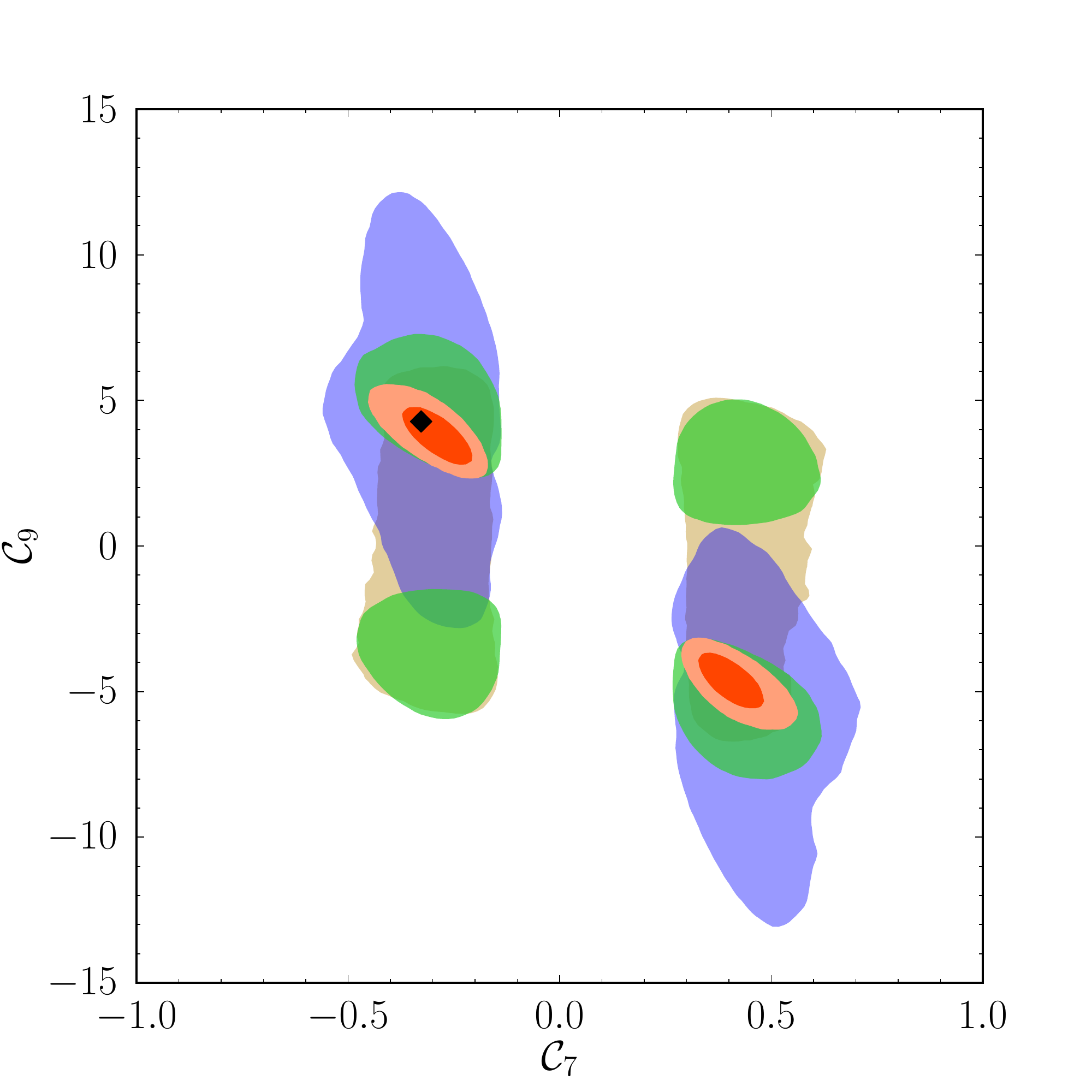}
  \end{minipage}
  \hspace{0.06\textwidth}
  \begin{minipage}[t]{0.45\textwidth}
     \caption{The marginalized 2-dimensional 95\% credibility regions of the
        Wilson coefficients $\wilson[]{7,9,10}$ for $\mu = 4.2\GeV$ are shown
        when applying the $B\to K^* \gamma$ constraints in combination with
        {\it i)} only low- and high-$q^2$ data from $B \to K\, \bar\ell\ell$ [brown];
        {\it ii)} only low-$q^2$ data from $B \to K^*\bar\ell\ell$ [blue];
        {\it iii)} only high-$q^2$ data from $B \to K^*\bar\ell\ell$ [green];
        and {\it iv)} all the data, including also $B_s \to \bar\mu\mu$
        [light red], showing as well the 68\% credibility interval [red].
        The SM values $\wilson[\rm SM]{7,9,10}$ are indicated by $\filleddiamond$.
    \label{fig:fit-all-data}}
   \end{minipage}
   \\
   \begin{minipage}[t]{0.49\textwidth}
    \includegraphics[width=1.05\textwidth]{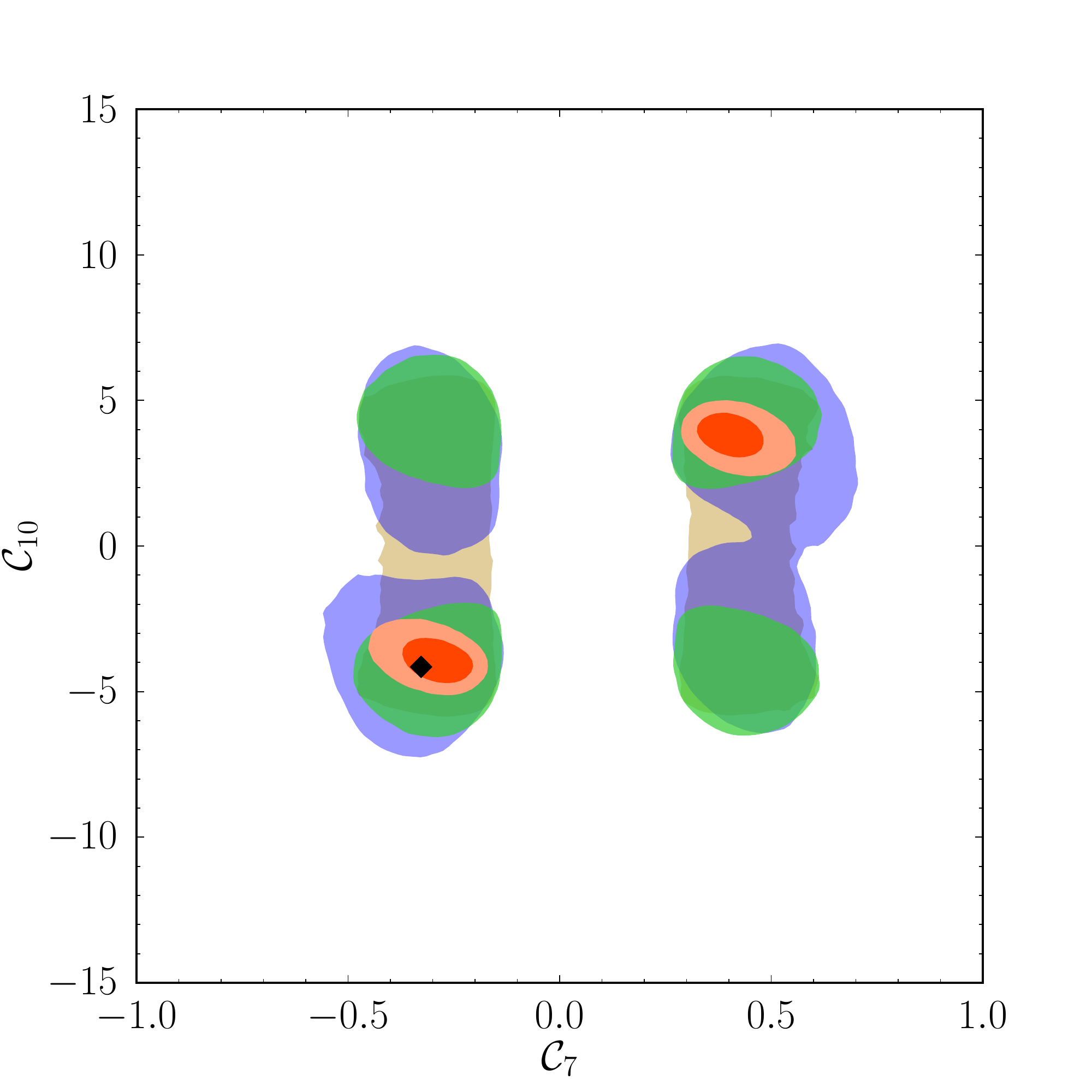}
  \end{minipage}
  \hspace{0.02\textwidth}
  \begin{minipage}[t]{0.49\textwidth}
    \includegraphics[width=1.05\textwidth]{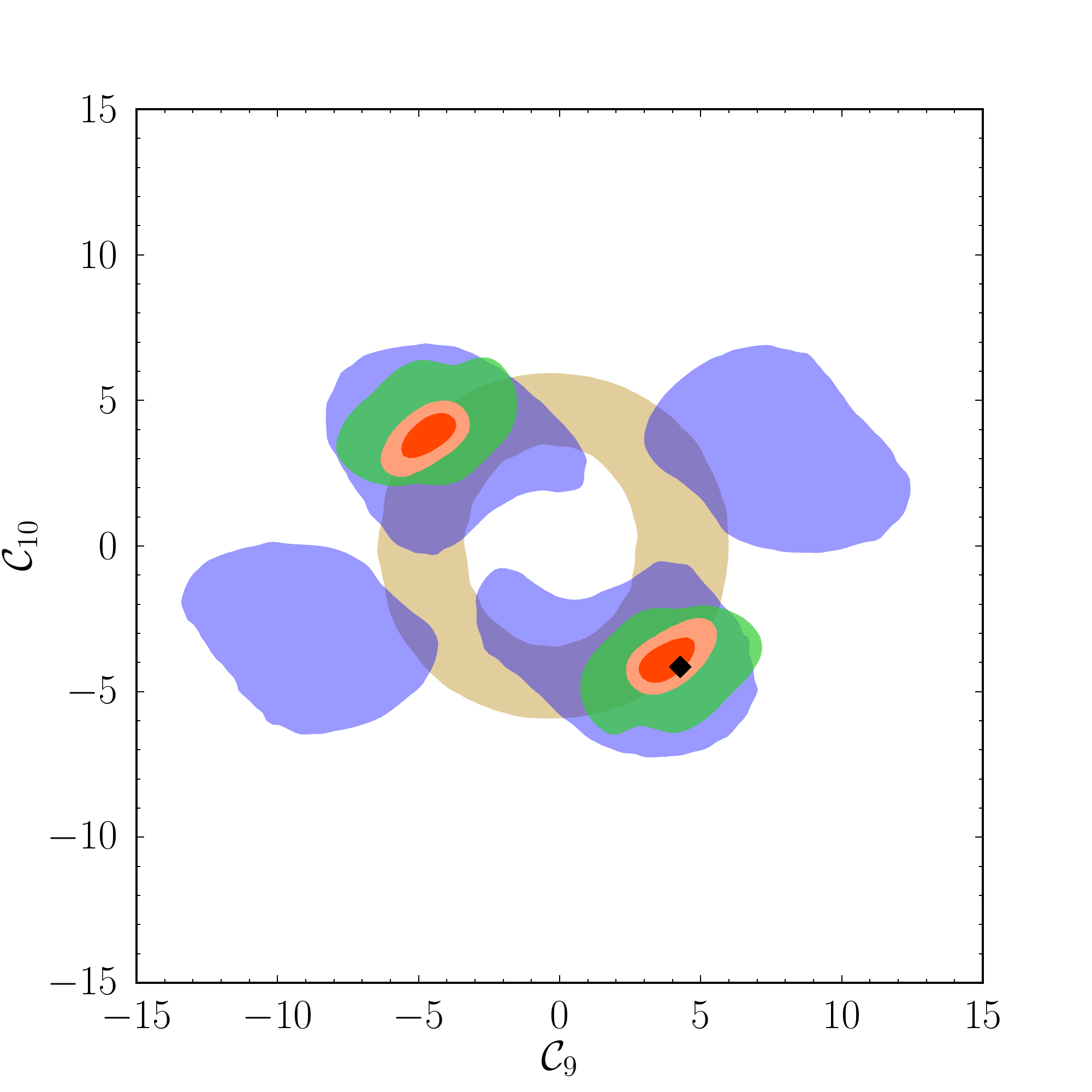}
  \end{minipage}
\end{figure}

Here, we summarize the main part of our work: the results of the fits of the
parameters of interest, the Wilson coefficients $\wilson[]{7,9,10}$, to the data
listed in \refsec{sec:expData}. Details of the Monte Carlo algorithm are given
in \refsec{sec:MCalgorithm}. The treatment of prior distributions is explained
in \refsec{sec:priors}, the priors of the nuisance parameters are specified
in \refapp{app:nuisance:pmrs}. For $\wilson[]{7,9,10}$, we use flat priors with
\begin{align}
\wilson[]{7} & \in [-1,1], &
\wilson[]{9,10} & \in [-10,10].
\label{eq:wc:ranges}
\end{align}
The fit not only constrains the Wilson coefficients $\wilson[]{7,9,10}$, but
updates our knowledge of the nuisance parameters. We discuss the most
significant changes.

The marginalized two-dimensional 95\% credibility regions are shown in
\reffig{fig:fit-all-data} when applying the $B\to K^* \gamma$ constraints
(\refsec{sec:BKstarGamma:constr}) in combination with {\it i)} only low- and
high-$q^2$ data from $B \to K\,\bar\ell\ell$ (\refsec{sec:BKll:constr}); {\it
  ii)} only low-$q^2$ data from $B \to K^*\bar\ell\ell$ \footnote{Here we
  enlarged the prior ranges of $\wilson[]{7,9,10}$ by a factor of 2, which is
  irrelevant for the remainder of our work.}; {\it iii)} only high-$q^2$ data
from $B \to K^*\bar\ell\ell$ (\refsec{sec:BKstarll:constr}); and finally {\it
  iv)} all the data, including also $B_s \to \bar\mu\mu$
(\refsec{sec:Bsmumu:constr}).

The most stringent constraints on $\wilson[]{9,10}$ come from the high-$q^2$
data of $B \to K^*\bar\ell\ell$, which should be taken with some caution since
the form factors are only available as extrapolations of LCSR results from
low $q^2$. In the near future, we expect more accurate lattice calculations of
form factors to close this weak point. Also shown are the SM predictions of
$\wilson[]{7,9,10}(\mu = 4.2\, \mbox{GeV})$ using NNLO evolution
\cite{Bobeth:2003at}.

We confirm the findings of previous analyses
\cite{Altmannshofer:2011gn, Bobeth:2011nj} that only two solutions make up
95\% of the probability: the first exhibits the same signs of
$\wilson[]{7,9,10}$ as the SM. The second solution corresponds to a
first order degeneracy of all observables under a simultaneous sign
flip $\wilson[]{7,9,10} \to -\wilson[]{7,9,10}$. There are two
additional local maxima that correspond to a sign flip of
$\wilson[]{7} \to -\wilson[]{7}$ of the former solutions. In
\reftab{tab:bfp:gof}, we list the properties of these four modes,
categorized by the signs of $\wilson[]{7,9,10}$. As witnessed by the
evidence, $Z$, the SM-like and sign-flipped solutions essentially make
up the whole posterior mass, with ratios of $52\%$ and $48\%$,
respectively. The other two solutions are suppressed by many orders of
magnitude, and thus do not appear at the $95\%$ level. For the two dominant
solutions, the goodness-of-fit results are nearly identical: both $p$-values
based on the statistics $T_{\rm like}$ and $T_{\rm pull}$ (see \refsec{sec:gof})
are large, indicating a good fit. In contrast, the suppressed solutions do not
seem to explain the data well.  We note that the MCMC revealed a handful of
additional modes with $ 6 \lesssim |\wilson[]{9,10}| \lesssim 9$.  We do not
consider these further because they are suppressed by a factor of roughly
$\exp(40)$ compared to the global maximum.

\begin{figure}[t]
  \includegraphics[width=0.51\textwidth]{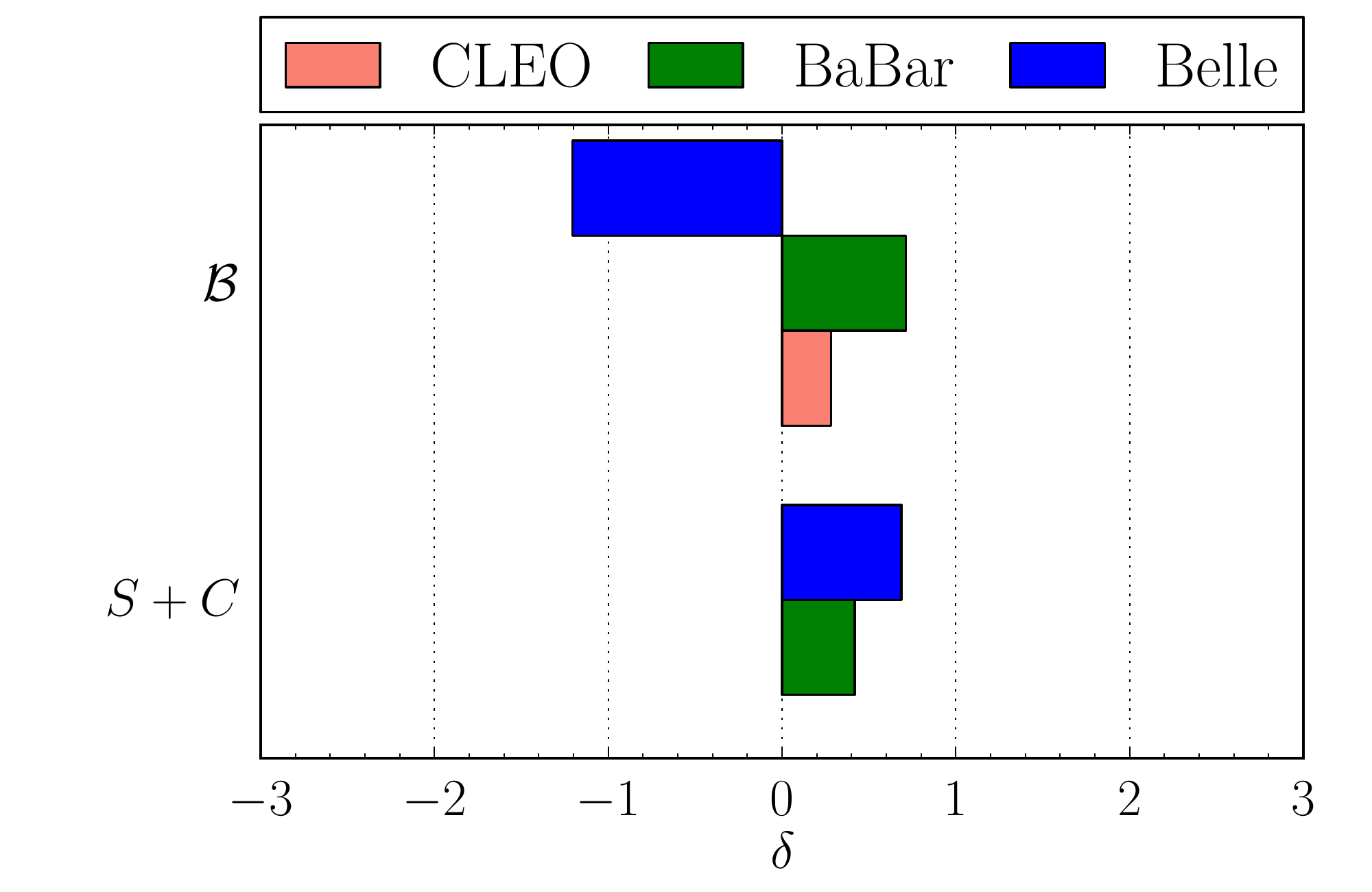}
  \includegraphics[width=0.51\textwidth]{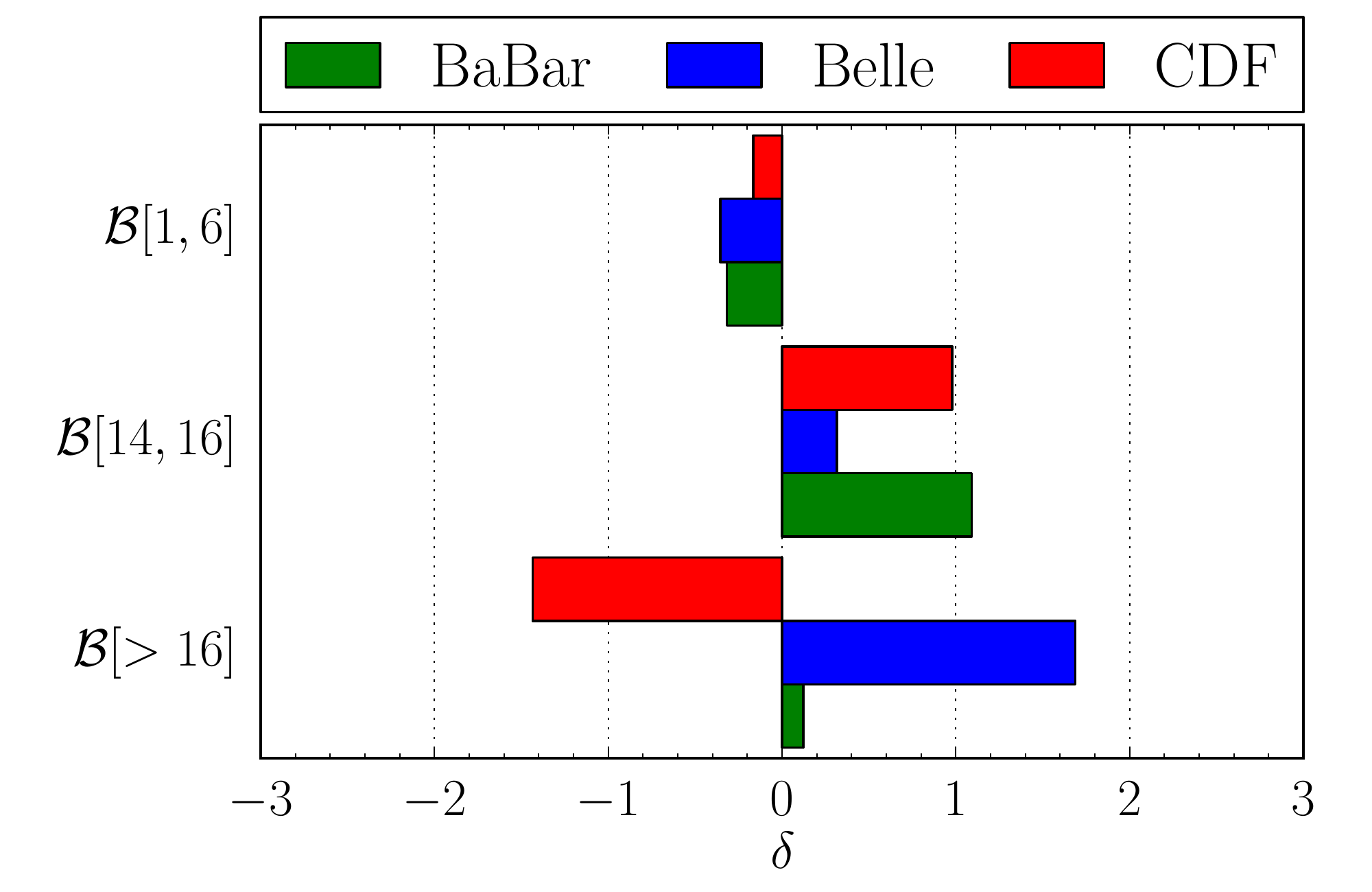}
  \caption{\label{fig:pull:BKstargamma:BKll} Pull values for observables
  in $B\to K^*\gamma$ [left] and $B\to K\,\bar\ell\ell$ [right] calculated
  at the best fit point. The pull definition for
  the correlated observables $S$ and $C$ permits only $ \delta \ge 0$;
  for details see \refsec{sec:gof}.}
\end{figure}

\begin{figure}[t]
  \includegraphics[width=0.51\textwidth]{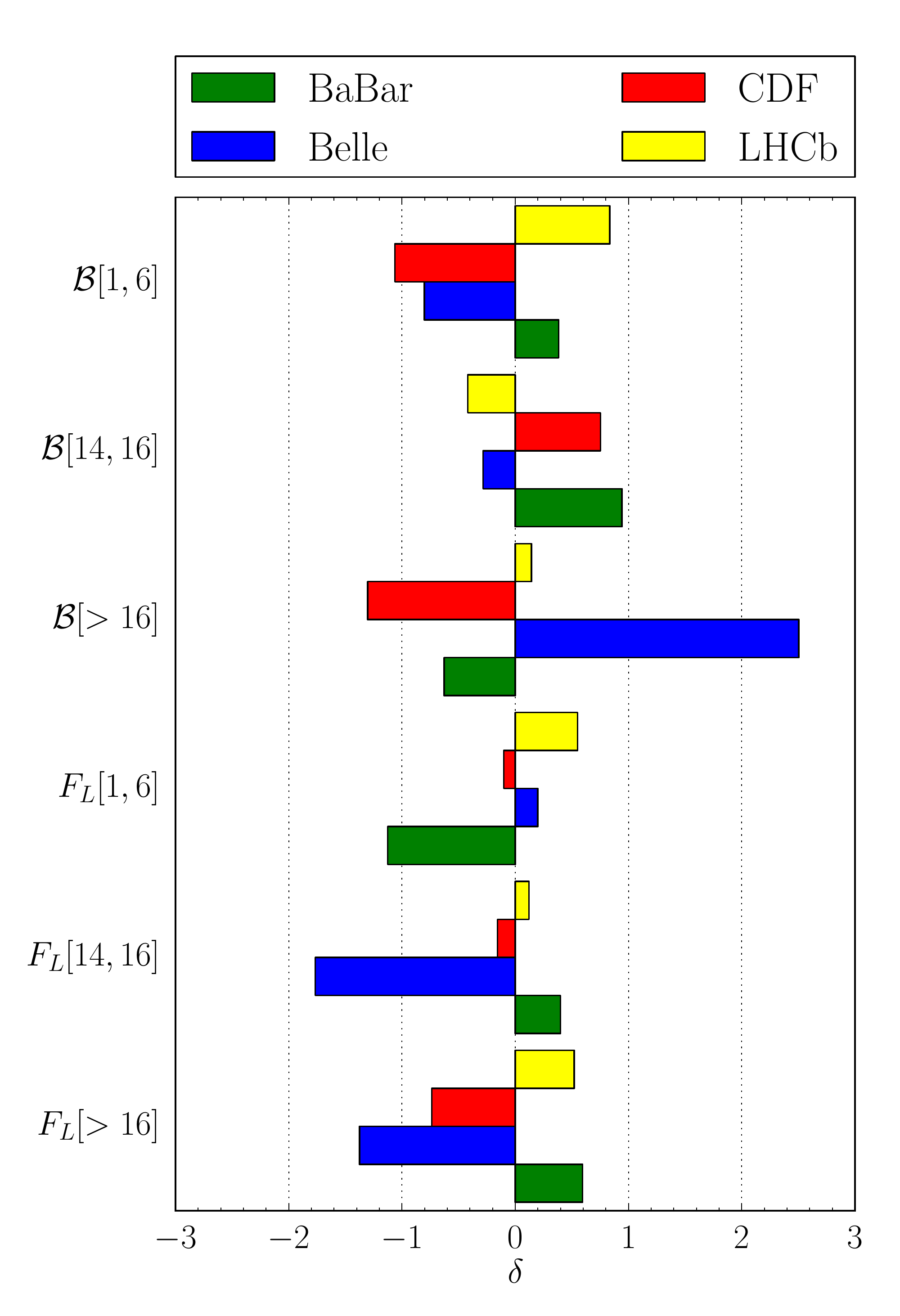}
  \includegraphics[width=0.51\textwidth]{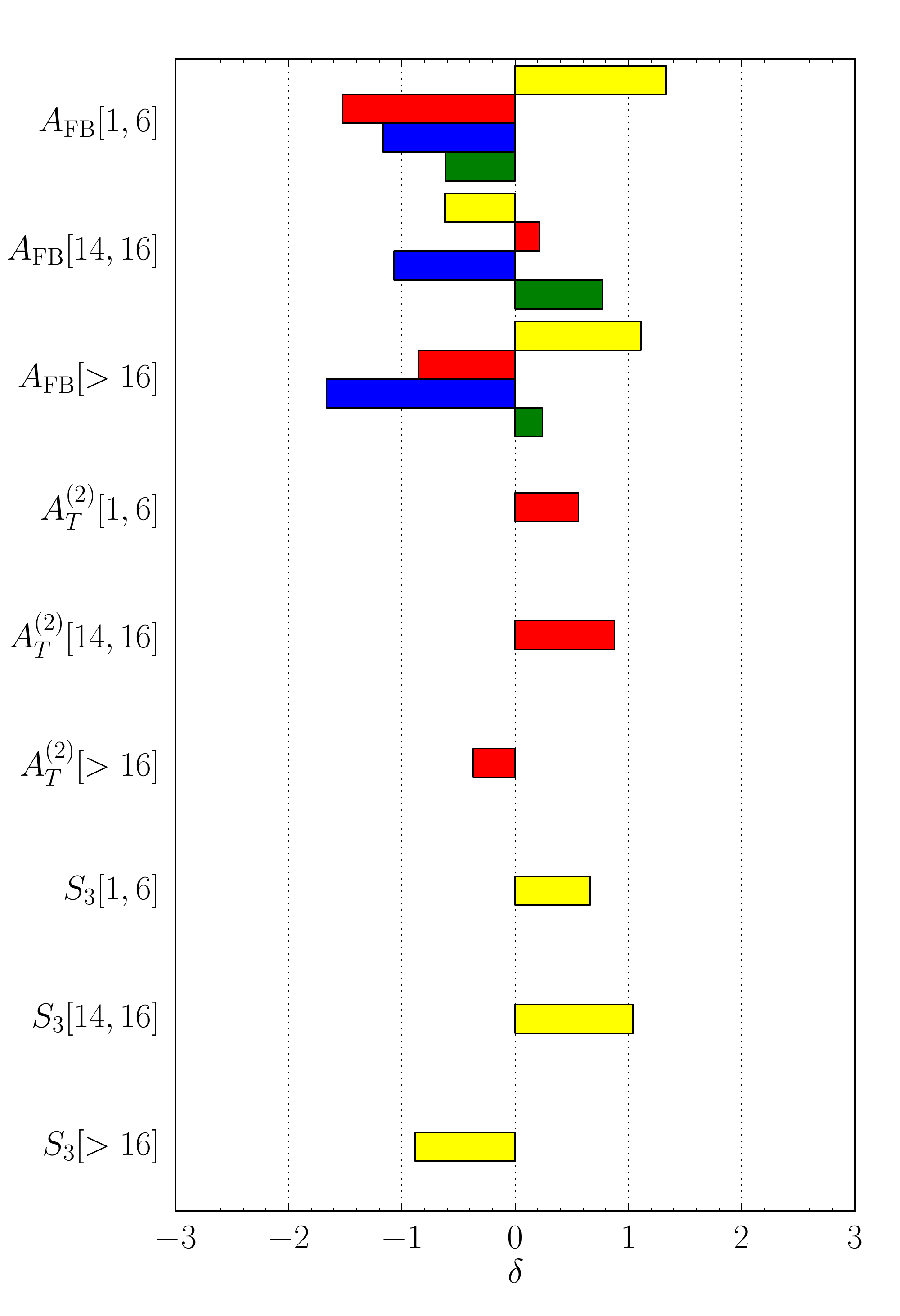}
  \caption{\label{fig:pull:BKstarll} Pull values for observables
  in $B\to K^*\bar\ell\ell$ calculated at the best fit point.}
\end{figure}

To highlight the fit results, we
present the pull \refeq{eq:pull:def} --- the normalized deviation between theory
prediction and measured value in units of Gaussian $\sigma$ --- for all 59
constraints.  Pulls for $B\to K^*\gamma$ [left] and $B\to K\,\bar\ell\ell$
[right] constraints are shown in \reffig{fig:pull:BKstargamma:BKll}; those for
$B\to K^*\bar\ell\ell$ in \reffig{fig:pull:BKstarll}; and the pull for LHCb's
result of $B_s \to \bar\mu\mu$ is $-1.1$; i.e., its most likely value from the
measurement is about $1 \sigma$ (in terms of the experimental uncertainty) lower
than the theory prediction.  Here, the theory parameters are chosen at the
global maximum of the posterior.  With the best-fit parameters with
$\wilson[]{7,9,10}$ fixed at SM values (see below), we obtain nearly identical
plots, and we therefore omit them. We observe the largest pull at $+2.5$ for the
Belle measurement of $\langle\mathcal{B}\rangle[16,19.21]$ for $B \to
K^*\bar\ell\ell$.  It is the only pull surpassing
$2.0$. \reffig{fig:pull:BKstarll} shows, for example, how the debate about the
existence of a zero crossing of $A_{\rm FB}$ at large recoil was settled: the
first published measurements by Belle and CDF deviated from the SM
  prediction, but when taken together with LHCb’s recent result that pulls the
  best fit point towards the SM, there is good agreement between the SM and the
experiments.

We also perform the global fit with $\wilson[]{7,9,10}$ fixed to the SM
values, varying only the nuisance parameters; see the bottom row in
\reftab{tab:bfp:gof}. The prior normalization then changes by $\log(800)=6.68$
due to omitting $\wilson[]{7,9,10}$ with ranges given in \refeq{eq:wc:ranges}.
The values of $T_{\rm like}$ and $T_{\rm pull}$ are just as good as for the two
dominant solutions, but the $p$-values are even larger, as the number of
degrees of freedom used in the $\chi^2$-distribution to calculate $p$ differs by
three.  We compute the Bayes factor of the SM vs the SM-like solution by
dividing their respective evidences:
\begin{equation}
  \label{eq:bayesfactor}
  B = \exp(392.6 - 385.3) \approx 1500.
\end{equation}
Assuming prior odds of one, the posterior odds are given by $B$, and thus are
clearly in favor of the simpler model. The effect persists if we cut the allowed
range of each $\wilson[]{i}$ in half to exclude all but the SM-like solution.
In conclusion, both the SM (with nuisance parameters allowed to vary) and our
extension with real floating $\wilson[]{7,9,10}$ fit the 59 experimental
observations of rare B decays well. Since the extension does not provide any
significant improvement, the simpler model should be preferred.

\begin{table}
\begin{center}
\resizebox{\textwidth}{!}{
\begin{tabular}{c|c|c|cc|cc|c}
\hline\hline
\tabvsptop
   \multirow{2}{*}{$\mbox{sgn}(\wilson[]{7},\, \wilson[]{9},\, \wilson[]{10})$} &
   \multirow{2}{*}{best-fit point} &
   \multirow{2}{*}{log(MAP)} &
   \multicolumn{4}{c|}{goodness of fit} &
   \multirow{2}{*}{$\log(Z)$}
\\[0.1cm]
\tabvspbot
  &
  &
  & $T_{\rm like}$
  & $p_{\rm like}$
  & $T_{\rm pull}$
  & $p_{\rm pull}$
  &
\\
\hline
\tabvsptop
$(-,\, +,\, -)$
  & $(-0.293, 3.69, -4.19)$ & $425.22$ & $402.59$ & $60\%$ & $48.4$  & $75\%$  & $385.3$
\\[0.1cm]
$(+,\, -,\, +)$
  & $(0.416, -4.59, 4.05)$  & $425.08$ & $402.49$ & $60\%$ & $48.5$  & $75\%$  & $385.2$
\\[0.1cm]
$(-,\, -,\, +)$
  & $(-0.393, -3.12, 3.20)$ & $404.67$ & $387.88$ & $0.9\%$ & $76.5$ & $4\%$   & $363.9$
\\[0.1cm]
$(+,\, +,\, -)$
  & $(0.558, 2.25, -3.24)$  & $400.91$ & $384.52$ & $0.2\%$ & $83.1$ & $1\%$   & $358.9$
\\[0.1cm]
\hline
\tabvsptop
SM: $(-,\, +,\, -)$
  & $(-0.327, 4.28, -4.15)$ & $431.46^{\dagger}$ & $402.53$ & $70\%$  & $48.5$ & $83\%$  & $392.6$
\\
\hline\hline
\end{tabular}
}
\end{center}
\caption{Best-fit point, log maximum a-posteriori (MAP) value, goodness of fit summary and $\log \mathrm{evidence}$
  for the four local modes
  (denoted by the signs of $(\wilson[]{7},\,\wilson[]{9},\, \wilson[]{10})$)
  of the posterior including all experimental constraints. The renormalization scale is
  fixed to $\mu = 4.2\GeV$. For comparison, we include the case
  with $(\wilson[]{7},\,\wilson[]{9},\, \wilson[]{10})$ fixed at the SM
  values for which only nuisance parameters are varied (denoted by SM).
  The nuisance parameters are discarded when counting the degrees of freedom
  to compute the shown $p$-values $[\%]$ based on the statistics
  $T_{\rm like}$ and $T_{\rm pull}$. $^{\dagger}$ When comparing the posterior of the SM with the other modes,
  it has to be noted that the prior volume of $(\wilson[]{7},\,\wilson[]{9},\, \wilson[]{10})$
  is $6.68$ in log units.
\label{tab:bfp:gof}}
\end{table}

\begin{figure}[t]
  \centerline{
  \includegraphics[width=0.34\textwidth]{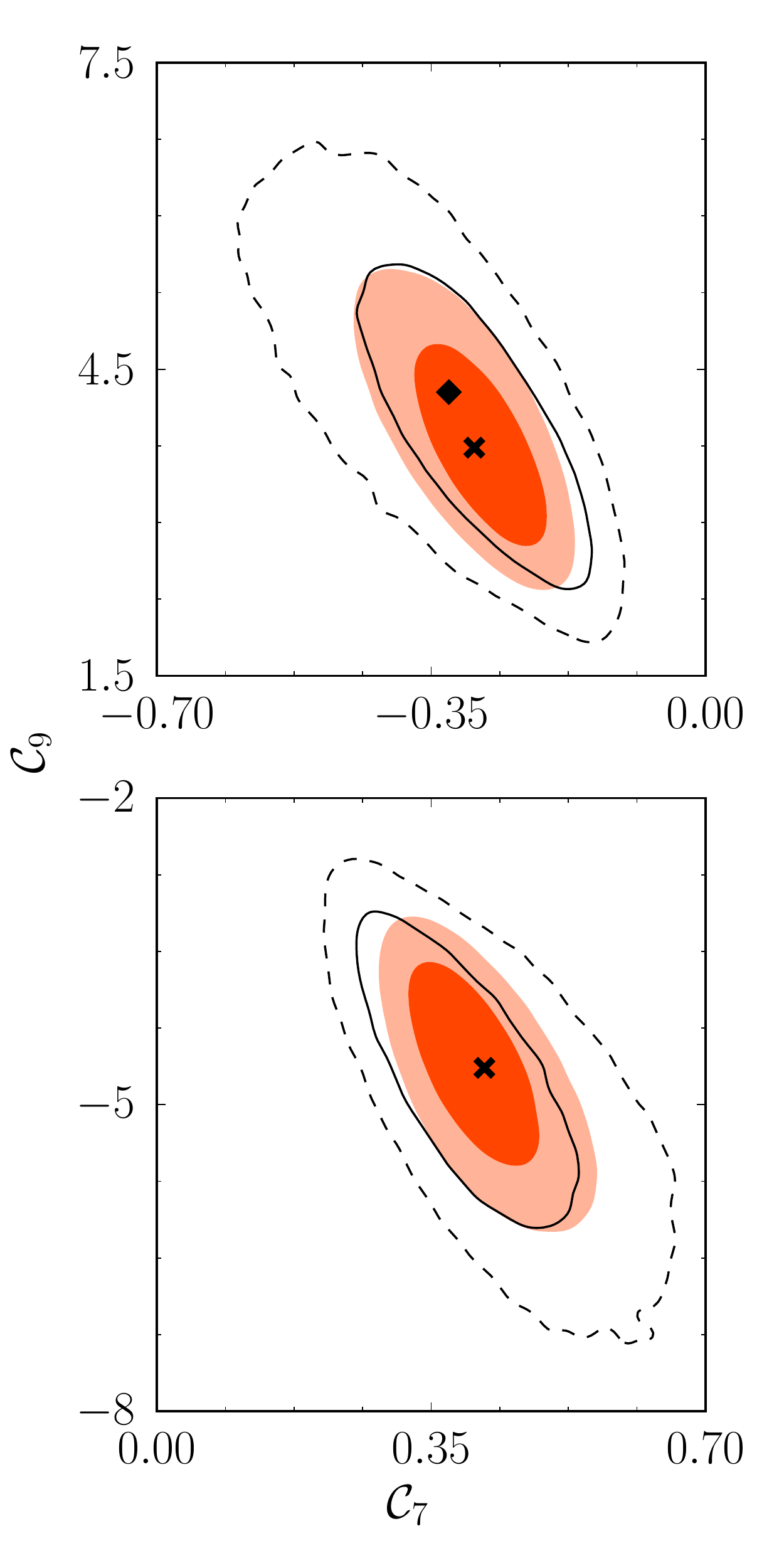}
  \includegraphics[width=0.34\textwidth]{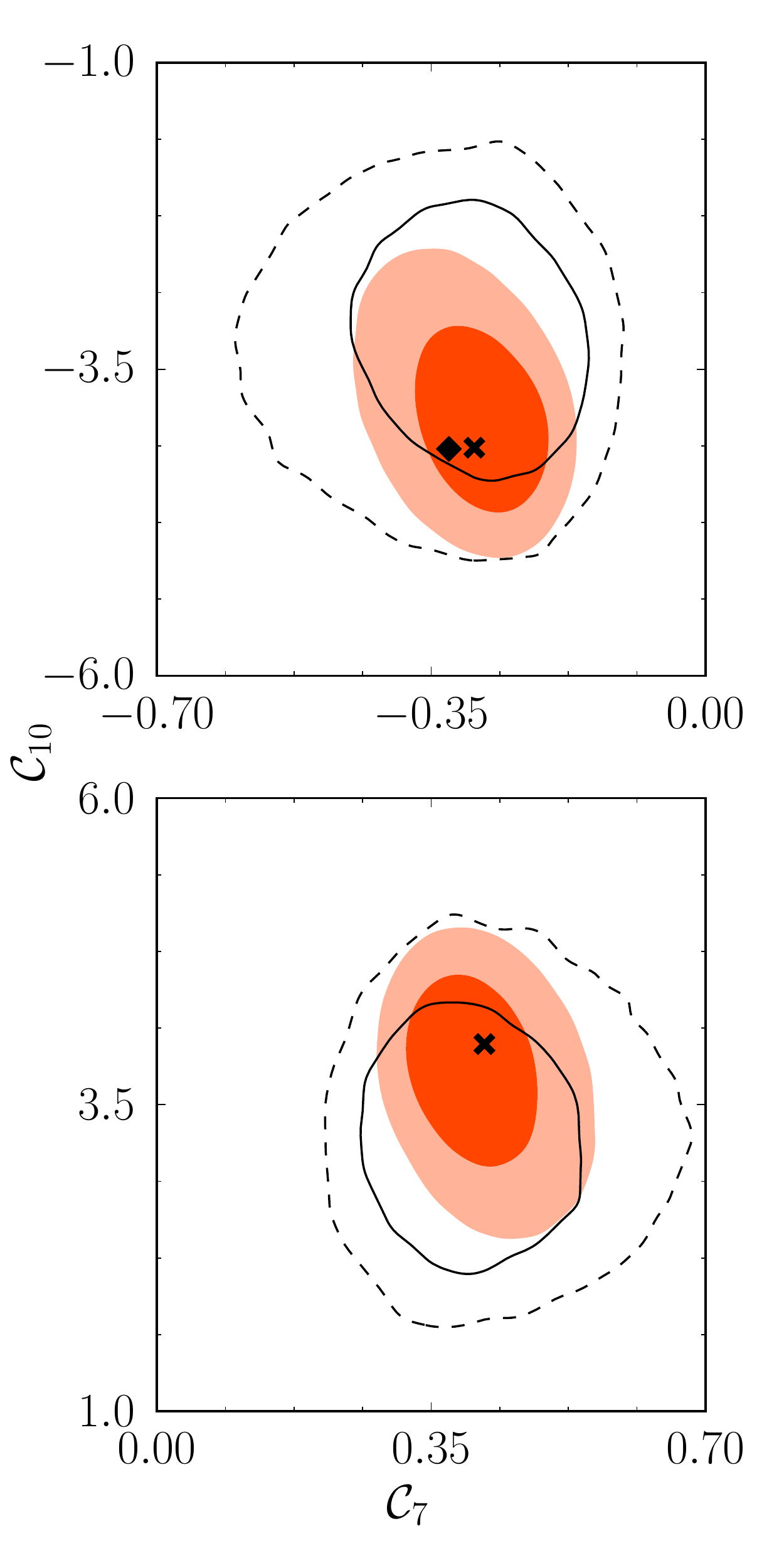}
  \includegraphics[width=0.34\textwidth]{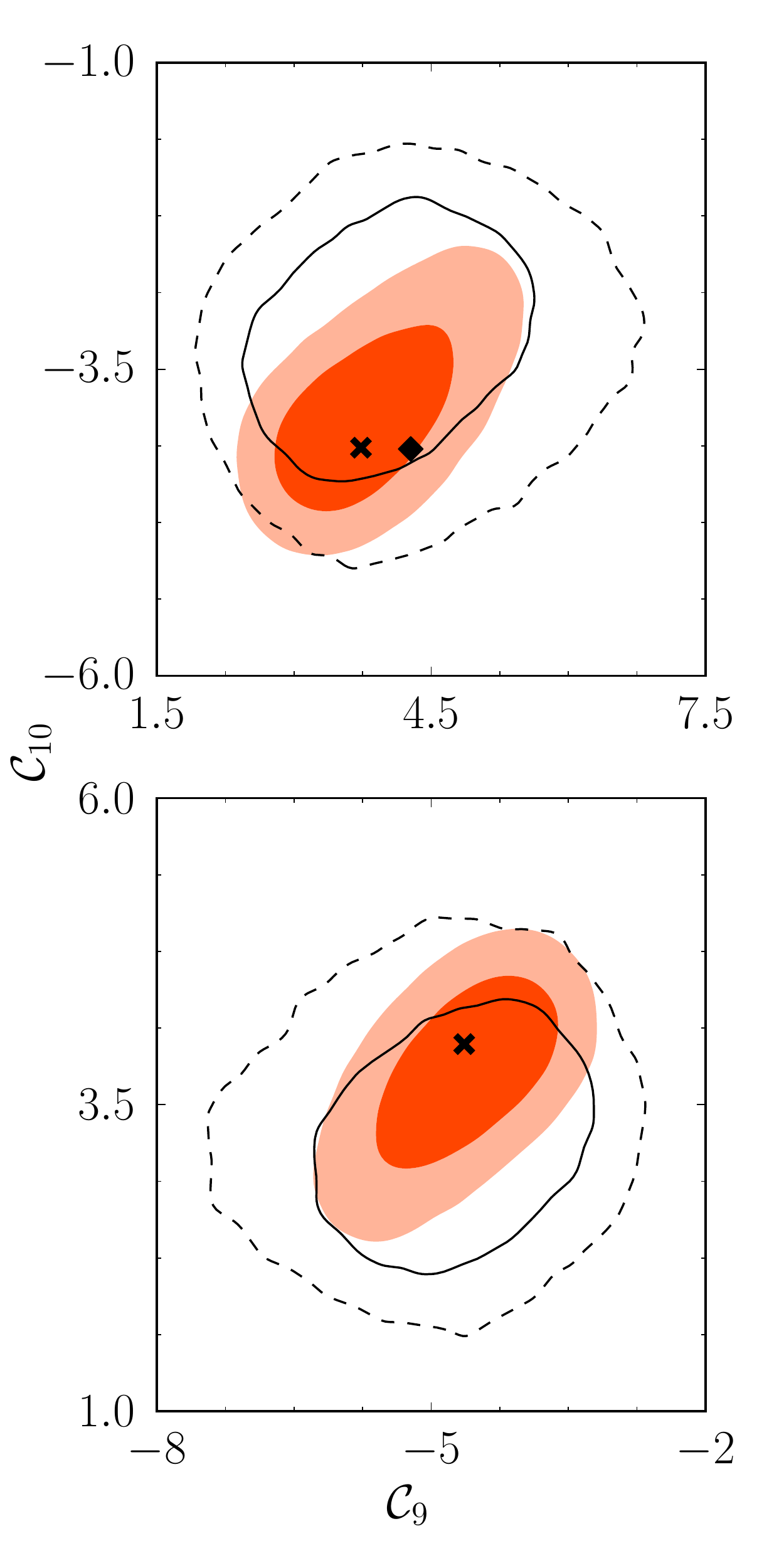}
  }
  \caption{The marginalized 2-dimensional 68\% (and 95\%) credibility regions of
    the Wilson coefficients $\wilson[]{7,9,10}$ at $\mu = 4.2\GeV$ for the SM-like [top row] and
    sign-flipped solution [bottom row], arising from nominal ranges of 
    \reffig{fig:fit-all-data} [red and light red, respectively] and wide ranges
    [solid and dashed contours, respectively] of the nuisance parameters.  We
    indicate the values of $\wilson[\rm SM]{7,9,10}$ in the SM
    [$\filleddiamond$] and at the local maximum of the posterior [\ding{54}]
    resulting from nominal prior ranges in the respective region.
    \label{fig:cmp-all-wide-nuisance}}
\end{figure}

\begin{table}
\begin{center}
\begin{tabular}{c|r@{ $\cup$ }l|r@{ $\cup$ }l|r@{ $\cup$ }l}
\hline\hline
\tabvsptop \tabvspbot

  & \multicolumn{2}{c|}{$\wilson[]{7}$}
  & \multicolumn{2}{c|}{$\wilson[]{9}$}
  & \multicolumn{2}{c}{$\wilson[]{10}$}
\\
\hline
\tabvsptop \tabvspbot
  68\%
  & $[-0.34, -0.23]$ & $[0.35,\, 0.45]$
  & $[-5.2, -4.0]$ & $[3.1,\, 4.4]$
  & $[-4.4, -3.4]$ & $[3.3,\, 4.3]$
\\
\tabvsptop \tabvspbot
  95\%
  & $[-0.41, -0.19]$ & $[0.31,\, 0.52]$
  & $[-5.9, -3.5]$ & $[2.6,\, 5.2]$
  & $[-4.8, -2.8]$ & $[2.7,\, 4.7]$
\\
\tabvsptop \tabvspbot
  max
  & $-0.28$ & $0.40$
  & $-4.56 $ & $3.64$
  & $-3.92 $ & $3.86$
\\
\hline
\tabvsptop \tabvspbot
  68\%
  & $[-0.39, -0.19]$ & $[0.30,\, 0.48]$
  & $[-5.6, -3.8]$ & $[2.9,\, 5.1]$
  & $[-4.0, -2.5]$ & $[2.6,\, 3.9]$
\\
\tabvsptop \tabvspbot
  95\%
  & $[-0.53, -0.13]$ & $[0.24,\, 0.61]$
  & $[-6.7, -3.1]$ & $[2.2,\, 6.2]$
  & $[-4.7, -1.9]$ & $[2.0,\, 4.6]$
\\
\tabvsptop \tabvspbot
  max
  & $-0.30$ & $0.38$
  & $-4.64 $ & $3.84$
  & $-3.24 $ & $3.30$
\\
\hline\hline
\end{tabular}
\end{center}
\caption{The 68\% and 95\% credibility intervals and the two local modes
  of the marginalized 1-dimensional posterior distributions
  of the Wilson coefficients $\wilson[]{7,9,10}$ at $\mu = 4.2\GeV$ for nominal [upper]
  and wide [lower] ranges of nuisance parameters (see \refapp{app:nuisance:pmrs}).
  \label{tab:wilson:coeff:1-dimCLs}
}
\end{table}

To study the dependence of our fit results on the priors, we use a second set of
priors (\emph{wide} priors). We scale the uncertainties of those parameters
associated with form factors and unknown subleading contributions in
$\Lambda/m_b$ (\reftab{tab:nuisance:FF}) by a factor of three and adjust the
parameter ranges accordingly. All other priors are kept the same.  This choice
includes the major sources of theory uncertainty and represents a
pessimist's view of a) the validity of form factor results based on LCSR at low
$q^2$, b) their extrapolation to high $q^2$ values, and c) subleading
corrections exceeding expectations from power counting. The results of the fit
at the low scale $\mu = 4.2\GeV$
to all data with these new priors is shown in \reffig{fig:cmp-all-wide-nuisance}
alongside the corresponding 68\%- and 95\%-credibility regions of
\reffig{fig:fit-all-data} for the two solutions in each of the three planes
$\wilson[]{7} - \wilson[]{9}$, $\wilson[]{7} - \wilson[]{10}$ and
$\wilson[]{9} - \wilson[]{10}$.  Most importantly, the fit is stable and gives
comparable results with both sets of priors thanks to the large number of
experimental constraints.  In all six planes, the area covered by the $68\%$
region with wide priors is similar to that of the $95\%$ region with nominal
priors. While the two sets of regions are concentric in the
$\wilson[]{7}-\wilson[]{9}$ plane, there appears a rather hard cut-off at
$|\wilson[]{10}| \approx 5$ in the $\wilson[]{7,9}$ -- $\wilson[]{10}$ planes. For
completeness, we list the set of smallest intervals and local maxima derived
from the one-dimensional marginalized distributions for $\wilson[]{7,9,10}$ for
both sets of priors in \reftab{tab:wilson:coeff:1-dimCLs}.  Our results for the
$95\%$ credibility intervals are compatible with those of
Ref. \cite{Altmannshofer:2011gn}. More specifically, we find a larger interval
for $\wilson[]{7}$, covering smaller values of $|\wilson[]{7}|$. This is due to
the use of $B\to X_s \gamma$ constraints that are used in
Ref.~\cite{Altmannshofer:2011gn}, but not included in our work. However, with
regard to $\wilson[]{9,10}$, we find that our credibility intervals are 
$10{-}40\%$ smaller. Compared to Ref.~\cite{Altmannshofer:2011gn}, we have added
the 2012 results by LHCb and BaBar. The question arises if the
inclusion of the inclusive decays $B \to X_s \gamma$ and $B\to X_s\bar\ell\ell$
could further shrink the $\wilson[]{9,10}$ credibility intervals. 

From the allowed ranges for $\wilson[]{7,9,10}$, we can estimate limits on the
scale of generic flavor-changing neutral currents at tree level, described by
\begin{align}
  \label{eq:Heff:gen}
  {\cal{H}}_{\rm eff} & =
   \sum_{i=7,9,10} \frac{\tilde{\mathcal{O}}_i}{(\Lambda^{\rm NP}_i)^2}\,,
\end{align}
\begin{align}
  \tilde{\mathcal{O}}_7 & =
     m_b \[\bar{s} \sigma_{\mu\nu} P_R b\] F^{\mu\nu}\,, &
  \tilde{\mathcal{O}}_{9,10} & =
    [\bar{s} \gamma_\mu P_L b] [\bar{\ell} \gamma^\mu (1, \gamma_5) \ell] \,.
\end{align}
Using $\wilson[]{i} = \wilson[SM]{i} + \wilson[NP]{i}$ and setting
$\wilson[]{i}$ to the boundary values of the $95\%$ intervals (nominal priors),
we extract $\wilson[NP]{i}$. By matching \refeq{eq:Heff:gen} with
\refeq{eq:Heff} and \refeq{eq:Heff:parts}, we extract the minimum scale
$\Lambda^{\rm NP}_i$ for both constructive and destructive interference with the
SM; see \reftab{tab:NP:scale}. The resulting scales above which NP ``is still
allowed'' are similar to those found in previous analyses \cite{Bobeth:2011nj}
and \cite{Altmannshofer:2011gn}.

\begin{table}
\begin{center}
\begin{tabular}{c|ccc}
\hline\hline
\tabvsptop
   &
  $\Lambda^{\rm NP}_7$ $[$TeV$]$ &
  $\Lambda^{\rm NP}_9$ $[$TeV$]$ &
  $\Lambda^{\rm NP}_{10}$ $[$TeV$]$
\\
\hline
\tabvsptop
  SM-like
  & 29, 38
  & 28, 37
  & 30, 44
\\
\tabvspbot
  SM-sign-flipped
  & 12, 13
  & 11, 13
  & 12, 13
\\
\hline\hline
\end{tabular}
\end{center}
\caption{Constraints on the NP scale $\Lambda^{\rm NP}_i$ ($i = 7, 9, 10$) assuming
  generic flavor violation at tree level using the 95\% credibility region from
  \reftab{tab:wilson:coeff:1-dimCLs}.
  Several possibilities arise from destructive and constructive interference
  of the SM with SM-like and SM-sign-flipped solutions.
  \label{tab:NP:scale}}
\end{table}

\begin{figure}[t]
  \includegraphics[width=0.50\textwidth]{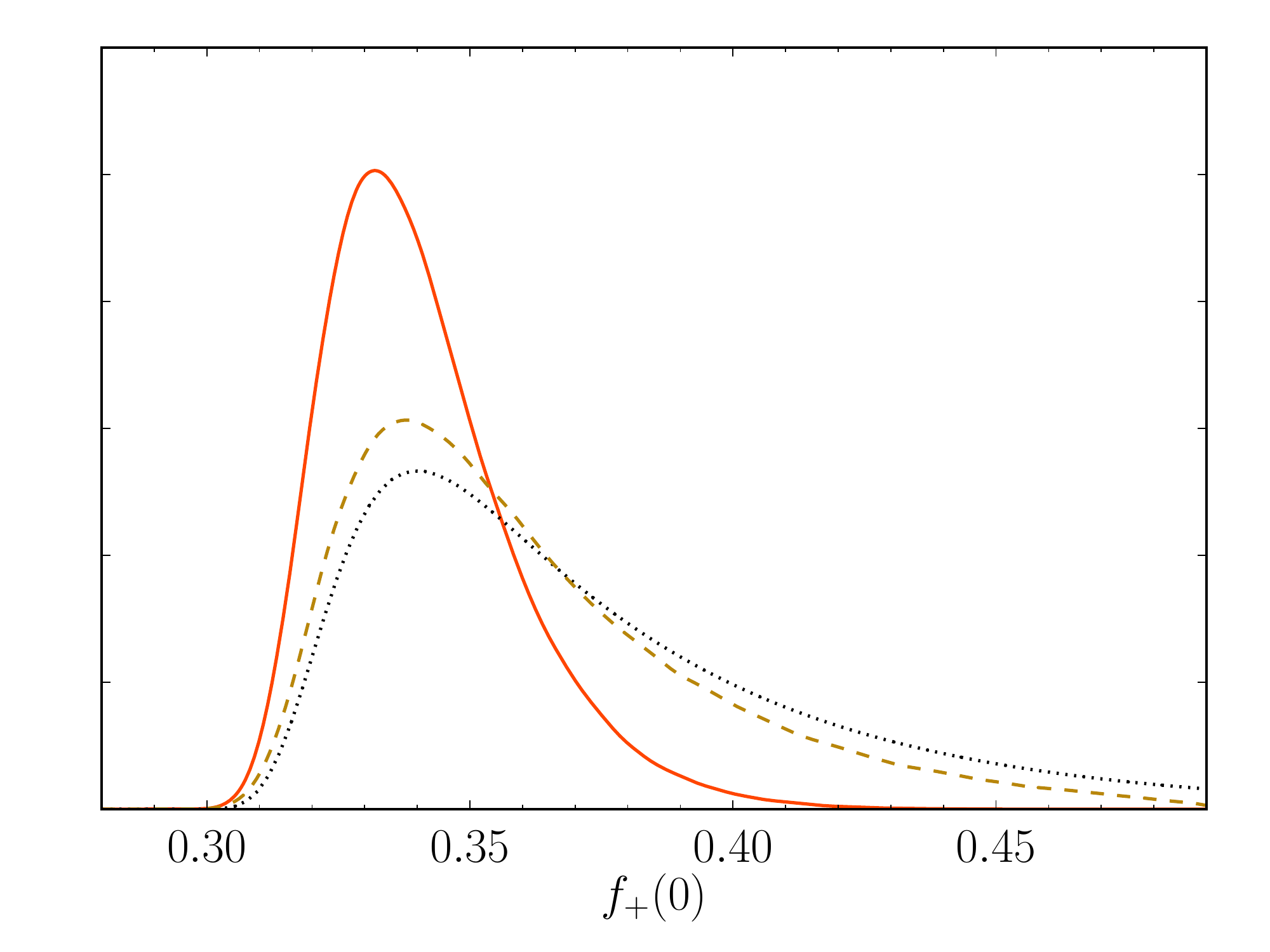}
  \includegraphics[width=0.50\textwidth]{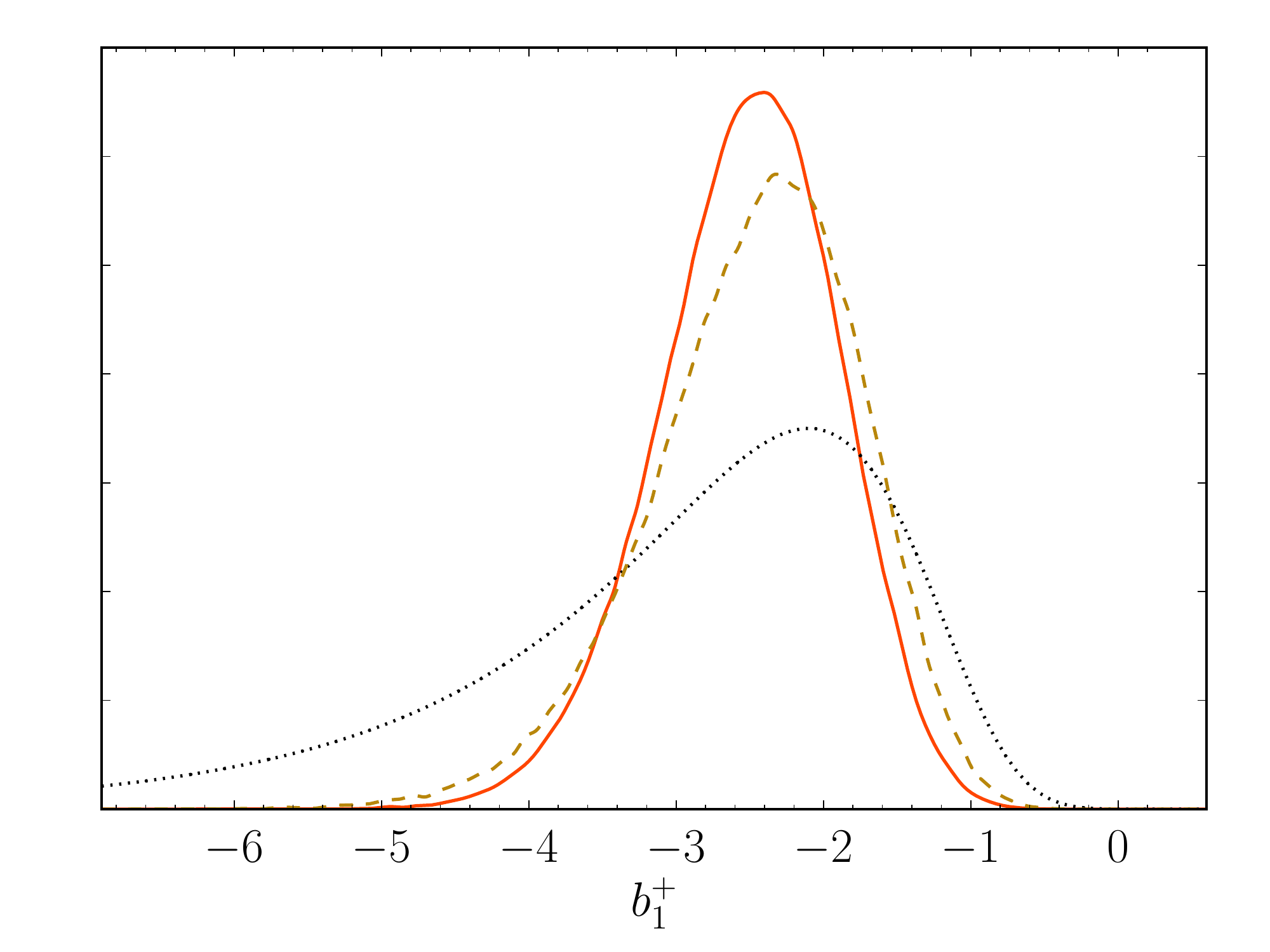}
  \caption{\label{fig:nuis:b1plus} Prior [dotted] and posterior distributions of
    the nuisance parameters $f_+(0)$ [left] and $b_+^1$ [right], governing the
    normalization and the $q^2$ shape of the $B\to K$ form factor $f_+(q^2)$,
    respectively. We show the posterior using $B\to K\bar{\ell}\ell$ data only
    [dashed] vs all data [solid].}
\end{figure}

So far, we discussed the fit results for the Wilson coefficients
$\wilson[]{7,9,10}$ that enter most, but not all of the observables. Exceptions
are those of $B\to K^* \gamma$, which depends only on $\wilson[]{7}$, and
$B_s\to \bar\mu\mu$, which depends only on $\wilson[]{10}$.  The marginalized
distributions in the $\wilson[]{9} - \wilson[]{10}$ plane of
\reffig{fig:fit-all-data} show that, compared to $B\to K^{*}$, the fit with
$B\to K$ only measurements prefers a smaller value of $|\wilson[]{9}|^2 +
|\wilson[]{10}|^2$; the marginal modes (not shown) are near $\wilson[]{9}=0,\;
\wilson[]{10} = \pm 5$.  Since the $B \to K^*$ constraints dominate the
combination, a ``tension'' arises.

Let us now discuss the role that the nuisance parameters play in the fit.
First, we note that the posterior distributions of the common nuisance
parameters --- those that are not specific to rare $b \to s$ decays, like the
CKM parameters and the $c\,$ and $b\,$ quark $\overline{\mbox{MS}}$ masses --- do
not deviate from their prior distributions given in
\reftab{tab:nuisance:num:input}. This is mainly due to the strong prior
knowledge from other measurements and the comparatively low precision of both
experimental and other, mostly hadronic, theory inputs in the rare $b \to
s$ decays.

Second, we consider the remaining hadronic nuisance parameters of form factors
and subleading corrections, for which the priors are based mostly on educated
guesses rather than precise knowledge. Because $B\to K$ and $B\to K^{*}$ form
factors enter observables at both low and high $q^2$, they are determined by all
the $B\to K\, \bar\ell\ell$ and $B\to K^* (\gamma,\, \bar\ell\ell)$ observables
respectively. In contrast, the parametrization of unknown subleading
$\Lambda/m_b$ corrections is different at low and high $q^2$ (and naturally in
$B\to K$ and $B\to K^*$ decays).
Since subleading
corrections at high $q^2$ receive further parametric suppression by either
$\wilson[]{7}/\wilson[]{9}$ or $\alpha_s$ \cite{Grinstein:2004vb,
  Bobeth:2010wg}, the corresponding observables at high $q^2$ are rather weakly
dependent on them. In contrast, at low $q^2$ large effects are not surprising.

Therefore, we expect a significant update to knowledge of form factors to
accommodate the tension between $B \to K$ and $B \to K^{*}$ constraints. Any
remaining tension should be visible in low-$q^2$ subleading corrections.

Let us first consider the posterior distributions of the two nuisance parameters
$f_+(0)$ and $b_1^+$, which enter the $q^2$ parametrization of the $B \to K$
form factor $f_+(q^2)$ (see \refeq{eq:BK-FF-param} and priors in
\reftab{tab:nuisance:FF} from LCSR results \cite{Khodjamirian:2010vf}).  The
$q^2$ shape of the form factor is controlled by $b_1^+$. The low- and high-$q^2$
data of the $B\to K\, \bar\ell\ell$ branching fraction
(\reftab{tab:BKll:expData}) give rise to a narrower posterior compared to the
prior distribution in \reffig{fig:nuis:b1plus}, which does not change much when
using only $B\to K\, \bar\ell\ell$ data or combining it with $B\to K^*
\bar\ell\ell$. This preference also appears when choosing the wide set of
ranges for the prior distributions of the nuisance parameters, demonstrating
that the data suppress the tails in the prior of $b_1^+$. Concerning $f_+(0)$,
which corresponds to the normalization of the form factor, we observe a strong
preference for low values in the posterior distribution in
\reffig{fig:nuis:b1plus}. However, this preference almost disappears when only
$B\to K\, \bar\ell\ell$ data is used in the fit.
This behavior persists even when allowing for wider prior ranges, and is easily
understood in terms of the above-mentioned tension.

We also find strong  modifications of the posterior with respect to prior distributions
for the three scale factors $\zeta_{A_1, A_2, V}$ entering the form factors
$A_1, A_2, V$ in $B \to K^{*}$. The posteriors are shown in \reffig{fig:nuis BV
  FF} along with the common prior distribution. Of the three, $A_1$ is known
most accurately after the fit, while $A_2$ and $V$ are simultaneously shifted
and compressed. Using all constraints, $A_1, A_2, V$ are shifted
towards higher values, but without $B \to K$ constraints, the shift actually
points in the opposite direction. Again, the positive shift serves to reduce the
tension and allows a good fit to all constraints with values of
$\wilson[]{9,10}$ smaller than required by the $B \to K^{*}$ constraints alone.

The parameters describing subleading phases are mostly unaffected by the
fit. All phases come out with a flat distribution, indicating that they could
have been omitted from the fit without any consequences.

The largest update to knowledge of subleading parameters occurs for the scale
factor of the transversity amplitudes $A^L_{0,\perp}$ (\ref{app:SL:corrections})
describing the $B \to K^*$ decays, with a downward shift of about $10\%$ and a
slight reduction of variance. We observe this effect only in the fit with all
observables. Neither $A^R_i$ nor $B \to K$ subleading parameters are updated
significantly in any of the fits. $A^R_i$ has little effect
compared to $A^L_i$ because the observables depend on $A^{L,R}_i \propto
\wilson[]{9} \mp \wilson[]{10}$, and $\wilson[]{9} \approx -\wilson[]{10}$.

In summary, we do not observe a drastic update of any nuisance parameter,
showing that the fit is stable \footnote{For the suppressed solutions, scale
  factors for $B \to K^*$ form factors and $A^L_{\perp}$ shift by
  $\mathcal{O}(15\%)$ and $A^L_{\parallel}$ even peaks at the left
  boundary.}. The uncertainty on the form factors and some subleading
corrections is reduced by the data, but the most likely values are shifted due
to the tension between $B \to K$ and $B \to K^*$ constraints. More theory input
is required to reduce the uncertainty on the remaining subleading corrections.

\begin{figure}[t]
  \begin{minipage}[t]{0.49\textwidth}
    \phantom{x} \vskip -0.8cm
    \includegraphics[width=1.05\textwidth]{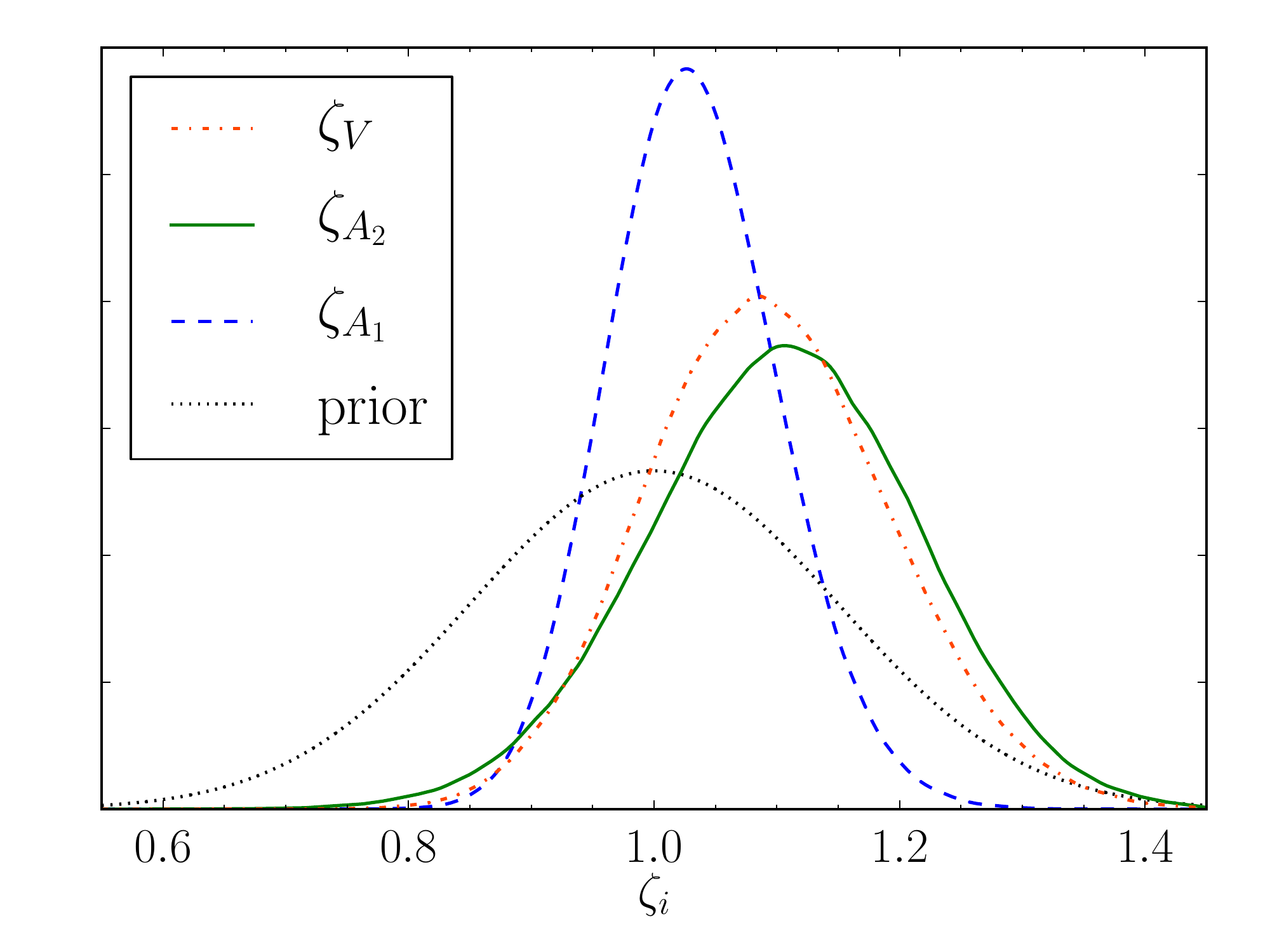}
  \end{minipage}
  \hspace{0.06\textwidth}
  \begin{minipage}[t]{0.45\textwidth}
     \caption{Posterior distributions of the fit with all data for the nuisance
       parameters $\zeta_{A_1,A_2,V}$ serving as scale factors to the
       corresponding $B \to K^*$ form factors. The common prior is indicated.
       \label{fig:nuis BV FF}}
   \end{minipage}
\end{figure}

%
\subsection{Predictions \label{sec:predictions}}

\begin{figure}[]
\centerline{
    \includegraphics[width=0.33\textwidth]{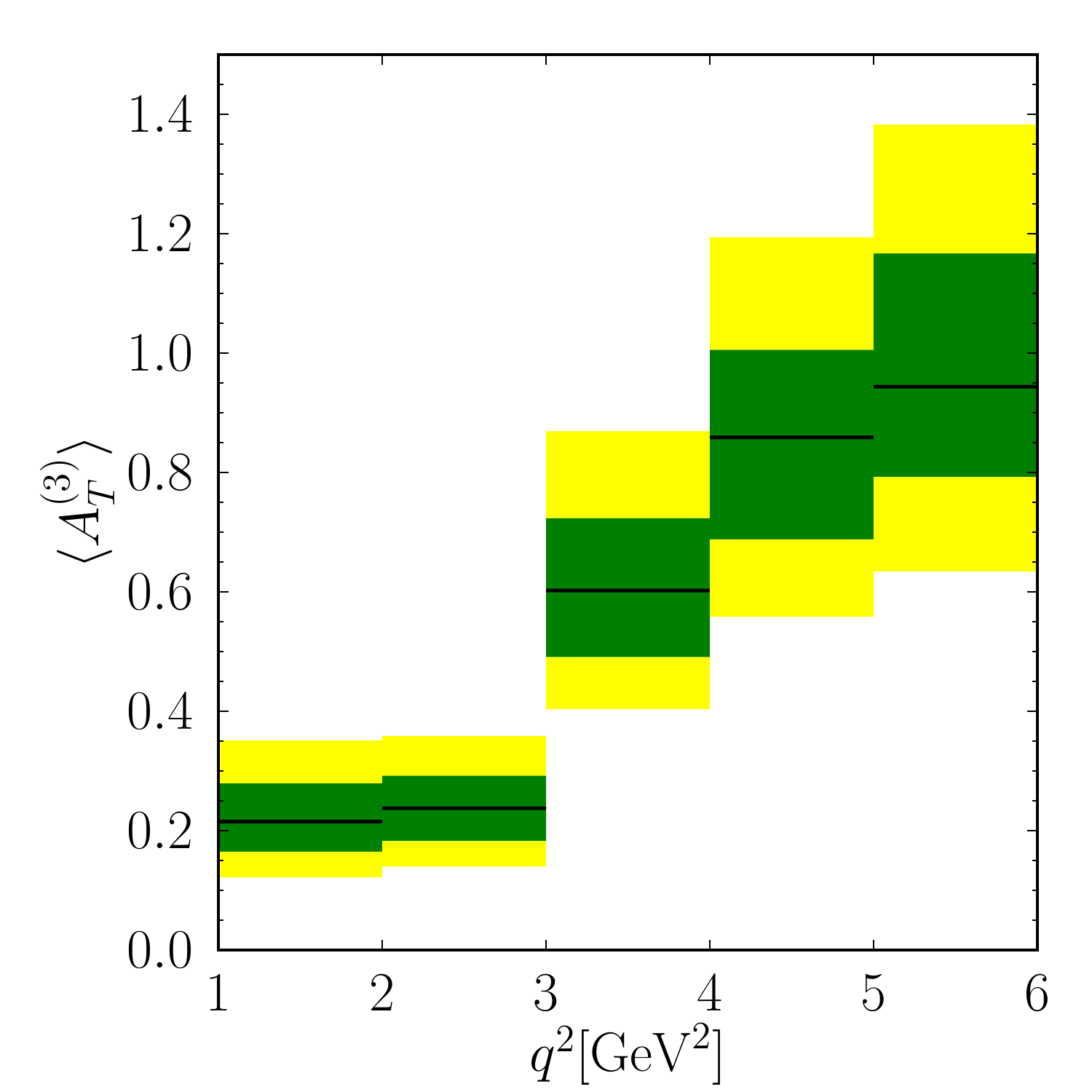}
    \includegraphics[width=0.33\textwidth]{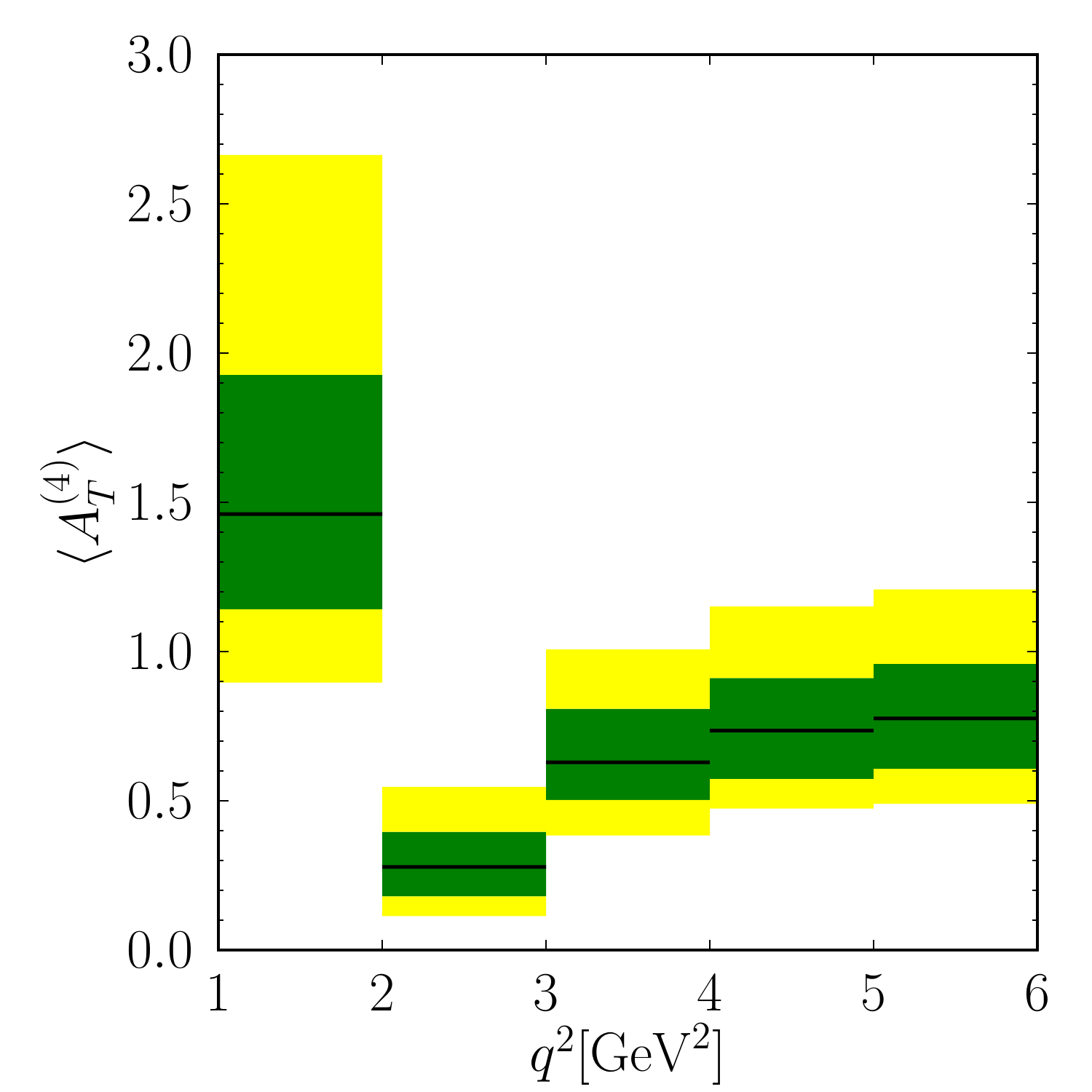}
    \includegraphics[width=0.33\textwidth]{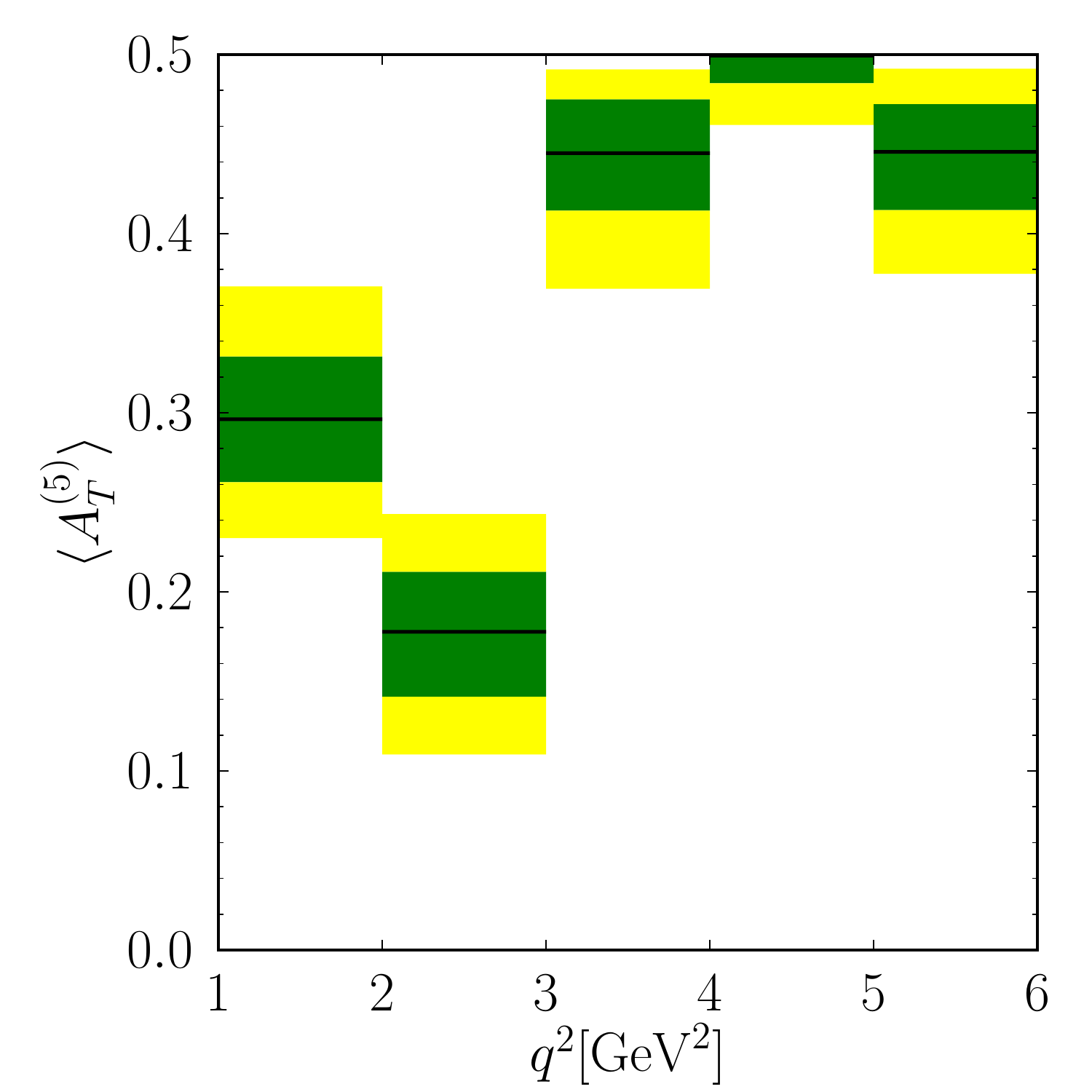}
}
\centerline{
    \includegraphics[width=0.33\textwidth]{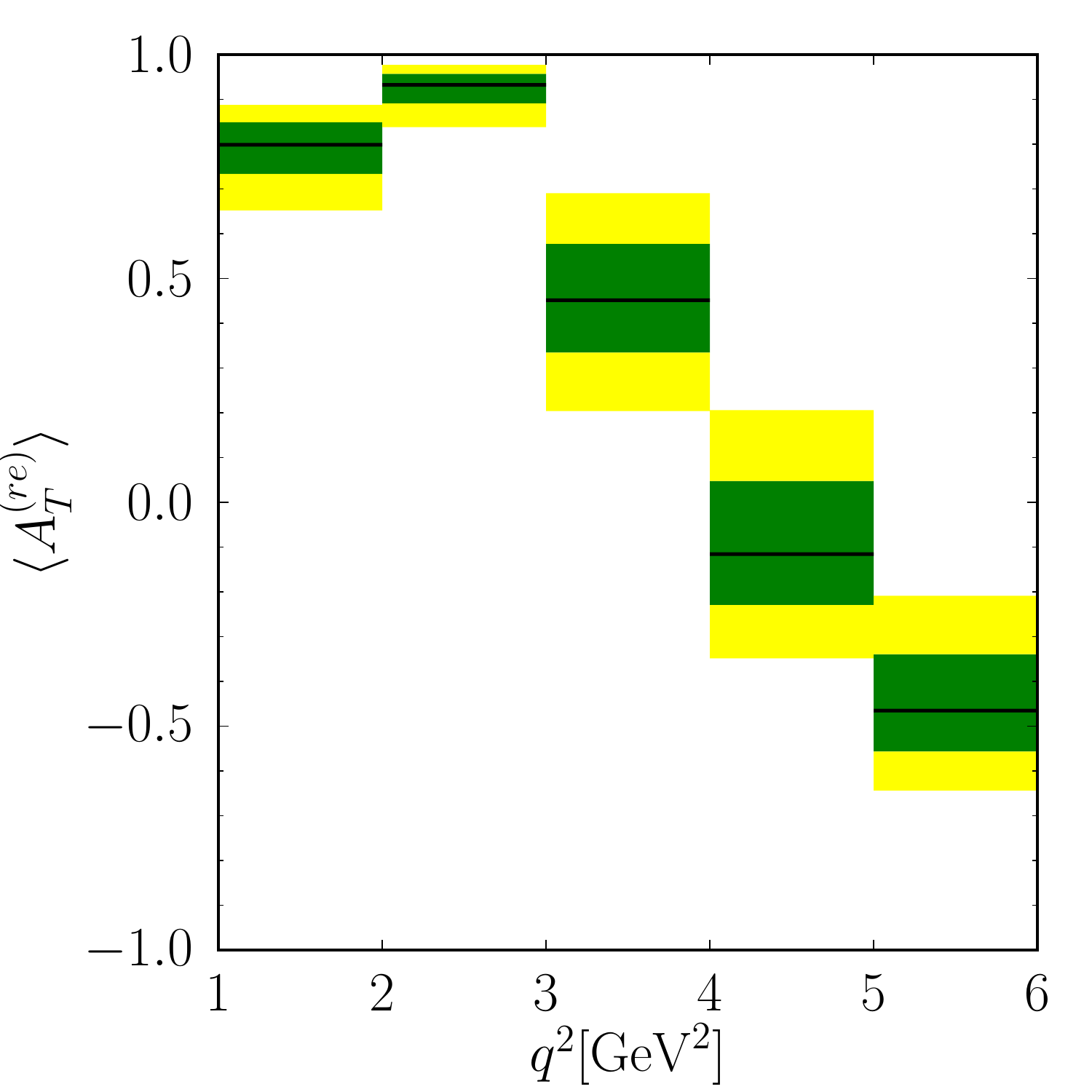}
    \includegraphics[width=0.33\textwidth]{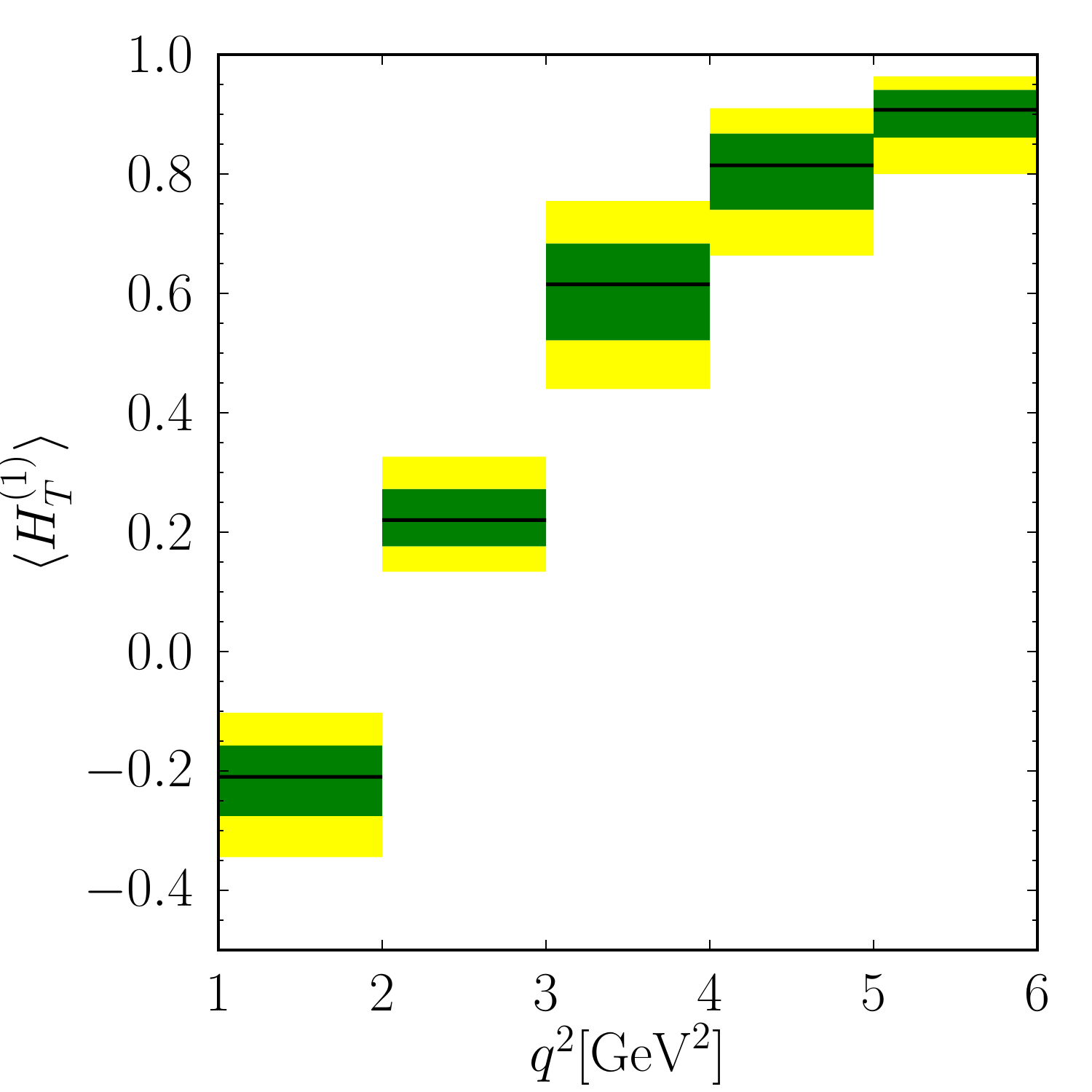}
    \includegraphics[width=0.33\textwidth]{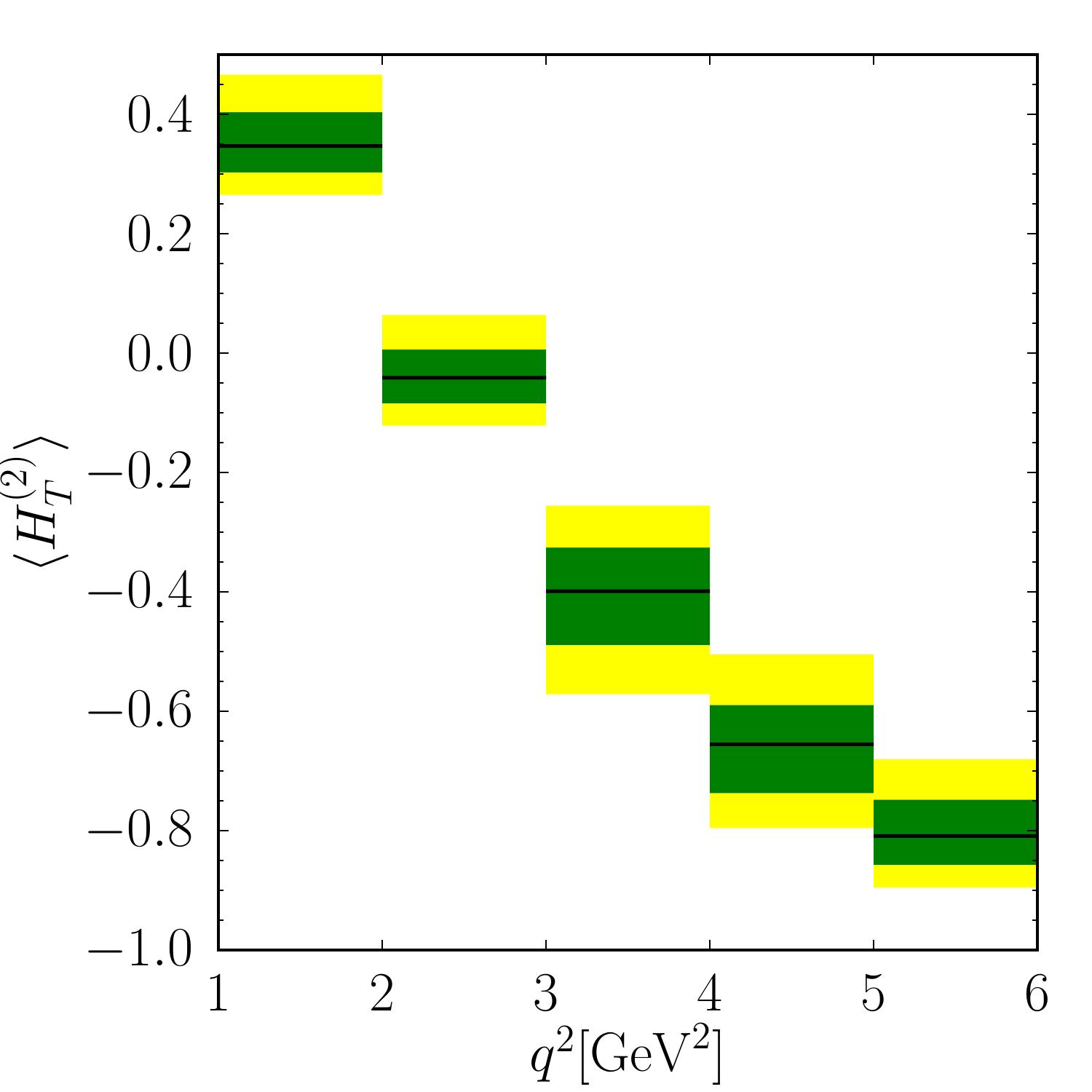}
}
\caption{Predictions of unmeasured optimized observables at large recoil based on the
  global fit output. We show the most probable value [solid black line] as well as
  the smallest $68\%$ (green) and $95\%$ (yellow) intervals of the $q^2$-integrated observables.
  \label{fig:uncert:q2-bins}}
\end{figure}
\begin{table}
\begin{center}
\begin{tabular}{c|ccc}
\hline\hline
\tabvsptop \tabvspbot
$q^2$-bin
& $\langle A_T^{(3)} \rangle $
& $\langle A_T^{(4)} \rangle $
& $\langle A_T^{(5)} \rangle $
\\
\hline
\tabvsptop \tabvspbot
$[1.0,\, 6.0]$
& $0.454 \; ^{+ 0.081} _{- 0.086} \; ^{+ 0.181} _{- 0.158}$
& $0.565 \; ^{+ 0.156} _{- 0.121} \; ^{+ 0.355} _{- 0.234}$
& $0.468 \; ^{+ 0.019} _{- 0.025} \; ^{+ 0.030} _{- 0.056}$
\\
\hline\hline
\tabvsptop \tabvspbot
$q^2$-bin
& $\langle A_T^{(\rm re)} \rangle $
& $\langle H_T^{(1)} \rangle $
& $\langle H_T^{(2)} \rangle $
\\
\hline
\tabvsptop \tabvspbot
$[1.0,\, 6.0]$
& $0.33 \; ^{+ 0.14} _{- 0.10} \; ^{+ 0.25} _{- 0.22}$
& $0.441 \; ^{+ 0.055} _{- 0.058} \; ^{+ 0.105} _{- 0.113}$
& $-0.271 \; ^{+ 0.057} _{- 0.060} \; ^{+ 0.117} _{- 0.117}$
\\
\hline\hline
\end{tabular}
\end{center}

\caption{Predictions of unmeasured, optimized observables based on
  global fit output integrated over the large recoil region.
  We list the most probable value, the smallest $68\%$ and $95\%$ intervals.
  \label{tab:NP:low-predictions}}
\end{table}

\begin{table}
\begin{center}
\begin{tabular}{c|ccc}
\hline\hline
\tabvsptop \tabvspbot
$q^2$-bin
& $\langle H_T^{(1)} \rangle $
& $\langle H_T^{(2)} \rangle $
& $\langle H_T^{(3)} \rangle $
\\
\hline
\tabvsptop \tabvspbot
$[14.18,\, 16]$
& $0.99969 \; ^{+ 0.00009} _{- 0.00011} \; ^{+ 0.00015} _{- 0.00026}$
& $-0.9843 \; ^{+ 0.0023} _{- 0.0022} \; ^{+ 0.0056} _{- 0.0039}$
& $-0.9837 \; ^{+ 0.0022} _{- 0.0019} \; ^{+ 0.0053} _{- 0.0033}$
\\[0.1cm]
$[16,19.21]$
& $0.99896 \; ^{+ 0.00025} _{- 0.00032} \; ^{+ 0.00044} _{- 0.00076}$
& $-0.9704 \; ^{+ 0.0018} _{- 0.0019} \; ^{+ 0.0042} _{- 0.0037}$
& $-0.9614 \; ^{+ 0.0015} _{- 0.0012} \; ^{+ 0.0037} _{- 0.0021}$
\\[0.1cm]
$[14.18, 19.21]$
& $0.99772 \; ^{+ 0.00058} _{- 0.00078} \; ^{+ 0.00105} _{- 0.00179}$
& $-0.9733 \; ^{+ 0.0027} _{- 0.0023} \; ^{+ 0.0057} _{- 0.0043}$
& $-0.9608 \; ^{+ 0.0019} _{- 0.0015} \; ^{+ 0.0045} _{- 0.0027}$
\\[0.2cm]
\hline\hline
\end{tabular}
\end{center}

\caption{Predictions of unmeasured, optimized observables based on
  global fit output for the two conventional bins and the entire low recoil region.
  We list the most probable value, the smallest $68\%$ and $95\%$ intervals.
\label{tab:NP:high-predictions}}
\end{table}

As outlined in \refsec{sec:BKstarll:constr}, the angular distribution of $B\to K^*
(\to K\pi)\, \bar\ell\ell$ allows one to form optimized observables, which have reduced
form factor uncertainties and may exhibit sensitivity to a particular type
of new physics. Currently, no measurements of these observables are
available. We provide predictions at low and high $q^2$ within the scenario
of the SM operator basis, taking into account the present data. Consequently, future
observations outside the predicted ranges would indicate physics beyond
the considered scenario.

The predictions of $A_T^{(3,4,5,{\rm re})}$ and $H_T^{(1,2)}$ at low $q^2$ are
given in $q^2$-integrated form for the bin $q^2 \in [1, 6]$ GeV$^2$ in
\reftab{tab:NP:low-predictions}. In addition, \reffig{fig:uncert:q2-bins} shows
the results of the 5 sub-bins with a bin width of $1$ GeV$^2$, as used in the
first measurement of the lepton $A_{\rm FB}$ of $B\to K^*\bar\ell\ell$ by LHCb
\cite{LHCbnote08}. The observables $A_T^{(3,4)}$ have been chosen due to their
sensitivity to the chirality-flipped $\wilson[']{7}$ \cite{Egede:2008uy}. The
large discontinuity of $A_T^{(4)}$ in $q^2 \in [1, 3]$ GeV$^2$ is caused by the
zero crossing of $J_4$ in its denominator \refeq{eq:def:AT34}. The observable
$A_T^{(5)}$ is restricted by construction to take values in $[-0.5,\, 0.5]$ and
reaches its maximal value at the zero crossing of the lepton $A_{\rm FB}$
in the bin $q^2 \in [4,5]$ GeV$^2$ \cite{Egede:2010zc}. Its shape is sensitive to new physics contributions of the
Wilson coefficients. Note that the theory uncertainty is at a minimum when
$A_T^{(5)}$ approaches $0.5$.

The observable $A_T^{(\rm re)}$ reaches its maximal value of about 1.0 in $q^2 \in
[2,3]$ GeV$^2$ and has the very same zero crossing as
the leptonic $A_{\rm FB}$. Our results are in qualitative agreement with
\cite{Becirevic:2011bp}, who stressed that the deviation of the maximal value
from 1.0 and its position are sensitive to new physics. The observables $H_T^{(1,2)}$
were first proposed for the high-$q^2$ region \cite{Bobeth:2010wg} as long-distance
free observables. In addition, $H_T^{(1)}$ is also short-distance free, with
$|H_T^{(1)}(q^2)| = 1$, depending only on the sign of a form factor. Recently it was
shown that at low $q^2$, form factors also cancel in $H_T^{(1,2)}$ \cite{Matias:2012xw}.
Each has a zero crossing in the region $q^2 \in [1,3]$ GeV$^2$ that is the very
same as in the CP-averaged normalized observables $J_4/\Gamma$ and $J_5/\Gamma$
\cite{Altmannshofer:2008dz, Bharucha:2010bb}. For $H_T^{(1)}$, one observes the
rise towards $\approx 1.0$ for rising $q^2$.

At high-$q^2$, the situation is more restrictive, and within the scenario of the
SM operator basis, there are only three optimized observables $H_T^{(1,2,3)}$
\cite{Bobeth:2010wg}. The predictions for three $q^2$ bins are given in
\reftab{tab:NP:high-predictions}. Besides $|H_T^{(1)}(q^2)| = 1$, we have the
additional relation $H_T^{(2)}(q^2) = H_T^{(3)}(q^2)$. Small deviations in the
predictions of $\aver{H_T^{(1,2,3)}}$ arise from separate $q^2$-integration of
$J_i$ (see \refeq{eq:aver:def} and below), such that the equality does not hold
exactly. Any large experimental deviation from the prediction $|H_T^{(1)}(q^2)|
= 1$ would signal a breakdown of the OPE \cite{Bobeth:2012??}.  The observables
$H_T^{(2,3)}(q^2)$ are given by the short-distance ratio \cite{Bobeth:2010wg}
\begin{align}
  H_T^{(2,3)}(q^2) &
  = \frac{2\, \mbox{Re} \left[\wilson[eff]{79}(q^2)\, \wilson[*]{10}\right]}
         {\left|\,\wilson[eff]{79}(q^2)\,\right|^2 + |\,\wilson[]{10}\,|^2}
  = \cos\left(\varphi_{79}(q^2) - \varphi_{10}\right) \frac{2\,r}{1 + r^2}
\end{align}
with
\begin{align}
  \wilson[eff]{79}(q^2) &
    = \wilson[eff]{9}(q^2) + \kappa \frac{2\, m_b^2}{q^2}\, \wilson[eff]{7}(q^2), &
  r(q^2) & = \frac{|\wilson[eff]{79}(q^2)|}{|\,\wilson[]{10}\,|}
\end{align}
and $\wilson[eff]{i}(q^2)$ and the factor $\kappa = 1 + \order{\alpha_s}$ of the
improved Isgur-Wise form factor relation defined in \cite{Bobeth:2010wg}.  In the SM,
$\wilson[SM]{10} \approx -4.2$ and therefore its phase $\varphi_{10} = \pi$. The
$q^2$ dependence of the sum of the effective Wilson coefficients
$\wilson[eff]{79}(q^2)$ is rather weak and its imaginary parts at NLO in QCD
small \cite{Bobeth:2011gi}, such that $\varphi_{79}(q^2) \approx 0$; whereas the
magnitudes of the Wilson coefficients are $\wilson[SM]{9} \approx +4.2$ and
$\wilson[SM]{7} \approx -0.3$, and lead to $r \approx 1$ and
$\cos\left(\varphi_{79}(q^2) - \varphi_{10}\right) \approx -1$. Therefore,
$H_T^{(2,3)}$ test roughly the ratio of $|\,\wilson[]{9}\,|/|\,\wilson[]{10}\,|$
within our scenario of the SM operator basis and real Wilson coefficients. The
results in \reftab{tab:NP:high-predictions} show that current data do not allow
for deviations from the SM prediction. 
We remark that the prediction of $\aver{H_T^{(1)}}$ is based on
the OPE and is expected to be 1 at any particular value of $q^2$. Therefore, our 
results just reflect how precisely the form factor and the modeled 
subleading corrections cancel for the $q^2$-integrated version when taking 
into account the update of our knowledge of the nuisance parameters due to
the experimental information. 

\begin{figure}[t]
\includegraphics[width=0.5\textwidth]{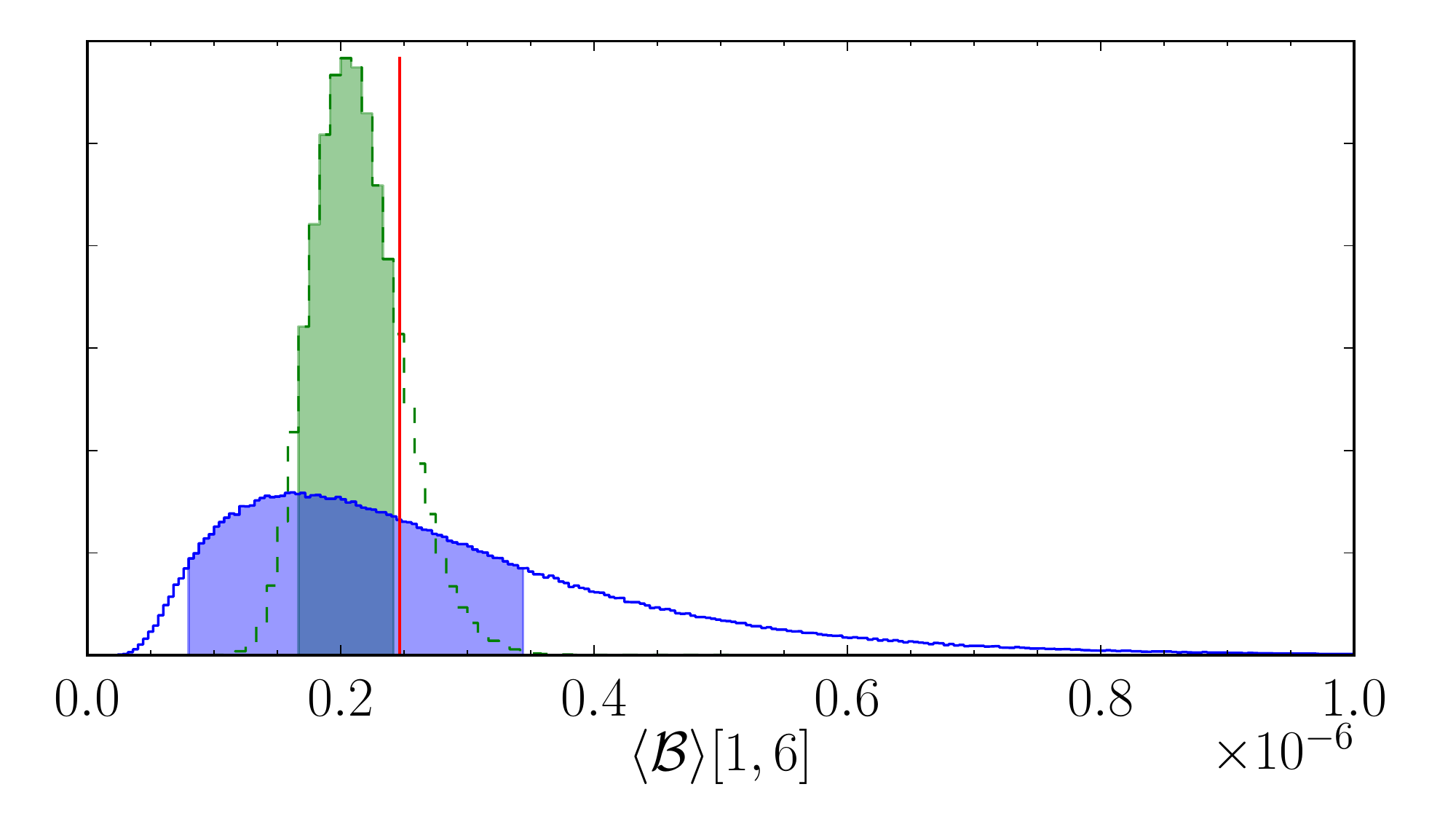}
\includegraphics[width=0.5\textwidth]{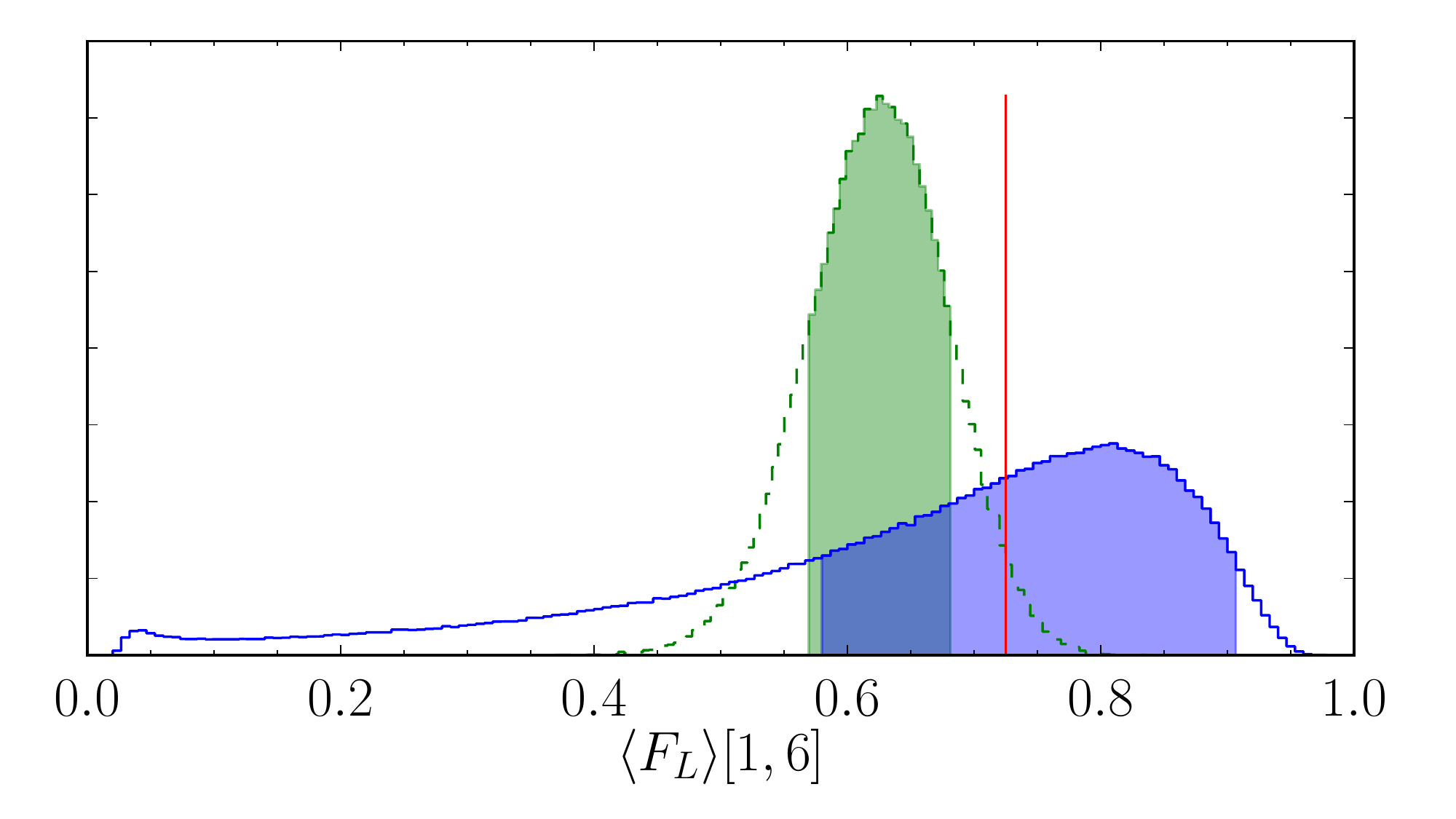}
\includegraphics[width=0.5\textwidth]{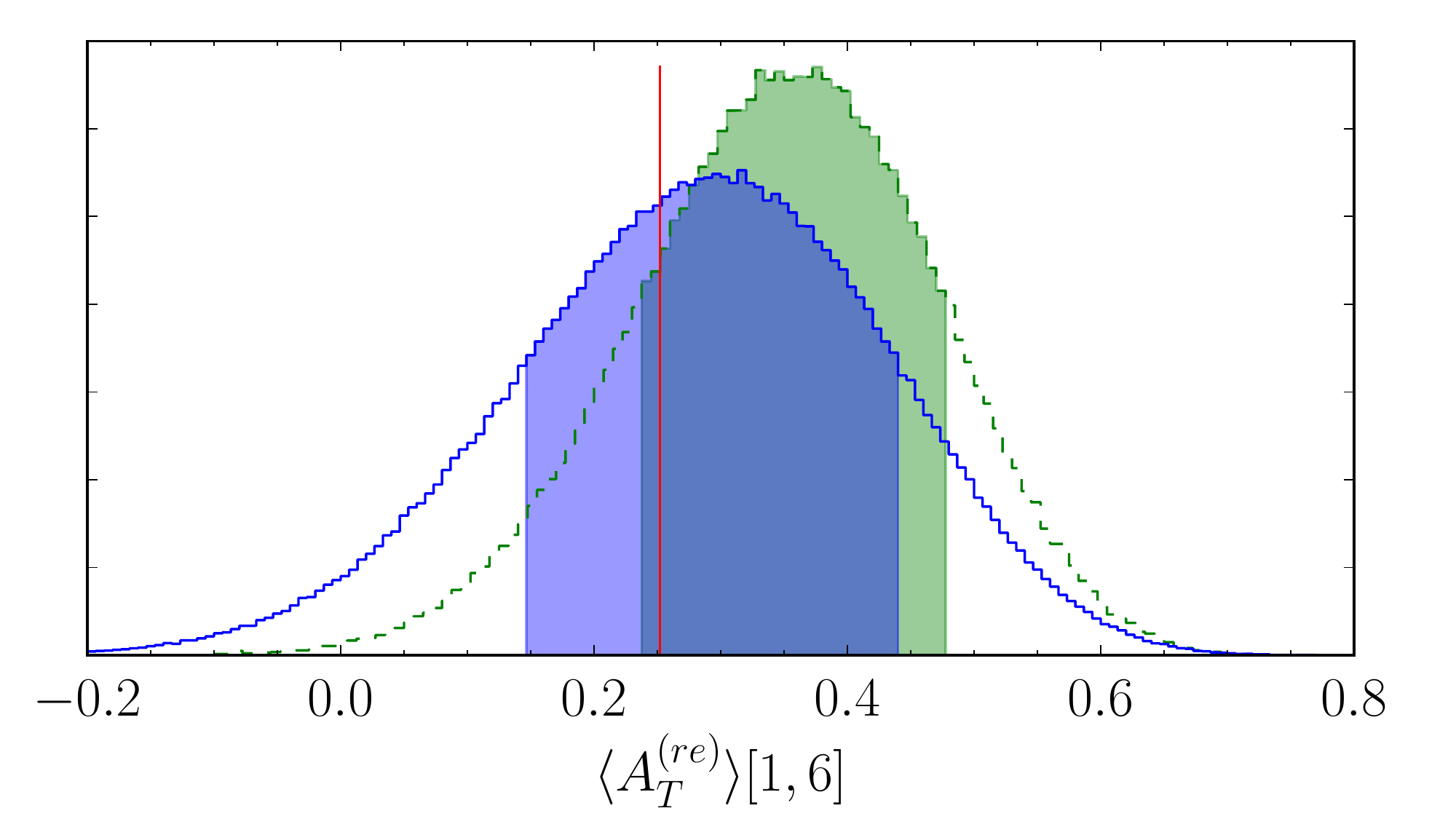}
\includegraphics[width=0.5\textwidth]{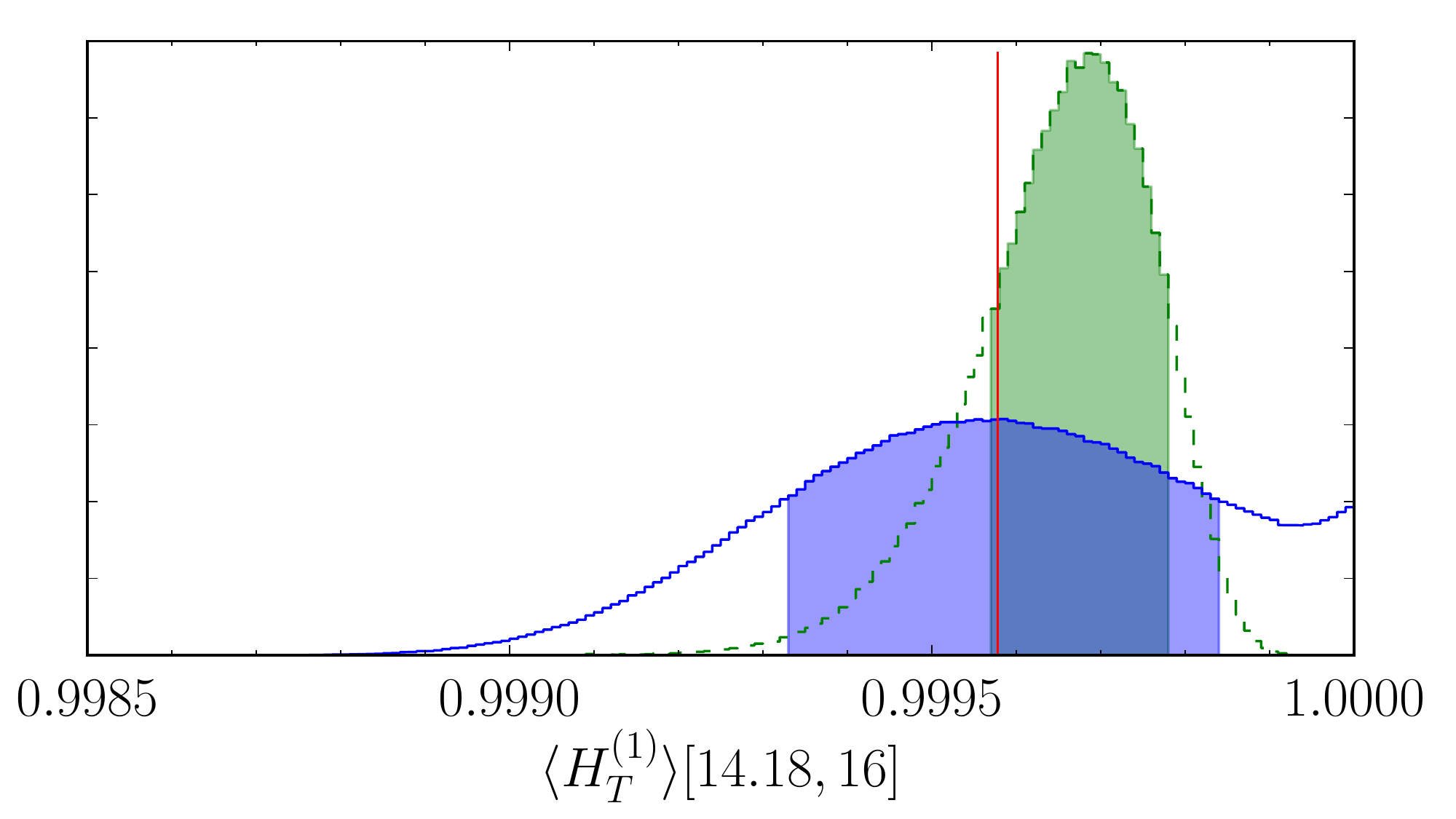}
\caption{ \label{fig:unc-prop:1}
  Probability distributions of the SM predictions of $q^2$-integrated observables in the
  $B\to K^*\bar\ell\ell$ decay, when
  varying nuisance parameters within their allowed prior ranges [solid, blue].
  The shaded region is the 68\% interval and the vertical (red) line indicates
  the prediction when using central values of nuisance parameters. Also shown
  are the predictions based on posterior distributions [dashed, green] determined
  by the experimental data, allowing also for NP in $\wilson[]{7,9,10}$ in the
  fit.
}
\end{figure}

Although SM predictions have been given previously \cite{Altmannshofer:2008dz,
  Altmannshofer:2011gn, Bobeth:2011nj}, our Monte Carlo approach described in
\refsec{sec:theo:unc:method} provides several improvements with respect to the
standard procedure to estimate theory uncertainties. Usually observables
$X(\nu)$ are computed at three values of a single parameter $\nu$: at the
central value $\nu_{\rm cen}$ and at $(\nu_{\rm cen})^{+b}_{-a}$.  The changes
in the predictions of $X$ are then interpreted as the associated uncertainty:
$\sigma_{+,-} = |X(\nu_{\rm cen}) - X(\nu_{\rm cen}\ ^{+b}_{-a})|$, and the
central value of $X$ is assumed to be $X(\nu_{\rm cen})$.  In the presence of
several parameters, the respective uncertainties are then combined either
linearly or in quadrature into a total uncertainty. In contrast to this
so-called min-max approach, we vary all parameters at the same time and thus
automatically take correlations into account.  Our intervals have a strict
probabilistic interpretation as Bayesian credibility intervals, and the
procedure automatically takes care of non-linearities.  As a simple example
consider the quadratic dependence of a branching ratio $\mathcal{B} \propto f^2$
on a decay constant or form factor $f$. Assuming a Gaussian prior distribution
of $f$, $p(\mathcal{B})$ is the (asymmetric) $\chi^2$-distribution with one
degree of freedom. Typical examples of asymmetry can be seen in in
\reffig{fig:unc-prop:1} for $\aver{{\cal B}}[1, 6]$ and $\aver{F_L}[1, 6]$
(blue, solid) of the decay $B\to K^*\bar\ell\ell$, where the maximum of the
distribution deviates from the vertical (red) line that indicates the
prediction obtained by using central values for all nuisance parameters;
i.e., the position of the maxima of their priors. This behavior is not present in
$\aver{A_T^{(\rm re)}}[1,6]$ since there, form factors cancel; likewise in
$\aver{H_T^{(1)}}[14.18,\,16]$. We list the modes and $68\%$ intervals for a
number of observables in \refapp{app:SM:predictions} in
\reftab{tab:SM:low-predictions} and \reftab{tab:SM:high-predictions}, but
stress that the uncertainty of an observable $X$ is best described by the
probability distribution $p(X)$. In the simplest case, $p(X)$ can be described
by the $68\%$ interval and the mode, but in general, it contains much more
information as demonstrated in \reffig{fig:unc-prop:1}.

Let us finally compare the SM predictions of observables based on the
prior information with predictions based on posterior distributions as
determined by experimental data and allowing also for NP in
$\wilson[]{7,9,10}$. Our posterior findings are overlaid on the SM
predictions for the examples in \reffig{fig:unc-prop:1}. Although NP
contributions to the Wilson coefficients are included, in all cases
the posterior distributions are narrower than the SM prediction based
on prior knowledge only. Obviously, the additional information from
data on the nuisance parameters updates our knowledge on quantities
$\aver{{\cal B}}[1, 6]$ and $\aver{F_L}[1, 6]$, which served as inputs
to the fit.

As described in the previous section, both fit solutions for $\wilson[]{7,9,10}$
give an overall good description of the experimental data.  For the optimized
observable $\aver{A_T^{(\rm re)}}[1,6]$, they yield a prediction of similar
range, shifted slightly towards larger values, compared to the SM prediction
based on prior knowledge alone. The same situation emerges for the other
optimized observables, which are free of form factor uncertainties --- compare
\reftab{tab:NP:low-predictions} and \ref{tab:SM:low-predictions} for low-$q^2$
as well as \reftab{tab:NP:high-predictions} and \ref{tab:SM:high-predictions}
for high $q^2$ --- and main uncertainties are due to lacking subleading
corrections. At this stage, better prior knowledge on the nuisance parameters is
needed. This will help to distinguish new physics from the SM with the help of
optimized observables, in the scenario of the SM operator basis with
real Wilson coefficients. However, any experimental observation
outside of the predicted range would point strongly to an extended scenario.

%
%
\section{Conclusion}

We perform a fit of the short-distance couplings $\wilson[]{7,9,10}$ appearing
in the effective theory of $\Delta B{=}1$ decays describing $b\to s\gamma$ and
$b\to s\,\bar\ell\ell$ transitions, assuming $\wilson[]{7,9,10}$ to be real
valued. For the first time, we include all relevant theory uncertainties in the
analysis by means of nuisance parameters.  Measurements of exclusive rare decays
$B\to K^* \gamma$, $B\to K^{(*)}\,\bar\ell\ell$ and $B_s \to \bar\mu\mu$
obtained by CLEO, BaBar, Belle, CDF, and LHCb serve as experimental inputs.
Besides presenting credibility intervals for the Wilson coefficients, we analyze
the goodness of fit of the obtained solutions. For the best-fit solution, we
show the pull values for all measurements in Figures
\ref{fig:pull:BKstargamma:BKll} and \ref{fig:pull:BKstarll}.  We use a novel
combination of Markov Chain Monte Carlo and adaptive importance sampling methods
in order to cope with the high dimensionality of the parameter space ($\sim 30$)
and the multimodal posterior distribution. With this approach, we can massively
parallelize the costly evaluation of the posterior. Our results should simplify
subsequent model-dependent studies; we are happy to provide the fit output in a
suitable format upon request.

The credibility intervals of the marginalized one- and two-dimensional posterior
distributions of $\wilson[]{7,9,10}$ are the main results of our fit, given in
\reftab{tab:wilson:coeff:1-dimCLs} and \reffig{fig:fit-all-data}.  Due to a
discrete symmetry, a SM-like and a flipped-sign solution remain with posterior
mass ratio of roughly 51\% over 49\%. Other local maxima exist, but their
posterior masses are negligible. The SM values $\wilson[\rm SM]{7,9,10}$ are
close to the best-fit point. Both solutions as well as the SM itself provide a
good fit of the data. Judging by the Bayes factor as model comparison criterion,
the data clearly favor the plain SM over a model with arbitrary real
$\wilson[]{7,9,10}$ --- a tribute to Occam's razor. Thus, from a purely
statistical point of view, even the simplest model-independent extension of the
SM is not necessary to describe the current data. We emphasize that the presence
of the sign-flipped solution still allows large NP contributions to the Wilson
coefficients. However, the degeneracy of the observables does not allow us to
distinguish them easily. This degeneracy is mildly broken by contributions of
4-quark operators, typically included in the effective Wilson coefficients
$\wilson[]{7,9} \to \wilson[eff]{7,9}$. Assuming improved theory uncertainties
and current experimental central values in $B\to K^*\gamma$, the fit suggests
that the additional information on $\wilson[eff]{7}$ enhances the SM-like
solution over the flipped-sign solution. We expect a reduced theory uncertainty
when including $B\to X_s \gamma$.

We provide updated predictions within the SM of selected observables in the
angular distribution of $B\to K^*(\to K\pi)\,\bar\ell\ell$. Based on prior
knowledge only, we obtain reduced theory uncertainties due the improved handling
of uncertainty propagation, observing that the central values of previous analyses
\cite{Bobeth:2010wg,Bobeth:2011gi,Altmannshofer:2011gn} are contained in the smallest $68\%$ regions.

Based on the fit output, we predict ranges for currently unmeasured observables
that exhibit a reduced form factor dependence.  Surprisingly, the predictions
based on the fit output yield smaller ranges than SM predictions based on prior
knowledge.  The extra variance due to Wilson coefficients is more than
compensated for by the reduced uncertainties as the fit constrains some of the
nuisance parameters and yields the correlation between all parameters.

We observe that a fit with current $B \to K\,\bar\ell\ell$ constraints prefers smaller
values of $\wilson[]{9,10}$ than a fit with the $B \to K^*\bar\ell\ell$
constraints. Including both sets of constraints, the fit accommodates this
tension by shifting the  $B\to K$ form factors towards
smaller values, and the $B\to K^*$ form factors towards larger values.

Future analyses can improve the fit by including results for the inclusive
decays $B\to X_s\gamma$ as well as $B \to X_s\bar\ell\ell$.  Besides the
inclusion of additional observables, further enhancements could arise when using
an alternative parametrization of $B \to K^*$ form factors; cf
\cite{Bharucha:2010im}.

\acknowledgments

We would like to thank our colleagues from LHCb Diego Martinez Santos and
Johannes Albrecht for providing help on the $B_s \to \bar\mu\mu$ results, as
well as Ulrik Egede, Thomas Blake, Mitesh Patel, and Will Reece for valuable
comments on the data of $B\to K^*\bar\ell\ell$. We are grateful to Martin
Kilbinger for his support on questions related to Population Monte Carlo and
``pmclib''.  We thank Gudrun Hiller and Allen Caldwell for many stimulating
discussions and remarks as well as their advice on the draft. Daniel Greenwald
provided innumerable corrections to enhance legibility and clarity of this
article. Finally, we thank Martin Gorbahn for his support of our
work.\\[0.2cm]

%
%
\appendix

%
%
\section{Numerical Input \label{app:num:input}}

\begin{table}
\centering
\begin{tabular}{llr|llr}
\hline \hline
\tabvsptop
$\alpha_s(M_Z)$          & $0.11762$                &                         &
$m_\mu$                  & $0.106~\GeV$             & \cite{Nakamura:2010zzi} \\
$\alpha_e(m_b)$          & $1/133$                  &                         &
$m_t^{\rm pole}$         & $173.3~\GeV$     & \cite{:2009ec}          \\
$\sin^2\!\theta_W$       & $0.23116$      & \cite{Nakamura:2010zzi} &
$M_W$                    & $80.399~\GeV$  & \cite{Nakamura:2010zzi} \\
\hline
\tabvsptop
$\tau_{B^+}$             & $1.638~\pico\second$     & \cite{Nakamura:2010zzi} &
$\tau_{B^0}$             & $1.525~\pico\second$     & \cite{Nakamura:2010zzi} \\
$M_{B^+}$                & $5.2792~\GeV$            & \cite{Nakamura:2010zzi} &
$M_{B^0}$                & $5.2795~\GeV$            & \cite{Nakamura:2010zzi} \\
$M_{K^+}$                & $0.494~\GeV$             & \cite{Nakamura:2010zzi} &
$M_{K^0}$                & $0.498~\GeV$             & \cite{Nakamura:2010zzi} \\
\tabvsptop
$M_{K^{*+}}$             & $0.892~\GeV$             & \cite{Nakamura:2010zzi} &
$M_{K^{*0}}$             & $0.896~\GeV$             & \cite{Nakamura:2010zzi} \\
\tabvsptop \tabvspbot
$\tau_{B_s}$             & $1.472~\pico\second$     & \cite{Nakamura:2010zzi} &
$M_{B_s}$                & $5.3663~\GeV$            & \cite{Nakamura:2010zzi} \\

\hline
\tabvsptop
$\lambda_{B,+}$          & $0.485~\GeV$              & &
$f_{B^{0,+}}$            & $0.212~\GeV$              & \cite{Simone:2010zz}\\
$f_{K}$                  & $0.1561~\GeV$             & &
& &\\
\tabvsptop \tabvspbot
$f_{K^*_\perp}(2\GeV)$   & $0.173\GeV$                     & &
$f_{K^*_\parallel}$      & $0.217\GeV$                     & \\
\tabvsptop
$a_1(K)$                 & $0.048$                         & &
$a_2(K)$                 & $0.174$                         & \\
\tabvsptop
$a_1(K^*_\perp)$         & $0.1$                           & &
$a_2(K^*_\perp)$         & $0.1$                           & \\
\tabvsptop \tabvspbot
$a_1(K^*_\parallel)$     & $0.1$                           & &
$a_2(K^*_\parallel)$     & $0.1$                           & \\
\hline \hline
\end{tabular}
\caption{The numerical input used in the analysis.
  The mass of the strange quark has been neglected throughout.
  $\tau_{B^0}$ ($\tau_{B^+}$) denotes the lifetime of the neutral (charged) $B$ meson.
  The following parameters appear in expressions of $B \to (K,\,K^*)\,\bar\ell\ell$
  at large recoil:
  $\lambda_{B,+}$ denotes the first inverse moment of the $B$-meson distribution amplitude,
  whereas $f_M$ the decay constants and $a_{1,2}(M)$ are the first two Gegenbauer moments
  of the LCDA's of the respective Kaon states $M = K,\, K^*_\perp,\, K^*_\parallel$.
  \label{tab:fixed:num:input}
}
\end{table}

The numerical input values of parameters are listed in \reftab{tab:fixed:num:input},
for which the uncertainties have not been included since they are either very small
or they enter in numerically subleading contributions to the observables of
interest.

The theory predictions of all the relevant semileptonic and radiative processes
at large recoil are based on the QCDF results \cite{Beneke:2001at,Beneke:2004dp}.
These include the usage of the Light Cone Distribution Amplitudes (LCDA) of the
involved kaons which are parametrized in terms of Gegenbauer moments $a_{n}(M)$
($M = K,\, K^*_\perp,\, K^*_\parallel$). In this work, we include terms in the
expansion in Gegenbauer moments up to $n = 2$, using the central values in
\reftab{tab:fixed:num:input}.

Since the $a_n(M)$ also enter the computation of the $B\to K^*$ form factors
via LC sum rules \cite{Ball:2004rg}, variation of the former would lead
to double counting. Furthermore, the residual influence of the $a_n(M)$ on the
observables is small compared to that of other parameters. We therefore do not vary the
Gegenbauer moments.

In addition, QCDF makes use of the decay constants $f_M$ ($M = K,\,
K^*_\perp,\, K^*_\parallel$), which enter in numerically subleading
contributions. The central values are listed in
\reftab{tab:fixed:num:input}.

%
%
\section{Nuisance Parameters \label{app:nuisance:pmrs}}

In this section we present the nuisance parameters that are considered in this
work and contribute the main uncertainties in theory predictions. All the priors
of these parameters are clipped to the parameter range that corresponds to their
respective $3\,\sigma$ interval. For the sake of readability, we categorize the
individual nuisance parameters.

%
\subsection{Common Nuisance Parameters}

\begin{table}
\centering
\begin{tabular}{llr|llr}
\hline \hline
\tabvsptop
$A$                      & $0.804\pm0.010$                 & \cite{Bona:2006ah}   &
$\lambda$                & $0.22535\pm0.00065$             & \cite{Bona:2006ah}   \\
\tabvspbot
$\bar{\rho}$             & $0.111\pm0.070$                 & \cite{Bona:2006ah}   &
$\bar{\eta}$             & $0.381\pm0.030$                 & \cite{Bona:2006ah}   \\
\hline
\tabvsptop \tabvspbot
$m_c(m_c)$               & $(1.27^{+0.07}_{-0.09})~\GeV$   & \cite{Nakamura:2010zzi}   &
$m_b(m_b)$               & $(4.19^{+0.18}_{-0.06})~\GeV$   & \cite{Nakamura:2010zzi}  \\
\hline \hline
\end{tabular}
\caption{Common nuisance parameters: The CKM Wolfenstein parameter values as obtained
  from the CKM \emph{tree-level fit}, \cf Sec. \ref{sec:priors}.
  \label{tab:nuisance:num:input}
}
\end{table}

The common nuisance parameters are those that enter most of the observables
and are not specific for rare $b\to s$ decays. These are the elements of the
Cabibbo-Kobayashi-Maskawa (CKM) quark-mixing matrix and the $b$ and $c$ quark
masses.

For the purpose of the fit of rare $b\to s$ decays, we take the CKM matrix
elements from other observables such as tree decays.  We parametrize the CKM
matrix elements using the Wolfenstein parametrization to
$\mathcal{O}(\lambda^9)$ \cite{Charles:2004jd} and use the results of the
tree-level fit of the UTfit collaboration \cite{Bona:2006ah} as priors in the
fit of $b\to s$ decays. In this way, we include non-SM effects, but assume they
do not affect tree-level decays. However, we use the results of the SM CKM fit
in order to determine the uncertainties of observables in the framework of the
SM in \refsec{sec:predictions}.  Note that the CKM matrix elements only enter
the branching ratios of $B\to K^{(*)} + (\gamma,\, \bar\ell\ell)$ decays in the
combination $V_{tb}^{} V_{ts}^*$. Although numerically negligible, the
combination $V_{ub}^{} V_{us}^*$ entering all observables is included in the
analysis. It becomes relevant only for CP-asymmetric observables. All priors are
Gaussian, with their $1\,\sigma$ ranges given in
\reftab{tab:nuisance:num:input}.

The values of the quark masses $m_b$ and $m_c$ enter most observables. In order
to account for the asymmetric errors, we use $\LogGamma$ distributions (see
\refsec{sec:logg-distr}) as priors whose modes and $68\%$-probability intervals
match the values given in \reftab{tab:nuisance:num:input}.

%
\subsection{$B\to K^{(*)}$ Form Factors and $f_{B_s}$
  \label{app:form:factors}}

The heavy-to-light form factors $f_{+,T,0}$ for $B\to K$ and $V,\, A_{0,1,2}$, and
$T_{1,2,3}$ for $B\to K^*$ transitions present a major source of uncertainty in
predictions of rare exclusive $B$ decays. They are functions of the dilepton
invariant mass $q^2$ and we adopt the definition used in \cite{Beneke:2000wa,
  Beneke:2004dp, Khodjamirian:2010vf, Ball:2004rg}. Due to the application of
form factor relations at large and low recoil, only $f_+$ enters $B\to K$ and
$V$ and $A_{1,2}$ enter $B\to K^*$ transitions\footnote{The form factors $f_0$
  and $A_0$ do not contribute within the framework of the SM operator basis, up
  to negligible terms suppressed by $m_\ell^2/q^2$.}. The application of form
factor relations introduces uncertainties of order $\Lambda_{\rm QCD}/m_b$ that
will be discussed in \refapp{app:SL:corrections}.

Currently, the form factors are only known from Light Cone Sum Rules (LCSR) which
are applicable at low $q^2$. Lattice QCD can provide results at high $q^2$, where
quenched results for some form factors \cite{Becirevic:2006nm, AlHaydari:2009zr}
are available and some preliminary unquenched results have been reported
in \cite{Liu:2009dj, Zhou:2011be, Liu:2011raa}. An extensive discussion of the
$q^2$-shape parametrization using series expansion and a fit to low-$q^2$ LCSR
combined with high-$q^2$ lattice results (when available) can be found in
\cite{Bharucha:2010im}.

\begin{table}
\centering
\begin{tabular}{c|cccc}
\hline \hline
\tabvsptop \tabvspbot
        & $r_1$    & $r_2$    & $m_R^2~[\GeV^2]$ & $m_{\rm fit}^2~[\GeV^2]$\\
\hline
\tabvsptop
  $V$   & $0.923$  & $-0.511$ & $5.32^2$         & $49.40$\\
  $A_1$ & --       & $0.290$  & --               & $40.38$\\
\tabvspbot
  $A_2$ & $-0.084$ & $0.343$  & --               & $52.00$
\\
\hline \hline
\end{tabular}
 \caption{The parameters of the form factors $V$ and $A_{1,2}$.
   \label{tab:ff-parameters}}
\end{table}

With regard to $B\to K^*$ form factors $V, A_{1,2}$, we use the LCSR results at
low-$q^2$ as given in \cite{Ball:2004rg}, where the extrapolation to high-$q^2$
is based on a (multi-)pole ansatz
\begin{align}
  V   & = \frac{r_1}{1 - q^2 / m_R^2} + \frac{r_2}{1 - q^2 / m_{\rm fit}^2},
\end{align}
\begin{align*}
  A_1 & = \frac{r_2}{1 - q^2 / m_{\rm fit}^2}, &
  A_2 & = \frac{r_1}{1 - q^2 / m_{\rm fit}^2} + \frac{r_2}{(1 - q^2/m_{\rm fit}^2)^2},
\end{align*}
and the numerical values of the parameters given in \reftab{tab:ff-parameters}.
We do not vary these parameters themselves as they strongly depend on the LCSR
analysis, but rather assign one multiplicative scaling factor $\zeta_i$ per form
factor ($i = V, A_1, A_2$) to model the respective uncertainty such that the value
$\zeta_i = 1.0$ corresponds to the central value of the form factor. A Gaussian
prior is assigned to these nuisance parameters, which has a width of $\sigma = 0.15$
(i.e., 15\% uncertainty) and its support extends up to $3\, \sigma$ (i.e., maximally
45\% uncertainty), outside of which the prior is set to zero (see \reftab{tab:nuisance:FF}).
Note that in this way we do not vary the $q^2$ shape of the form factors. At large
recoil, two universal form factors \cite{Beneke:2004dp} appear
\begin{align}
  \xi_\perp & \equiv
     \frac{M_B}{M_B + M_{K^*}} V\,, &
  \xi_\parallel & \equiv
     \frac{M_B + M_{K^*}}{2 E_{K^*}} A_1 - \frac{M_B - M_{K^*}}{M_B} A_2\,,
\end{align}
and their variation is obtained by the uncorrelated variation of $V$ and $A_{1,2}$
as described above.

Since we calculate the $B\to K^*\gamma$ matrix element within QCDF for $q^2 = 0$,
all nuisance parameters that affect the process $B\to K^*\,\bar\ell\ell$
in the large recoil region likewise affect the radiative process, as far as
they are applicable.

With regard to the $B\to K$ form factor $f_+$, we use the BCL
parametrization \cite{Bourrely:2008za} of the LCSR results \cite{Khodjamirian:2010vf}
\begin{align}
  \label{eq:BK-FF-param}
  f_+(q^2) & = \frac{f_+(0)}{1 - q^2/M^2_{{\rm res},+}} \left[ 1 +
    b_1^+ \left( z(q^2) - z(0) + \frac{1}{2} \left[z(q^2)^2 - z(0)^2\right] \right)
    \right],
\end{align}
\begin{align*}
  z(s) & = \frac{\sqrt{\tau_+ - s} - \sqrt{\tau_+ - \tau_0}}
                {\sqrt{\tau_+ - s} - \sqrt{\tau_+ - \tau_0}}, &
  \tau_0 & = \sqrt{\tau_+} \left(\sqrt{\tau_+} - \sqrt{\tau_+ - \tau_-}\right), &
  \tau_\pm & = \left(M_B \pm M_K\right)^2.
\end{align*}
This parametrization depends on the central value of the form factor at $q^2 = 0$,
$f_+(0)$, and the slope parameter $b_1^+$ (and $M_{{\rm res},+} = 5.412$ GeV).
At large recoil, the dipole form factor $f_T$ is replaced by the large-energy
universal form factor $\xi_{P} \equiv f_+$ \cite{Beneke:2000wa, Bobeth:2007dw}.
At low recoil, the dipole form factor $f_T$ is substituted for by means of the improved
Isgur-Wise relation \cite{Bobeth:2011gi}.

In addition, we vary the decay constant $f_{B_s}$ of the $B_s$ meson, since it 
constitutes the dominant uncertainty in the decay $B_s\to\bar\mu\mu$.
The most recent lattice results \cite{McNeile:2011ng, Bazavov:2011aa} have been 
averaged \cite{Laiho:2009eu}, yielding the number listed in \reftab{tab:nuisance:FF}.

In order to assess the dependence of the fit on the choice of priors, we adopt
two sets of priors. The first set reflects the uncertainties as reported by the
authors of \cite{Ball:2004rg, Khodjamirian:2010vf, Laiho:2009eu}, thereby
assuming the extrapolation of form factors to high $q^2$ has the same
uncertainties as predicted by LCSR's at low $q^2$. In the second set we triple
the uncertainties. Both sets are given in \reftab{tab:nuisance:FF}.

\begin{table}
\centering
\begin{tabular}{c|c|cc|cc}
\hline \hline
 \multirow{2}{*}{parameter} & \multirow{2}{*}{central}
 & \multicolumn{2}{c|}{nominal} & \multicolumn{2}{c}{wide}
\\
 & & $1\,\sigma$ & support & $1\,\sigma$ & support
\\
\hline\hline
\tabvsptop\tabvspbot
  $\zeta_{V,A_1,A_2}$
  & $1.0$
  & $0.15$              & $3\,\sigma$
  & $0.45$              & $3\,\sigma$
\\
\hline
\tabvsptop
  $f_+(0)$
  & $0.34$
  & $[0.32 ,\, 0.39]$   & $[0.28 ,\, 0.49]$
  & $[0.28 ,\, 0.49]$   & $[0.0  ,\, 0.79]$
\\
\tabvspbot
  $b_1^+$
  & $-2.1$
  & $[-3.7 ,\, -1.2]$   & $[-6.9 ,\, 0.6]$
  & $[-6.9 ,\, 0.6]$    & $[-10  ,\, 3.7]$
\\
\hline\hline
\tabvsptop \tabvspbot
  $f_{B_s}$
  & $227.7 \MeV$
  & $6.2 \MeV$        & $3\,\sigma$
  & $18.6 \MeV$        & $3\,\sigma$
\\
\hline\hline
\tabvsptop\tabvspbot
  $\zeta_{K^*}^{ij}$, $\zeta_K^{}$
  & $1.0$
  & $0.15$              & $3\,\sigma$
  & $0.45$              & $[0.0,\, 2.0]$
\\
\hline
\tabvsptop\tabvspbot
  $|r_{0,\perp,\parallel}|$, $|r_K|$
  & $0.0$
  & $0.15$              & $3\,\sigma$
  & $0.45$              & $3\,\sigma$
\\
\hline \hline
\end{tabular}
\caption{Priors of the nuisance parameters of the $B \to K^{(*)}$ form factors,
  the $B_s$ decay constant $f_{B_s}$, and parametrization of lacking subleading
  corrections at low $q^2$ ($i=L,R$ and $j=0,\perp,\parallel$) and high $q^2$,
  specified for the nominal and wide set. All priors are Gaussian
  and we give the central value, its standard deviation $\sigma$, and the
  support of the prior. The nominal $1\,\sigma$ ranges of $V$ and $A_{1,2}$ correspond
  to uncertainties quoted in \cite{Ball:2004rg}, whereas, $f_+(0)$ and $b_1^+$ are
  taken from the LCSR analysis \cite{Khodjamirian:2010vf}; however, possible
  correlations among both are not available.
\label{tab:nuisance:FF}
}
\end{table}

%
\subsection{Subleading $\Lambda/m_b$ Corrections \label{app:SL:corrections}}

There are several distinct sources of $\Lambda/m_b$ corrections arising in
exclusive $B \to K^{(*)}\bar\ell\ell$ decays. Here $\Lambda$ is assumed to be of
the order of the strong scale, however the particular physical meaning depends
on the framework. When using power counting we use the generic value of 500 MeV.

The first type is due to the form factor relations in the limit of heavy quark
masses \cite{Isgur:1990kf}, which is valid for the whole $q^2$-kinematic region.
At the leading order in $\Lambda/m_b$, they relate the $B\to K^*$ ($B\to K$)
tensor form factors $T_{1,2,3}$ ($f_T$) to vector $V$ ($f_+$) and axial-vector
$A_{1,2}$ form factors\footnote{The authors \cite{Altmannshofer:2008dz}
  take the viewpoint, that such corrections can be accounted for at low $q^2$,
  if form factor relations are not used in the leading-order contribution (in
  $\Lambda/m_b$ and $\alpha_s$) to the amplitude.}. This approximation receives
a further numerical suppression due to $\wilson[]{7} / \wilson[]{9} \sim {\cal
  O}(0.1)$. The additional large enery limit \cite{Charles:1998dr, Beneke:2000wa}
at low $q^2$ allows us to eliminate another $B\to K^*$ form factor, introducing
an additional subleading uncertainty not suppressed by $\wilson[]{7} /
\wilson[]{9}$. Besides subleading corrections due to the use of form factor
relations, the two distinct expansions in $\Lambda/m_b$, QCDF at low $q^2$ and
the OPE at high $q^2$, introduce a second type at the amplitude level, when
truncating the expansion after the leading order in $\Lambda/m_b$.

At low $q^2$, QCDF (or equivalently SCET) provides a possibility to calculate
such corrections, which are in general suppressed by a factor of
$\Lambda/m_b$\footnote{In some subleading corrections one encounters infrared
  divergences \cite{Feldmann:2002iw, Kagan:2001zk}.}. In principle, the partially known corrections
\cite{Feldmann:2002iw, Beneke:2004dp} could be included as an estimate of the
lacking corrections, but here we model them by 6 real scale factors for
each of the transversity amplitudes $A_{\perp,\parallel,0}^{L,R}$
in the case of $B\to K^*\bar\ell\ell$ and one
for $B\to K\,\bar\ell\ell$. These scale factors $\zeta^{ij}_{K*}$ ($i=L,R$ and
$j=0,\perp,\parallel$) and $\zeta_K$ have Gaussian prior distributions each with
central value 1 and a $1\,\sigma$ range of $0.15 \approx \Lambda/m_b$ with a
support up to $3\,\sigma$ and include the subleading corrections due to form
factor relations discussed above. A $1\,\sigma$ range of $0.45 \approx \Lambda/m_b$
with a support $[0.0,\, 2.0]$ is chosen for the wide-prior scenario.

At high $q^2$, the interaction of the 4-quark operators and the electromagnetic
current, which couples to the pair of leptons, might be treated within a local
operator product expansion either in full
QCD \cite{Beylich:2011aq} or with subsequent matching on HQET
\cite{Grinstein:2004vb}. In both approaches, subleading corrections to the decay
amplitudes arise at $(\Lambda/m_b)^2$ and $\alpha_s\, \Lambda/m_b$, respectively,
which are of similar numerical size. The additional
suppression factor of $\Lambda/m_b$ or $\alpha_s$, yields smaller theory
uncertainties due to omission of subleading corrections at high $q^2$ in
contrast to the low-$q^2$ region. This is also not spoiled by the use of form
factor relations \cite{Grinstein:2004vb, Bobeth:2010wg} for tensor form factors
$T_{1,2,3}$ ($f_T$) due to the accompanying numerical suppression by
$\wilson[]{7} / \wilson[]{9}$, which depends on the new physics contributions.
Note that for both approaches, full QCD and HQET, the subleading corrections
are known in part, and in the future it is conceivable that they can be included
completely.  For example, the unknown subleading form factor arising in
\cite{Beylich:2011aq} could be calculated on the lattice. We follow
\cite{Grinstein:2004vb}, using $\alpha_s(m_b) \sim 0.3$. This gives rise to 3
complex $r_a \sim \Lambda/m_b$ ($a = 0,\perp, \parallel$) for $B\to
K^*\bar\ell\ell$ \cite{Bobeth:2011gi} and one complex $r_K\sim \Lambda/m_b$ for
$B\to K\,\bar\ell\ell$ \cite{Bobeth:2011nj}, which are additive at the level of
the amplitude. We treat the complex-valued subleading nuisance parameters $r_a$
with eight additional real-valued degrees of freedom, using Gaussian priors each
with central value 0, a $1\,\sigma$ range of $0.15 \approx \Lambda/m_b$, and a
support up to $3\,\sigma$ for its magnitude. The accompanying phases have
uniform priors in $[-\pi/2, \pi/2]$.  A three-times-wider $1\,\sigma$ range of $0.45
\approx \Lambda/m_b$ and a support up to $3\,\sigma$ is chosen for the
wide-prior scenario.

The choices are also listed in \reftab{tab:nuisance:FF}.

%
%
\section{Standard Model Predictions \label{app:SM:predictions}}

In this appendix we provide $q^2$-integrated SM predictions for measured and
unmeasured observables, focusing on those low- and high-$q^2$ bins that are
currently used in experimental analysis and are also accessible
to theoretical methods. All quantities are CP averaged and
lepton-mass effects have been taken into account using $\ell = \mu$. The
theory uncertainties have been obtained using the (nominal) prior distributions
of the nuisance parameters. The results are listed in \reftab{tab:SM:low-predictions}
and \reftab{tab:SM:high-predictions} for low and high $q^2$.
The central value corresponds to the mode and the errors to the smallest
$68\%$ interval of the probability distribution obtained with the Monte Carlo
method. The value in parentheses is obtained when setting all nuisance parameters
to the most probable prior value.

At low $q^2$, we do not predict $J_{3,9}$ and associated optimized observables
$A_T^{(2,{\rm im})}$, since they vanish at leading order in QCDF (including
the $\alpha_s$ corrections), although we obtain non-vanishing values due to
the implementation of subleading terms of kinematic origin ($\sim M_{K^*}/M_B$).

At high $q^2$, $J_{7,8,9}$ is zero at leading order in the OPE and when
applying form factor relations, so is $A_T^{({\rm im})}$.
Furthermore, we recall that $F_L$ and $A_T^{(2,3)}$ become short-distance
independent \cite{Bobeth:2010wg} within the framework of the SM operator
basis, and predictions are strongly dependent on the extrapolation of the
form factor results from low $q^2$ obtained using LCSR.

We do not predict $J_{6c}$ since it vanishes in the absence of scalar and
tensor operators.

\begin{table}
\begin{center}
\begin{tabular}{c|cc}
\hline\hline
\tabvsptop \tabvspbot
Observable & $[2.0,\, 4.3]$  & $[1.0,\, 6.0]$
\\
\hline
\tabvsptop \tabvspbot
$\langle \mathcal{B}_{K} \rangle \times 10^{7} \ ^{\dagger}$
& $0.85 \; ^{+ 0.25} _{- 0.13} \; (0.81)$
& $1.85 \; ^{+ 0.54} _{- 0.28} \; (1.75)$
\\
\hline
\tabvsptop
$\langle \mathcal{B}_{K^*} \rangle \times 10^{7} \ ^{\ddagger}$
& $0.69 \; ^{+ 0.77} _{- 0.41} \; (1.05)$
& $1.64 \; ^{+ 1.80} _{- 0.83} \; (2.46)$
\\
$\langle A_{\rm FB} \rangle $
& $0.055 \; ^{+ 0.087} _{- 0.033} \; (0.086)$
& $0.03 \; ^{+ 0.07} _{- 0.02} \; (0.05)$
\\
$\langle F_L \rangle $
& $0.85 \; ^{+ 0.08} _{- 0.20} \; (0.78)$
& $0.81 \; ^{+ 0.09} _{- 0.22} \; (0.73)$
\\[0.2cm]
$\langle J_{1s} \rangle \times 10^{8}$
 & $1.18 \; ^{+ 0.48} _{- 0.35} \; (1.26)$
 & $3.43 \; ^{+ 1.37} _{- 0.95} \; (3.66)$
\\[0.1cm]
$\langle J_{1c} \rangle \times 10^{7}$
& $0.31 \; ^{+ 0.57} _{- 0.29} \; (0.63)$
& $0.83 \; ^{+ 1.07} _{- 0.76} \; (1.37)$
\\[0.1cm]
$\langle J_{2s} \rangle \times 10^{8}$
& $0.39 \; ^{+ 0.16} _{- 0.12} \; (0.42)$
& $1.13 \; ^{+ 0.45} _{- 0.31} \; (1.21)$
\\[0.1cm]
$\langle J_{2c} \rangle \times 10^{7}$
& $-0.30 \; ^{+ 0.28} _{- 0.56} \; (-0.61)$
 &$-0.79 \; ^{+ 0.75} _{- 1.05} \; (-1.33)$
\\[0.1cm]
$\langle J_4 \rangle \times 10^{8}$
& $0.57 \; ^{+ 0.39} _{- 0.24} \; (0.77)$
& $1.43 \; ^{+ 0.82} _{- 0.62} \; (1.82)$
\\[0.1cm]
$\langle J_5 \rangle \times 10^{8}$
 & $-0.69 \; ^{+ 0.37} _{- 0.64} \; (-1.07)$
 & $-1.80 \; ^{+ 0.88} _{- 1.37} \; (-2.58)$
\\[0.1cm]
$\langle J_{6s} \rangle \times 10^{8}$
 & $0.84 \; ^{+ 0.45} _{- 0.29} \; (0.90)$
 & $1.19 \; ^{+ 0.87} _{- 0.74} \; (1.21)$
\\[0.1cm]
$\langle J_7 \rangle \times 10^{9}$
& $2.52 \; ^{+ 1.50} _{- 1.06} \; (2.78)$
& $5.86 \; ^{+ 3.03} _{- 2.62} \; (6.21)$
\\[0.1cm]
$\langle J_8 \rangle \times 10^{9}$
& $-0.89 \; ^{+ 0.49} _{- 0.57} \; (-0.97)$
& $-1.79 \; ^{+ 0.94} _{- 1.36} \; (-2.14)$
\\[0.2cm]
$\langle A_T^{(3)} \rangle $
& $0.45 \; ^{+ 0.12} _{- 0.08} \; (0.50)$
& $0.42 \; ^{+ 0.11} _{- 0.08} \; (0.47)$
\\[0.1cm]
$\langle A_T^{(4)} \rangle $
& $0.63 \; ^{+ 0.17} _{- 0.17} \; (0.69)$
& $0.64 \; ^{+ 0.18} _{- 0.15} \; (0.71)$
\\[0.1cm]
$\langle A_T^{(5)} \rangle $
& $0.41 \; ^{+ 0.03} _{- 0.05} \; (0.42)$
& $0.48 \; ^{+ 0.01} _{- 0.03} \; (0.48)$
\\[0.1cm]
$\langle A_T^{(\rm re)} \rangle $
& $0.61 \; ^{+ 0.10} _{- 0.13} \; (0.54)$
& $0.29 \; ^{+ 0.14} _{- 0.14} \; (0.25)$
\\[0.2cm]
$\langle H_T^{(1)} \rangle $
& $0.45 \; ^{+ 0.08} _{- 0.08} \; (0.48)$
& $0.42 \; ^{+ 0.07} _{- 0.07} \; (0.45)$
\\[0.1cm]
$\langle H_T^{(2)} \rangle $
& $-0.29 \; ^{+ 0.08} _{- 0.08} \; (-0.34)$
& $-0.29 \; ^{+ 0.07} _{- 0.07} \; (-0.33)$
\\[0.2cm]
\hline\hline
\end{tabular}
\end{center}

\caption{SM predictions of $q^2$-integrated observables at 
  low-$q^2$ in the bins $q^2 \in [q^2_{\rm min}, q^2_{\rm max}]$ for 
  $^{\dagger} B^-\to K^-\bar\mu\mu$ and 
  $^{\ddagger} \bar{B}^0\to \bar{K}^{*0}\bar\mu\mu$.
  We list the mode and the smallest $68\%$ interval of the probability
  distribution, along with the value obtained by the conventional method of
  setting all nuisance parameters to the prior modes (in parentheses).
  \label{tab:SM:low-predictions}}
\end{table}

\begin{table}
\begin{center}
\resizebox{\textwidth}{!}{
\begin{tabular}{c|ccc}
\hline\hline
\tabvsptop \tabvspbot
Observable & $[14.18,\, 16.0]$  &  $[>16.0] $ & $[>14.18]$
\\
\hline
\tabvsptop \tabvspbot
$\langle \mathcal{B}_{K} \rangle \times 10^{7} \ ^{\dagger}$
& $0.39 \; ^{+ 0.22} _{- 0.09} \; (0.37)$
& $0.73 \; ^{+ 0.43} _{- 0.22} \; (0.68)$
& $1.11 \; ^{+ 0.66} _{- 0.28} \; (1.04)$
\\
\hline
\tabvsptop
$\langle \mathcal{B}_{K^*} \rangle \times 10^{7} \ ^{\ddagger}$
& $1.19 \; ^{+ 0.37} _{- 0.31} \; (1.26)$
& $1.41 \; ^{+ 0.40} _{- 0.38} \; (1.46)$
& $2.57 \; ^{+ 0.80} _{- 0.68} \; (2.72)$
\\
$\langle A_{\rm FB} \rangle $
& $-0.44 \; ^{+ 0.07} _{- 0.07} \; (-0.44)$
& $-0.37 \; ^{+ 0.06} _{- 0.07} \; (-0.38)$
& $-0.40 \; ^{+ 0.06} _{- 0.07} \; (-0.41)$
\\
$\langle F_L \rangle $
& $0.38 \; ^{+ 0.04} _{- 0.06} \; (0.36)$
& $0.35 \; ^{+ 0.02} _{- 0.03} \; (0.34)$
& $0.36 \; ^{+ 0.04} _{- 0.05} \; (0.35)$
\\[0.2cm]
$\langle J_{1s} \rangle \times 10^{8}$
 & $4.44 \; ^{+ 0.96} _{- 1.00} \; (4.51)$
 & $5.10 \; ^{+ 1.48} _{- 1.11} \; (5.44)$
 & $9.70 \; ^{+ 2.31} _{- 2.21} \; (9.96)$
\\[0.1cm]
$\langle J_{1c} \rangle \times 10^{8}$
& $3.23 \; ^{+ 1.31} _{- 1.37} \; (3.43)$
& $3.40 \; ^{+ 1.41} _{- 1.07} \; (3.72)$
& $6.64 \; ^{+ 2.75} _{- 2.43} \; (7.14)$
\\[0.1cm]
$\langle J_{2s} \rangle \times 10^{8}$
& $1.48 \; ^{+ 0.32} _{- 0.33} \; (1.50)$
& $1.70 \; ^{+ 0.49} _{- 0.37} \; (1.81)$
& $3.23 \; ^{+ 0.77} _{- 0.74} \; (3.31)$
\\[0.1cm]
$\langle J_{2c} \rangle \times 10^{8}$
& $-3.21 \; ^{+ 1.36} _{- 1.31} \; (-3.41)$
 &$-3.38 \; ^{+ 1.07} _{- 1.41} \; (-3.70)$
 &$-6.61 \; ^{+ 2.42} _{- 2.74} \; (-7.11)$
\\[0.1cm]
$\langle J_3 \rangle \times 10^{8}$
& $-0.99 \; ^{+ 0.59} _{- 0.71} \; (-1.11)$
& $-2.12 \; ^{+ 0.89} _{- 0.82} \; (-2.19)$
& $-3.06 \; ^{+ 1.44} _{- 1.57} \; (-3.29)$
\\[0.1cm]
$\langle J_4 \rangle \times 10^{8}$
& $2.47 \; ^{+ 0.95} _{- 0.85} \; (2.65)$
& $3.10 \; ^{+ 1.08} _{- 0.96} \; (3.27)$
& $5.49 \; ^{+ 2.06} _{- 1.77} \; (5.92)$
\\[0.1cm]
$\langle J_5 \rangle \times 10^{8}$
 & $-3.36 \; ^{+ 0.87} _{- 0.87} \; (-3.54)$
 & $-2.95 \; ^{+ 0.63} _{- 0.80} \; (-3.17)$
 & $-6.23 \; ^{+ 1.34} _{- 1.79} \; (-6.72)$
\\[0.1cm]
$\langle J_{6s} \rangle \times 10^{7}$
 & $-0.52 \; ^{+ 0.10} _{- 0.12} \; (-0.55)$
 & $-0.53 \; ^{+ 0.11} _{- 0.12} \; (-0.56)$
 & $-1.05 \; ^{+ 0.22} _{- 0.24} \; (-1.11)$
\\[0.2cm]
$\langle A_T^{(2)} \rangle $
& $-0.38 \; ^{+ 0.17} _{- 0.18} \; (-0.37)$
& $-0.64 \; ^{+ 0.15} _{- 0.10} \; (-0.60)$
& $-0.51 \; ^{+ 0.16} _{- 0.16} \; (-0.50)$
\\[0.1cm]
$\langle A_T^{(3)} \rangle $
& $1.45 \; ^{+ 0.29} _{- 0.31} \; (1.47)$
& $1.95 \; ^{+ 0.42} _{- 0.40} \; (2.01)$
& $1.67 \; ^{+ 0.36} _{- 0.34} \; (1.72)$
\\[0.1cm]
$\langle A_T^{(4)} \rangle $
& $0.66 \; ^{+ 0.14} _{- 0.14} \; (0.67)$
& $0.48 \; ^{+ 0.10} _{- 0.10} \; (0.48)$
& $0.56 \; ^{+ 0.12} _{- 0.11} \; (0.57)$
\\[0.1cm]
$\langle A_T^{(5)} \rangle $
& $0.085 \; ^{+ 0.008} _{- 0.008} \; (0.081)$
& $0.111 \; ^{+ 0.014} _{- 0.014} \; (0.109)$
& $0.123 \; ^{+ 0.012} _{- 0.012} \; (0.120)$
\\[0.1cm]
$\langle A_T^{(\rm re)} \rangle $
& $-0.982 \; ^{+ 0.110} _{- 0.003} \; (-0.915)$
& $-0.777 \; ^{+ 0.099} _{- 0.089} \; (-0.767)$
& $-0.843 \; ^{+ 0.075} _{- 0.087} \; (-0.834)$
\\[0.2cm]
$\langle H_T^{(1)} \rangle $
& $0.9996 \; ^{+ 0.0002} _{- 0.0003} \; (0.9996)$
& $0.9986 \; ^{+ 0.0008} _{- 0.0007} \; (0.9986)$
& $0.9970 \; ^{+ 0.0017} _{- 0.0018} \; (0.9969)$
\\[0.1cm]
$\langle H_T^{(2)} \rangle $
& $-0.9844 \; ^{+ 0.0027} _{- 0.0020} \; (-0.9853)$
& $-0.9719 \; ^{+ 0.0034} _{- 0.0024} \; (-0.9722)$
& $-0.9748 \; ^{+ 0.0040} _{- 0.0031} \; (-0.9751)$
\\[0.1cm]
$\langle H_T^{(3)} \rangle $
& $-0.9837 \; ^{+ 0.0024} _{- 0.0018} \; (-0.9845)$
& $-0.9614 \; ^{+ 0.0017} _{- 0.0011} \; (-0.9618)$
& $-0.9606 \; ^{+ 0.0018} _{- 0.0016} \; (-0.9613)$
\\[0.2cm]
\hline\hline
\end{tabular}
}
\end{center}

\caption{SM predictions of $q^2$-integrated observables
  at high-$q^2$ in the bins $q^2 \in [q^2_{\rm min}, q^2_{\rm max}]$ for
  $^{\dagger} B^-\to K^-\bar\mu\mu$ and 
  $^{\ddagger} \bar{B}^0\to \bar{K}^{*0}\bar\mu\mu$.
  We list the mode and the smallest $68\%$ interval of the probability
  distribution, along with the value obtained by the conventional method of
  setting all nuisance parameters to the prior modes (in parentheses).
\label{tab:SM:high-predictions}}
\end{table}

%
%
\section{Distributions \label{app:distributions}}

%
\subsection{$\Amoroso$ Distribution \label{sec:amoroso-distribution}}

Consider the posterior $P(x|D)$, describing the search
for a decay whose existence has not been established yet,
with $x$ representing the branching ratio.
Suppose we know  $P(x|D)$ at a number of
of data points, $(x_i,P_i)$.
Using the cumulative distribution function
\begin{equation}
F(x_a|D) = \int_{-\infty}^{x_a} \mbox{d}x P(x|D),
\end{equation}
we can determine the limit $x_a$
at level $a$ from $F(x_a|D) = a$.
For convenience, we seek an analytical
expression $g(x)$ interpolating the data points.
We constrain $g(\cdot)$ by requiring that it vanish for
negative branching ratios and that it yield
the same $10 [50, 90]\%$ limits as obtained from $F(\cdot|D)$:
\begin{align}
   g(x \le 0) &= 0 \label{eq:1}\\
   \int_0^{x_{a}} \mbox{d}x\ g(x) &= a, \quad  a=0.1,0.5,0.9 \, .\label{eq:2}
\end{align}
We choose  $g(x) =  \Amoroso(x| l, \lambda, \alpha, \beta)$.
The Amoroso family \cite{Crooks:2010} is a continuous unimodal
four-parameter family of probability distributions that easily
accommodates the constraints and provides an accurate approximation.
Many well known distributions are direct members or appear as limits of the
Amoroso family. Its functional form is
\begin{align}
\label{Amoroso}
 \Amoroso(x| l, \lambda, \alpha, \beta)
&=
\frac{1}{\Gamma(\alpha)}
 \left|\frac{\beta}{\lambda}\right|
 \left(\frac{x-l}{\lambda}\right)^{\alpha \beta -1}
 \exp \left[ -  \left(\frac{x-l}{\lambda}\right)^{\beta} \right]
\\ \notag
& \mbox{for } x,\ l,\ \lambda,\ \alpha,\ \beta\  \in \mathbb{R},
\ \alpha>0, \
\\ \notag
& \mbox{support } x \geq l \ \mbox{if}\ \lambda > 0,  \ x\leq l  \ \mbox{if}\  \lambda < 0 .
\end{align}
We set the location parameter $l$ to the minimum physical value,
$l=0$, and ensure that $\lambda>0$ to satisfy \eqref{eq:1}.
The scale parameter $\lambda$ and the shape parameters $\alpha$ and $\beta$ are found
by numerically solving the set of three equations
\eqref{eq:2}.
In the limit of $\alpha \to \infty$ and $\beta = 1$, $\Amoroso(\cdot)$ converges to a Gaussian distribution \cite{Crooks:2010}.

%
%
\subsection{$\LogGamma$ Distribution \label{sec:logg-distr}}

Consider a nuisance parameter $\nu$ whose reported uncertainties are asymmetric,
$
\nu = \mu^{+b}_{-a}, a \ne b.
$
In this case, we use the LogGamma distribution \cite{Crooks:2010} to
obtain a continuous prior over the given range of $\nu$. The LogGamma family is
a continuous unimodal three-parameter family of probability distributions
\begin{align}
 \label{LogGamma}
 \LogGamma(\nu|l, \lambda, \alpha)
 &=
 \frac{1}{ \Gamma(\alpha) |\lambda|}
 \exp\left( \alpha \left(\frac{\nu-l}{\lambda}\right) - \exp\left(\frac{\nu-l}{\lambda}\right)  \right)
\\ \notag
& \qquad \text{for } \nu,\ l,\ \lambda,\ \alpha\ \in \mathbb{R},
\ \alpha>0, \
\\ \notag
& \qquad \text{support } -\infty \leq \nu \leq \infty.
\notag
\end{align}
The three parameters are uniquely fixed by demanding that the mode of $P(\nu)$
be at $\mu$, that the interval $[\mu - a, \mu + b]$ contain $68\%$, and that the density
be identical at $\mu - a$ and $\mu + b$.
 More concisely, we have three conditions:
\begin{align}
  \arg \max_{\nu} \ P(\nu) & = \mu \\
  \int_{\mu - a}^{\mu + b} {\rm d}\nu P(\nu) & = 0.68 \label{lg_constr:2} \\
  P(\mu - a) & = P(\mu + b). \label{lg_constr:3}
\end{align}
While the first constraint is used to fix the location parameter $l$,
the scale parameter $\lambda$ and the shape parameter $\alpha$ must be extracted numerically
by solving the coupled equations \eqref{lg_constr:2} and \eqref{lg_constr:3}.
For a finite range of $\nu$, say $[\nu_{min}, \nu_{max}]$, the resulting density
is normalized such that $\int_{\nu_{min}}^{\nu_{max}} {\rm d}\nu P(\nu)=1$.

The asymmetry is governed by $\alpha$: $\LogGamma(\cdot)$ approaches a Gaussian distribution
in the limit $\alpha \to \infty$.

%
%

\end{document}